\documentstyle[prl,aps,epsfig,twocolumn]{revtex}
\begin{document}

\twocolumn[\hsize\textwidth\columnwidth\hsize\csname
@twocolumnfalse\endcsname

%\maketitle
\title{\LARGE \bf Neutron scattering investigations on \\ methyl group dynamics in polymers.\vspace{1.5 mm}}


\author{\Large Juan Colmenero$^{1,2,3}$ \and Angel J. Moreno$^{1}$
\and Angel Alegr\'{\i}a$^{2,3}$ \vspace{6 mm}}
%\vspace{6 mm}

\address{$^1$Donostia International Physics Center, Paseo Manuel de Lardizabal 4,
20018 San Sebasti\'{a}n, Spain.\\
$^2$Departamento de F\'{\i}sica de Materiales, Universidad del Pa\'{\i}s Vasco (UPV/EHU),\\ 
Apdo. 1072, 20080 San Sebasti\'{a}n, Spain.\\
$^3$Unidad de F\'{\i}sica de Materiales, Centro Mixto CSIC-UPV, 
Apdo. 1072, 20080 San Sebasti\'{a}n, Spain. \\}
%
%
%\affiliation{Donostia International Physics Center, Paseo Manuel de Lardizabal 4,
%20018 San Sebasti\'{a}n, Spain}
%
%\author{}
%\affiliation{Departamento de F\'{\i}sica de Materiales, Universidad del Pa\'{\i}s Vasco (UPV/EHU), 
%Apdo. 1072, 20080 San Sebasti\'{a}n, Spain}
%\affiliation{Unidad de F\'{\i}sica de Materiales, Centro Mixto CSIC-UPV 
%Apdo. 1072, 20080 San Sebasti\'{a}n, Spain}
%\maketitle
%
\maketitle
%
%\begin{minipage}
\begin{abstract}
%\onecolumn
%\nopagebreak[2]
%\onecolumn
%
{\bf Abstract}
\\\\
Among the different dynamical processes that take place in polymers, methyl group
rotation is perhaps the simplest one, since all the relevant interactions on the methyl
group can be condensed in an effective mean-field one-dimensional potential. Recent
experimental neutron scattering results have stimulated a new revival of the interest on
methyl group dynamics in glasses and polymer systems. The existence of quantum rotational
tunnelling of methyl groups in polymers was expected for a long time but only very
recently (1998), these processes have been directly observed by high-resolution neutron
scattering techniques. This paper revises and summarizes the work done on this topic over
last ten years by means of neutron scattering. It is shown that the results obtained in
many chemically and structurally different polymers can be consistently
described in the whole temperature range  --- from the quantum tunnelling limit
to the classical hopping regime --- as well as in the librational spectrum,
in terms of the Rotation Rate Distribution Model (RRDM), which was first
proposed in 1994. This model introduces a distribution of potential barriers for methyl
group rotation, which is associated to the disorder present in any structural glass. The
molecular and structural origin of the barrier distribution in polymers is discussed on the basis of a
huge collection of investigations reported in the literature, including recent fully
atomistic molecular dynamics simulations.
%This article reviews in a comprehensive form recent advances in the application of
%neutron scattering techniques to investigate one of the, in principle, simplest dynamic processes one
%can define in polymer systems: methyl group rotation. In particular we emphasize the progress
%achieved in this topic over last 10 years after the proposal in 1994 of the
%Rotation Rate Distribution Model (RRDM) for methyl group dynamics in polymers,
%and the first direct experimental observation in 1998 of quantum rotational tunnelling
%of methyl groups in polymers. The grounds of the RRDM,
%which introduces a distribution of potential barriers for methyl group rotation,
%are here exposed in detail. It is shown that the RRDM provides a consistent description, 
%at all the temperature range, for the experimentally observable features of methyl group dynamics:
%the crossover from quantum rotational tunnelling to thermally activated hopping, and the
%librational energies. The molecular and structural origin of the barrier distribution is discussed
%on the basis of a huge collection of investigations reported in the literature. 
\vspace{2 mm}
%\twocolumn
%\nopagebreak[2]
\end{abstract}
\small
\begin{center}
\end{center}
]
\vspace{3 mm}
\small
\twocolumn[\hsize\textwidth\columnwidth\hsize\csname
@twocolumnfalse\endcsname

{\bf Contents} 
\\\\
\ref{sec:intro}. Introduction \\
\\
\ref{sec:neutscat}. Neutron scattering techniques and instrumentation \\
\mbox{  }\ref{sec:neutscat}A.  Neutron scattering techniques \\
\mbox{  }\ref{sec:neutscat}B.  Neutron scattering instrumentation \\
\mbox{  }\ref{sec:neutscat}C.  Final experimental considerations \\
\\
\ref{sec:theorcryst}. Methyl group dynamics in crystalline systems: 
Theory and scattering functions \\
\mbox{  }\ref{sec:theorcryst}A.  Rotational tunnelling \\
\mbox{  }\ref{sec:theorcryst}B.  Crossover from rotational tunnelling to classical hopping\\
\mbox{  }\ref{sec:theorcryst}C.  Classical hopping \\
\mbox{  }\ref{sec:theorcryst}D.  Additional remarks \\
\\
\ref{sec:rrdm}. Methyl group dynamics in polymer systems \\
\mbox{  }\ref{sec:rrdm}A. Rotation Rate Distribution Model (RRDM) \\
\mbox{  }\ref{sec:rrdm}B. Other approaches \\
\\
\ref{sec:pvac}. Application of the RRDM to a showcase: poly(vinyl acetate) \\
\\
\ref{sec:summexper}. Summary of experimental results by neutron scattering on polymers \\
\mbox{  }\ref{sec:summexper}A. Classical hopping \\
\mbox{  }\ref{sec:summexper}B. Rotational tunnelling and crossover to hopping\\
\mbox{  }\ref{sec:summexper}C. Librational levels\\
\\
\ref{sec:origin}. Molecular origin of the barrier distribution for methyl group dynamics \\
\mbox{  }\ref{sec:origin}A. Influence of the chain conformation \\
\mbox{  }\ref{sec:origin}B. Influence of mixing with other materials \\
\mbox{  }\ref{sec:origin}C. Molecular glasses and comparison with the crystalline state \\
\mbox{  }\ref{sec:origin}D. Molecular dynamics simulations \\
\\
\ref{sec:conclusions}. Conclusions and outlook \\
\\
Acknowledgements \\
\\
References
\vspace{1 cm}
]

\normalsize
%\end{minipage}
%\maketitle
%
%\begin{keyword}
%\PACS 
%\end{keyword}
%\end{frontmatter}
%
% main text
%
%\maketitle
\newpage
{\bf Nomenclature}
\begin{table}
\begin{tabular}{ll}
BS  &  Backscattering \\
EISF  & Elastic incoherent structure factor \\
FEW &  Fixed elastic window \\
HWHM & Half-width at half-maximum \\
INS & Inelastic neutron scattering \\
MDS  &  Molecular dynamics simulations \\
NMR & Nuclear magnetic resonance \\
QENS & Quasielastic neutron scattering \\
RRDM  &  Rotation Rate Distribution Model \\
TOF  & Time of flight \\
VDOS &  Vibrational density of states \\
$A,E_a, E_b$  & Sublevels of the librational ground state \\
$A(Q)$ & Elastic incoherent structure factor \\
$B$  & Rotational constant \\
$E_A$ & Classical activation energy \\
$\langle E_A \rangle$ &  Average barrier of $f(E_A)$\\
$E_{01}$   & First librational energy \\
%\end{tabular}
%
%\begin{tabular}{ll}
%$e^{-2W(Q)}$ & Debye-Waller factor \\
$F(E_{01})$  &  Distribution of $E_{01}$ \\
$f(E_A)$   &  Distribution of $E_A$ \\
$G(T_c)$  &  Distribution of $T_c$ \\
$g(V_3)$  & Distribution of $V_3$ \\
$H(\log\Gamma)$  &  Distribution of $\log\Gamma$ \\
$h(\hbar\omega_t)$  & Distribution of $\omega_t$ \\
$I_{inc}(Q,t)$ & Intermediate incoherent scattering function \\
$L(\omega,\Gamma)$  & Normalized Lorentzian function \\
$\log\Gamma_0$  &  Average of $H(\log\Gamma)$  \\
$S(Q)$ &  Elastic coherent structure factor \\
$S_{coh}(Q,\omega)$ & Coherent scattering function \\
$S_{inc}(Q,\omega)$ & Incoherent scattering function \\
$S_{inc}^{MG}(Q,\omega)$ & Incoherent scattering function \\ \ & for methyl group rotation \\
$T_c$  & Crossover temperature \\
$V_3$   & Threefold barrier height \\
$\langle V_3 \rangle$  & Average barrier of $g(V_3)$\\
$W_3^c$ &  Coupling potential between methyl groups \\ 
$Z(\omega)$ & Generalized VDOS \\
%\end{tabular}
%
%\begin{tabular}{ll}
$\hbar{\bf Q}$ & Neutron momentum transfer \\
$\hbar\omega$ & Neutron energy transfer \\
$\Gamma$  &  Lorentzian HWHM for classical hopping \\
$\Gamma_{AE}$  &  Inelastic crossover Lorentzian HWHM \\
$\gamma_{AE}$  &  Preexponential factor for $\Gamma_{AE}$ \\
$\Gamma_{E_a E_b}$  & Quasielastic crossover Lorentzian HWHM \\
$\gamma_{E_aE_b}$  & Preexponential factor for $\Gamma_{E_a E_b}$ \\
$\gamma_b$  &  RRDM preexponential factor for $\Gamma_{AE}$ and $\Gamma_{E_a E_b}$ \\
$\gamma_s$  & RRDM preexponential factor for $\Delta\omega_t$ \\
$\Gamma_{\infty}$ &  Preexponential factor for classical hopping \\ 
$\Delta\omega_t$ &  Shift of the tunneling frequency \\
$\sigma$  &  Standard deviation of $H(\log\Gamma)$  \\
$\sigma_E$  & Standard deviation of $f(E_A)$\\
$\sigma_V$  & Standard deviation of $g(V_3)$\\
$\sigma_{coh}$ & Coherent cross-section \\
$\sigma_{inc}$  & Incoherent cross-section \\
$\tau$  & Residence time for classical hopping \\
$\omega_t$  & Tunneling frequency \\
%\end{tabular}
%
%\begin{tabular}{ll}
%
PC   & Polycarbonate \\
PDMS & Poly(dimethyl siloxane) \\
PEO  & Poly(ethylene oxide) \\
PEP  & Poly(ethylene propylene) \\
PH   & Phenoxy \\
PI   & Polyisoprene \\
PIB  & Polyisobutylene \\
PMMA & Poly(methyl methacrylate) \\
PMPS & Poly(methyl phenyl siloxane) \\
PP   & Polypropylene \\
hh-PP & Head-to-head Poly(propylene) \\
PPO  & Poly(propylene oxide) \\
PS   & Polystyrene \\
PSF  & Polysulfone \\
PVAc & Poly(vinyl acetate) \\
PVME & Poly(vinyl methyl ether) \\
P$\alpha$MS & Poly($\alpha$-methylstyrene) \\
SCPE & Solution chlorinated polyethylene \\

\end{tabular}
\end{table}
\newpage

\section{Introduction}
\label{sec:intro}

\begin{figure}
\begin{center}
\includegraphics[width=0.46\textwidth]{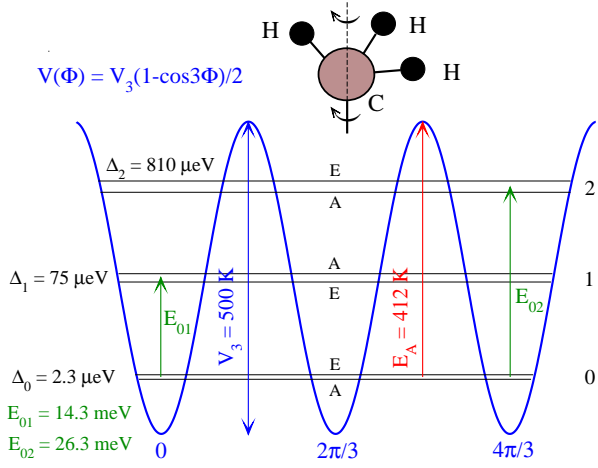}
\newline
\newline
\caption{Picture of methyl group rotation about its $C_3$-axis,
and level scheme for a purely threefold potential with barrier height $V_{3} =$ 500 K.}
\label{fig:levels}
\end{center}
\end{figure}

The different dynamic processes present in amorphous polymers
cover a extremely wide time scale, spanning from ca. $10^{-13}$ s to years
(see, e.g., Refs. \cite{mccrum,bailey,doiedwards,spinecho}). These processes include terminal relaxations,
conformational rearrangements, segmental dynamics, localized Johari-Goldstein relaxations,
rotations of side groups, or fast vibrational dynamics.  
One of the simplest processes among the latters is methyl group, -CH$_3$, rotation.
Many natural and synthetic macromolecular chains contain this simple molecular unit as a side
group or as a part of more complicated side groups. In the glassy state of polymeric materials, at
temperatures well below the glass transition region, one can assume as a good approximation that
the main-chain dynamics are completely frozen-in and only small side
units, as the methyl groups, can still remain mobile. In such conditions, 
the interaction between the methyl group and its environment is often well approximated by an effective
mean-field or "single-particle" rotational potential \cite{pragerreview,pressbook,carlilereview}.
The methyl group can be regarded as a rigid rotor
because the strength of the C-H covalent bonds allows one to neglect the
internal degrees of freedom in comparison with both
translational and rotational motions of the group as a whole.
Hence, the single-particle potential only depends on one characteristic angular coordinate $\phi$,
which is measured in the plane perpendicular to the C$_3$-symmetry axis of the methyl group.
The latter corresponds to the bond joining the carbon to the rest
of the molecule (see Fig. \ref{fig:levels}). 

In the simplest approximation (see Section \ref{sec:theorcryst}) 
the rotational potential is threefold: 

\begin{equation}
V(\phi)=\frac{V_{3}}{2}(1-\cos3\phi) .
\end{equation}
Note that, in this form, $V_3$ corresponds to both the maximum and the amplitude
of the potential $V(\phi)$.
Figure \ref{fig:levels} shows the quantized energy levels of a threefold potential
with barrier height $V_{3}=500$ K (in the following $V_3$ will be given as $V_3/k_B$, with $k_B$
the Boltzmann constant). Such levels are obtained \cite{pragerreview,pressbook,carlilereview} by solving
the corresponding stationary Schr\"{o}dinger equation, which in the case
of a threefold potential takes the form of the well-known Mathieu equation
(see, e.g., Ref. \cite{handmath}). The energy levels $i=0,1,2...$
of the individual potential wells are named {\it torsional} or {\it librational} levels, 
and are split (with energy splitting $\Delta_i$) due to the coupling between the single-well wavefunctions.
Three quantities characterizing the methyl group rotational
potential are experimentally accessible (see below):
i) the energies $E_{0i}$ of the transitions between the librational levels, 
specially the transition $E_{01}$ between the ground and first excited level, 
ii) the energy splitting of the ground librational state, $\Delta_0 =\hbar\omega_t$, with $\omega_t$ 
the quantum {\it tunnelling frequency}, iii) the activation energy $E_A$ for
methyl group classical hopping between adjacent wells, defined as the diference between the
top of the barrier and the ground state (see Figure \ref{fig:levels}). 
The two former quantities can be directly accessed
by inelastic spectroscopic techniques. The latter is indirectly derived from the temperature
dependence of the hopping rate, which can be determined by quasielastic spectroscopy
or by relaxation techniques (see below).

According to the scheme shown in Figure \ref{fig:levels},
methyl group dynamics at very low temperature (typically $T \lesssim 20$ K) 
is dominated by quantum rotational tunnelling, whereas classical hopping
processes over the rotational barrier control the non-vibrational dynamics at high
temperature (typically $T \gtrsim 80$ K). Due to the one-dimensional
character of the rotational potential, methyl group dynamics is one of
the simplest examples of barrier penetration phenomena involving the interaction
of a quantum system with lattice vibrations. For that reason, it has attracted a great
theoretical interest as an ideal case for the understanding of the fundamental problem
of the transition from quantum tunnelling to classical hopping (see, e.g,
Refs. \cite{huller80,clough82jpc,hewson1,hewson2,wurger89zp,clough93pra}).
Moreover, due to the high sensivity of the rotational potential
to the chemical and geometrical nature of the environment, and to the relative
simplicity of their dynamics, methyl groups can be used as built-in probes
for exploring structural and dynamic properties of the host material.

Dynamics, and in particular rotational tunnelling, of methyl groups has been widely
investigated in molecular crystals by different techniques over last decades.
(see, e.g., \cite{pragerreview,pressbook,carlilereview}). 
Measurements of tunnelling frequencies of methyl groups in about 200 systems
of very different chemical nature have been compiled in Ref. \cite{pragerreview}.
The used techniques include mostly neutron scattering and nuclear magnetic resonance (NMR) methods
(see, e.g., Ref. \cite{reviewnmr} for a review on NMR measurements),
but also calorimetry, electron nuclear double resonance, infrared or hole burning optical spectroscopy. 
However, no method yields such a direct model independent insight as inelastic neutron scattering (INS),
where the tunnelling frequency is directly observed as two resolution-width
inelastic peaks at $\pm\hbar\omega_t$. Nevertheless, 
it must be stressed that this observation is only possible for moderate
and weak potentials ($V_{3}\lesssim$ 700 K). For stronger potentials $\hbar\omega_t$ is smaller
than the current available instrumental resolution ($\approx 0.3$ $\mu$eV) and cannot be detected 
by neutron scattering spectroscopy. In such cases NMR is the mostly employed technique.

The lowest librational transitions $E_{01}$, $E_{02}$ take values typically in the range
$10-60$ meV and can also be directly detected by INS as resolution-width inelastic peaks.
However, they are observed in the same energy range that the other intramolecular vibrational modes
and the lattice phonons. Though this fact often complicates the identification
of methyl group librational transitions, measurements of the latters are useful
for an accurate determination of the rotational potential, since as the tunnelling
frequency, they strongly depend on the height and shape of the barrier.

Neutron scattering techniques have also been applied over the last decades to investigate
methyl group dynamics in polymers.
It is worth emphasizing that polymer systems are in general highly disordered materials. 
This implies that, in principle, one might expect the existence of
a distribution of potential barriers for methyl group rotation, resulting from the
different local environments present in the system.
Indeed, the experimental librational peaks reported in the first INS investigations
\cite{hig72,allen74} in polymers showed a broad shape, as might be expected
for an inherent distribution of potential barriers. However, this aspect of the problem,
which had already been considered in NMR investigations \cite{con64,sch85}, was ignored for a long time.
Hence, first quasielastic neutron scattering data (QENS, see Section \ref{sec:neutscat})
on classical hopping of methyl groups in polymers 
were analyzed, as for the case of crystalline systems, by assuming
a unique value of the rotational potential. However, this procedure provided, in contrast to 
the result observed in crystalline systems, a non-Arrhenius temperature dependence of the hopping rate,
and an apparent temperature dependence of the geometry of the motion, with large deviations from 
threefold rotation \cite{gab84,gab85,flo92,arrighi95,arrighi96,chahid94,frickpip}.

The presence of quantum tunnelling effects at unusually high temperature was considered
as an unreliable explanation of the observed features, and a possible interpretation, by mantaining the picture
of a unique rotational barrier, was discussed in terms of sixfold symmetry contributions to the
rotational potential \cite{gab84,gab85}. Another explanation was introduced in terms
of temperature dependent fractions of rotating and non-rotating methyl groups \cite{flo92,chahid93}.
Though a reasonable description of the experimental data was achieved, a physical justification 
of the latter hypothesis was lacking. Finally, it became clear that the basic 
assumption of a unique barrier was wrong for amorphous polymers
when experiments on a same system performed in different spectrometers, i.e., 
in different time/energy windows, provided uncompatible results
for the measured hopping rate \cite{arrighi95,arrighi96}.
In the seminal works of Refs. \cite{chahid94} and \cite{frickpip}, QENS data for methyl group 
hopping in polymers were successfully interpreted by introducing a broad distribution of rotational barriers,
resulting from the highly disordered nature of the glassy state.
In last years, a noticeable amount of experiments in different systems have supported this picture 
(see Sections \ref{sec:pvac} and \ref{sec:summexper}).
  
%This effect was clearly seen in ethylbenzene, where the methyl group librational mode was found to
%get drastically broadened for a quenched sample \cite{fri95}.
%Distributions of molecular correlation times were used by Connor \cite{con64}, among others, to describe
%the results obtained by NMR measurements and in particular by Schmidt et al. to interpret deuteron
%NMR data of methyl group reorientation in polycarbonate \cite{sch85}.
%In the case of QENS measurements considering the  disorder effects has been demonstrated crucial
%to provide a consistent description of the experiments looking for the methyl group hopping performed using
%different spectrometers on the same polymer (see below for more details).

Another aspect of the problem of methyl group dynamics in polymers is the 
dynamic behavior at very low temperature. From the early times, there was
in the literature a number of claims or predictions about low temperature quantum effects.
In particular, NMR measurements by Hoch {\it et al.} \cite{hoch71} showed
methyl group dynamics in poly(vinyl acetate), PVAc, to be active down to very low temperature. 
Rotational tunnelling of methyl groups was also claimed to be present in poly(methyl methacrylate),
PMMA, which would account for the reported non-Arrhenius behavior of the mechanical relaxation
time at very low temperature \cite{wil75,wil78}. However, a quantitative and unambiguous
interpretation of the experimental data was not provided.
As mentioned above, the rotational tunnelling picture
has been evidenced in a vaste collection of crystalline systems by different techniques,
and specially by neutron scattering, where the tunnelling frequency is directly observed
as two resolution-width inelastic peaks at $\pm\hbar\omega_t$. 
For the case of polymers, one might expect to observe broad peaks as a consequence
of the distribution effects. However, observation of peaks, or of any scattered intensity 
over the instrumental resolution at very low temperature was not reported for a long time.

Finally, high resolution measurements in PVAc \cite{pvacprl} and PMMA \cite{isotprb} 
at $T\approx 1$ K showed, instead of well-defined peaks, a broad scattered intensity similar
to that commonly observed in the high temperature hopping regime. 
This result, which might have been interpreted as a signature
of hopping events at unusually low temperature, was indeed explained in terms of 
{\it rotational tunnelling}, namely as the result of the presence of a strongly assymetric
distribution of tunnelling frequencies, with the maximum located well beneath the instrumental resolution.
The latter distribution just followed from the distribution of potential barriers for methyl group 
rotation. This interpretation was supported by exploting the expected isotope effect
on the tunnelling frequencies \cite{isotprb} (see Section \ref{sec:summexper}).
The consistency of the distribution picture, formalized in the Rotation Rate Distribution
Model (RRDM, see Section \ref{sec:rrdm}) was supported by the fact that the same barrier distribution
was able to quantitatively account for the experimental spectra
in all the temperature range \cite{crossprb,pmmamacro}, covering the crossover from rotational
tunnelling to classical hopping, as well as the librational peaks observed by INS.

The grounds of the RRDM can, in principle, be extended to non-polymeric disordered systems,
as has been successfully checked in the glassy state of toluene \cite{tolujcp}
and sodium acetate trihydrate \cite{sodacetprb} (see Section \ref{sec:origin}). 
Contrary to the case of polymer systems, results for methyl group dynamics in molecular glasses
can be directly compared with those for the crystalline state, allowing one to get insight into the
molecular origin of the barrier distribution. 

The goal of this review is to summarize the progress in the topic
of methyl group dynamics in polymers that has been achieved over last 10 years by means
of neutron scattering techniques, together with the development of the RRDM. 
The molecular origin of the distribution of potential barriers,
which is an essential ingredient of the RRDM, is discussed on the basis of the
experimental and computational investigations reported in the literature. 
A brief introduction to the theoretical and experimental grounds
of neutron scattering techniques is also given in next Section.

%The review is organized as follows: In Section we expose \ref{sec:neutscat} basic theoretical and experimental
%grounds on neutron scattering spectroscopy. In Section \ref{sec:theorcryst} we summarize the theoretical grounds
%of methyl group dynamics in crystalline systems, as well as the RRDM for disordered systems.
%Section \ref{sec:pvac} shows the application of the RRDM to PVAc as a showcase. Section \ref{sec:summexper}
%summarizes the work done on methyl group dynamics in last decades by means of neutron scattering techniques.
%In Section \ref{sec:origin}, on the basis of experimental and simulation results 
%some insight is provided on the molecular origin of the barrier distribution for methyl group dynamics
%in polymers, and in general, in highly disordered systems.
%Conclusions and outlook on future trends are given in Section \ref{sec:conclusions}. 
  
\section{Neutron scattering techniques and instrumentation}
\label{sec:neutscat}
\subsection{Neutron scattering techniques}

Neutrons are adequate probes for the investigation of both
structure and dynamics of condensed matter.
There are three major features for that: i) their
typical wavelengths are of a few {\rm \AA}, which correspond to
the typical interatomic distance, ii)  their typical energies
are of the order of some meV, which correspond to the energy scale of the typical excitations in 
condensed matter, iii) neutrons being particles without
electrical charge, they can penetrate the sample and provide
information about bulk properties, in contrast to charged
particles like electrons, which mainly probe surface properties.

The range of the nucleus-neutron interaction is $\sim 1.5 \times 10^{-5}$ {\rm \AA},
much smaller than the neutron wavelength and the nuclear radius.
Hence, scattering can be approximated as isotropic, and the interaction is modelled
by the Fermi pseudopotential:

\begin{equation}
V_F =\frac{2\pi \hbar^{2}}{m}b\delta({\bf r}-{\bf R}) ,
\label{eq:fermi}
\end{equation}
where $m$ and ${\bf r}$ are respectively the neutron mass and position, 
and ${\bf R}$ is the nucleus position.
The scattering length operator, $b$, is given by

\begin{equation}
b = b_{coh} + \frac{2 b_{inc}}{\sqrt{I(I+1)}}{\bf S \cdot I} ,
\end{equation}
where ${\bf S}$ and ${\bf I}$ are respectively the neutron and nuclear spins, and
$b_{coh}$ and $b_{inc}$ are respectively the coherent and incoherent scattering lengths.

\begin{figure}
\begin{center}
%\vspace{4 mm}
\includegraphics[width=0.40\textwidth]{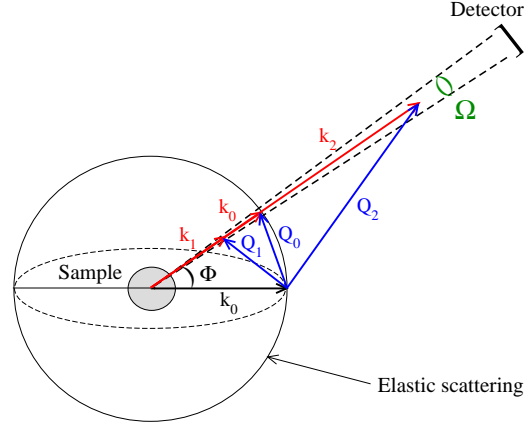}
\newline
\newline
\caption{Scheme of a scattering experiment. The sphere of radius $k_0$
represents elastic interactions.
Indexes 1 and 2 respectively correspond
to inelastic processes with neutron energy loss and gain.}
\label{fig:scattfig}
%\vspace{2 mm}
\end{center}
\end{figure}

As in any scattering technique, there are two basic quantities to be measured
by neutron scattering: the energy transfer,
$\hbar\omega = \hbar^{2}(k^{2} - k^{2}_0)/2m$, which is the difference between the final and the initial neutron
energy, and the momentum transfer, $\hbar{\bf Q}=\hbar{\bf k}-\hbar{\bf k}_{0}$, where
${\bf k}$ and ${\bf k}_0$ are respectively the final and initial neutron wavevectors 
(see Fig. \ref{fig:scattfig}). The number of neutrons scattered within a solid angle
between $\Omega$ and $\Omega + d\Omega$, which have
changed their energy in $\hbar \omega$ and their momentum in $\hbar{\bf Q}$, is proportional to
the double-differential cross-section \cite{bee,springer,lovesey,squires}. The latter
can be split in two parts, the coherent and the
incoherent double-differential cross-sections:
\begin{equation}
\frac{\partial^{2}\sigma}{\partial\Omega\partial\omega} =
\left(\frac{\partial^{2}\sigma}{\partial\Omega\partial
\omega}\right)_{coh}+
\left(\frac{\partial^{2}\sigma}{\partial\Omega\partial
\omega}\right)_{inc} ,
\label{doble2}
\end{equation}
which can be expressed as a function of the scattering functions:

\begin{eqnarray}
\nonumber
\left(\frac{\partial^{2}\sigma}{\partial\Omega\partial\omega}\right)_{coh}
\propto \frac{k}{k_{0}}\sigma_{coh}S_{coh}(Q,\omega) \\
\left(\frac{\partial^{2}\sigma}{\partial\Omega\partial\omega}\right)_{inc}
\propto \frac{k}{k_{0}}\sigma_{inc}S_{inc}(Q,\omega) .
\label{eq:inccoh}
\end{eqnarray}
In this equation $\sigma_{coh}$ and $\sigma_{inc}$ are respectively 
the coherent and incoherent cross-sections.
$S_{coh}(Q,\omega)$ and $S_{inc}(Q,\omega)$ are respectively the coherent and incoherent
scattering functions. These functions give account for the amplitude transitions,
mediated by the Fermi pseudopotential, between the eigenstates of the sample. 
While the coherent contribution results from interferences between the neutron waves dispersed
by the nuclei, the incoherent contribution is equivalent to the dispersion of the isolated nuclei.
In the classical limit, $S_{coh}(Q,\omega)$ and $S_{inc}(Q,\omega)$
are related via double Fourier transform, respectively,
with the pair-correlation and self-correlation functions of the atomic positions. 

Protium hydrogen (H) has a large incoherent scattering cross-section
as compared to its coherent part and to the total cross-section,
$\sigma_{tot}=\sigma_{inc}+\sigma_{coh}$, of 
the other elements typically present in polymer materials
(see Table \ref{table:scs}).
Hence, a neutron scattering experiment on a protonated
polymer sample provides mainly information on
the self-correlation function of the protium hydrogens.
Moreover, due to the large difference between the incoherent cross-section of
protium and the total cross-section of deuterium (D), the intensity scattered
by a subset of the protium hydrogens can be strongly enhanced if selective deuteration 
of the other hydrogens is possible by means of chemical methods. 

When neutron scattering is used for dynamic investigations, one finds
three different types of scattering processes: elastic, quasielastic and inelastic scattering.
Inelastic neutron scattering (INS) corresponds to the case where the neutron exchanges energy with the sample
excitations, and manifests in the experimental spectrum
as two resolution-width lines of finite energy (for neutron energy gain and loss).
On the contrary, elastic scattering corresponds to the case where the neutron energy remains
constant, yielding a resolution-width elastic line.
Finally, quasielastic neutron scattering (QENS) is that associated with the energy transfer
resulting from the Doppler effect occurring when a neutron interacts
elastically with a moving particle, and manifests as a broad line around the elastic line.
Of course, inelastic or quasielastic processes are
no detectable when the involved energy change is smaller than the instrumental 
energy resolution, yielding an additional contribution to the experimental elastic line.
Quasielastic and inelastic processes typically cover a range $|\hbar\omega| < 2$ meV
and $|\hbar\omega| > 2$ meV respectively, the latter corresponding to typical
vibrational excitations in condensed matter. The terms QENS and INS are often used
to refer to these energy ranges and the dynamic window of the corresponding spectrometers.
However, it must be stressed that excitacions in the $\mu$eV range can also be present in the system,
as the case of methyl group tunnelling. Hence, such excitations, which strictly correspond to INS,
are accessible by means of instruments specifically designed for QENS. 
  
In the liquid state, molecules are able to diffuse and loss memory from their initial positions.
Molecular diffusion yields no elastic component in neutron scattering spectra. 
On the contrary, if the motion of a group of particles
is constrained into a localized region, as the case of methyl group rotation
in a crystalline or glassy system, in addition to the eventual inelastic and quasielastic
contributions, such a motion provides an elastic componet in the experimental spectrum. 
The elastic peak is modulated by a $Q$-dependent quantity named the 
elastic-incoherent-structure-factor (EISF), $A(Q)$.
The EISF gives account, via Fourier transform, for the average positions of the individual particles
performing the localized motion, i.e, it provides information about the geometry of the motion.

Integration of the coherent scattering function provides the elastic coherent
structure factor, $S(Q)$. Fourier transformation of the latter gives account for the probability
distribution of the configurations of all the ensemble of particles in the system, i.e.,
it provides static structural information.

\subsection{Neutron scattering instrumentation}

Neutron scattering investigations on methyl group dynamics in polymers 
are carried out mainly by means of time-of-flight (TOF) and high-resolution
backscattering (BS) techniques. In addition, in most of the cases it is convenient
to evaluate the total coherent and incoherent contributions to the scattered
intensity. For this purpose, diffraction experiments with spin polarization analysis
are neccessary. The showcase presented here for PVAc (see Section \ref{sec:pvac})
corresponds to experiments performed 
at the spectrometers IN5, IN6, IN16 and D7 of the Institute Laue Langevin,
(ILL, Grenoble, France) and TOSCA at the source
ISIS of the Rutherford Appleton Laboratory (RAL, Chilton, United Kingdom).
Next we summarize the technical characteristics of these spectrometers.
A detailed description can be found in Refs. \cite{in5,in6,in16,d7,scharpf,tosca}. 

IN5 \cite{in5} and IN6 \cite{in6} are TOF spectrometers. 
By means of a monochromator, different wavelengths, $\lambda$, can be selected
in the ranges $1.8 - 16$  {\rm\AA} for IN5 and $4 -6$ {\rm \AA} for IN6.
Most of the data presented here were obtained with $\lambda =
6.5$ {\rm \AA} at IN5 and $\lambda= 5.1$ {\rm \AA} at IN6. In this situation
the elastic energy resolution (half-width at half maximum, HWHM) is
$\approx 25$ $\mu$eV for IN5 and $\approx 50$ $\mu$eV for IN6.
The explored energy window (in neutron energy gain) is 
-0.6 meV $\lesssim \hbar\omega \lesssim$ 19.5 meV
for IN5, and -2 meV $\lesssim \hbar\omega \lesssim$ 1600 meV for IN6.
The scattering function $S(Q,\omega)$ at constant $Q$ can be obtained
from the experimental quantity which is directly measured, namely $S(\Phi,\omega)$, 
where $\Phi$ is the scattering angle, by an interpolation procedure 
according to the trivial geometrical relation:
\begin{equation}
\hbar^2Q^2/2m = 2E_0+\hbar\omega-2\sqrt{E_0(E_0+\hbar\omega)}\cos\Phi ,
\label{eq:qfi1}
\end{equation}
where $m$ is the neutron mass and $E_0 =\hbar^{2}k^{2}_0/2m$ is the incoming neutron energy.
The interpolation procedure provides reliable spectra in the 
quasielastic energy range $|\hbar\omega| \lesssim$ 2 meV
for a $Q$-range of $0.6 {\rm \AA}^{-1} \lesssim Q \lesssim 1.8 {\rm \AA}^{-1}$,
both ranges being of interest for investigations on methyl group rotation. 

IN5 and IN6 are suitable to investigate methyl group rotational hopping at
intermediate and high temperature, where the corresponding time scale 
is typically of the order of 10 ps, but usually are not
adequate for the investigation of rotational tunnelling.
Only when the rotational barrier is very low
($V_{3}\lesssim 200 K$) the tunnelling lines are accessible in
the energy range covered by TOF spectrometers.
For higher barriers, the use of backscattering instruments is needed. 

IN16 is a high-resolution BS spectrometer\cite{in16}
which can explore excitations in the $\mu$eV-range, and correspondingly,
time scales of the order of 1 ns. The high resolution of IN16 makes it adequate for the
investigation of methyl group dynamics in, both, the classical hopping
regime at intermediate temperature, and the quantum tunnelling
regime at very low temperature. By using a Si (111) monochromator the
wavelength of the incident neutrons is $\lambda =6.27$ {\rm \AA}. The
corresponding accessible $Q$-range is $0.2 - 1.9$  ${\rm\AA}^{-1}$.
When deformed Si (111) crystals are used as analyzers a resolution of nearly
Gaussian shape with HWHM $\approx 0.45$ $\mu$eV is attained.
In the standard operation mode the monochromator is subject to an
oscillatory motion at a frequency of $\approx$ 14 Hz, which allows one to
explore an energy transfer range up to 15 $\mu$eV. As in backscattering spectrometers
the incoming beam energy is much larger than the maximum
energy transfer, Eq. (\ref{eq:qfi1}) can be approximated 
by a direct relation between the position of the detector and the momentum transfer, 
given by:
\begin{equation}
Q = (4\pi/\lambda)\sin(\Phi/2) .
\label{eq:qfi2}
\end{equation}
BS spectrometers as IN16 can also be used in a different operation mode,
known as the fixed-elastic-window (FEW) technique.
In this technique the monochromator is held fixed and the sample
temperature is varied at a moderate rate. 
In this way one measures the elastic intensity as a function of the
temperature, since the FEW technique only detects the fraction of neutrons with
an energy exchange within the instrumental resolution.
Hence, what is considered as elastic intensity depends on the resolution of the used spectrometer.
For IN16, with a resolution function of HWHM $\approx 0.45$ $\mu$eV, 
the dynamics of scattering particles being
slower than a characteristic time $\tau \approx 10$ ns 
are observed as elastic scattering. When increasing temperature, $T$, the fraction of particles
slower than $\tau$ decreases, yielding a decrease of the elastic intensity.
This decrease is exponential if dynamics in that time scale is purely vibrational,
and it is given by the $Q$-dependent Debye-Waller
factor $e^{-2W(Q)}$ \cite{bee,springer,lovesey,squires},
with $W(Q)=Q^2\langle u^2\rangle/3$. In this expression $\langle u^2\rangle$ is the mean-squared
displacement per particle. For vibrational motions $\langle u^2\rangle \propto T$.
The presence of other dynamic processes is evidenced by deviations from exponential behavior,
yielding a step-like decay of the elastic intensity (see Section \ref{sec:summexper}). 

The spectrometer D7 combines TOF methods with
spin polarization analysis capabilities\cite{d7,scharpf}.
Among the different types of experiments which are possible
with D7, that of particular interest for the investigation of methyl group dynamics
in polymers is the measurement of the coherent and incoherent
contributions to the total scattered intensity (without energy analysis).
These measurements are usually performed at low temperature to
ensure that no significant intensity is lost due to the finite
range of energy transfer covered by the instrument.
By using a neutron incident wavelength of 4.8 {\rm \AA}, the relevant accessible $Q$-range,
$Q \lesssim 2.5 {\rm \AA}^{-1}$, corresponds to that covered in dynamic experiments
with TOF and BS instruments. 

In TOF instruments as IN5 or IN6, high values of vibrational frequencies 
are only accessible in neutron energy gain.
Since the thermal statistical weight of high-energy vibrational excitations
is negligible at low temperature, such excitations can only be detected by the former instruments
at high temperature. This becomes a problem for the observation of methyl group librational peaks,
which, due to anharmonicity effects, show a strongly damped
intensity at high temperature \cite{bat77,prager83,cavagnat84,heidemann89acet}. 
This problem can be avoided by using an instrument which has access to vibrational frequencies in
neutron energy loss, i.e., which is able to create the corresponding vibrational excitations in the sample
even at very low temperature. An example of a TOF spectrometer that is optimized for
vibrational spectroscopy is TOSCA \cite{tosca}, which allows one to measure neutron
energy loss in a wide range of values, typically 1 meV $\lesssim \hbar\omega \lesssim$ 250 meV. 
TOSCA design is such that there is a unique value of $Q$ for each value of the energy transfer,
so all the detectors are summed up to obtain the vibrational spectrum. 
The arrangement of the spectrometer is such that the neutrons are detected at a fixed 
scattering angle $\Phi=135^{\rm o}$. In this situation 
Eq. (\ref{eq:qfi1}) reduces approximately to the relation $Q^{2}\propto \omega$ for low values
of the neutron energy transfer ($\hbar\omega \lesssim 40$ meV). Hence, TOSCA arrangement
provides a nearly direct access to the generalized vibrational
density of states (VDOS), $Z(\omega)$, given by \cite{bee,springer,lovesey,squires}:
\begin{equation}
Z(\omega) = \frac{S(Q,\omega)\omega}{Q^{2}[n(\omega)+1]} ,
\label{eq:zw}
\end{equation}
where $n(\omega)=[\exp(\hbar\omega/k_{B}T)-1]^{-1}$ is the statistical Bose occupation factor.
It must stressed that the generalized VDOS $Z(\omega)$ is not proportional to the real VDOS,
but a "deformed" function, where excitations are weighted by scattering cross-sections, molecular masses,
and coordinates of the normal modes. Only for a strictly monoatomic sample $Z(\omega)$ is proportional 
to the real VDOS. In the general case, though the experimental peaks will correspond to the real excitations,
their amplitudes will be controlled by the mentioned factors, being larger for the modes involving
motions of protium hydrogens.

\subsection{Final experimental considerations}

\begin{figure}
\begin{center}
\includegraphics[width=0.38\textwidth]{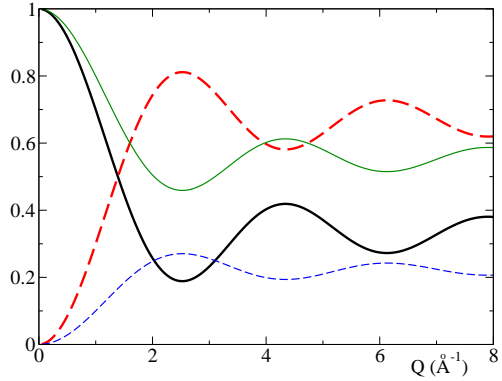}
\newline
\caption{Thick lines: $A(Q)$ (solid) and $1-A(Q)$ (dashed) for methyl group classical hopping.
Thin lines: $A(Q)$ (solid) and $(1-A(Q))/2$ (dashed) for methyl group rotational tunnelling.}
\label{fig:eisf}
\end{center}
\end{figure}

Instrumental resolution functions are usually determined
from measurements on a vanadium sample, which provides solely
elastic scattering. The same measurements can be used to
calibrate detectors. In order to obtain the actual experimental
scattering function $S(Q,\omega)$, raw data must be corrected for detector efficiency,
sample container and absorption. Due to its small scattering cross-section,
aluminium is the usual material for sample holders. Slab geometry is usually preferred.
The typical area of the neutron beam is of $10 - 15$ cm$^{2}$. The sample thickness
is selected to provide a transmission of  $90 - 95$ \%.
For fully protonated polymers this condition is attained with sample thickness of $\sim 0.1$ mm.
High transmission allows one to neglect multiple scattering effects, provided that the
very low $Q$-range is not considered in the data analysis.
This is the case for methyl group dynamics, since the EISF at low $Q$ is very close to unity
(see Fig. \ref{fig:eisf}), and correspondingly, the quasielastic intensity --- which is
weighted by $1-A(Q)$, see Section \ref{sec:theorcryst}--- is very weak and difficult to analyze.

\section{Methyl group dynamics in crystalline systems: theory and scattering functions}
\label{sec:theorcryst}
\subsection{Rotational tunnelling}

As mentioned in the Introduction, methyl group motion
at very low temperature ($T\sim $ 1 K) can be modelled
as that of a rigid rotor in an effective single-particle one-dimensional
rotational potential $V(\phi)$. The corresponding
Hamiltonian is given by \cite{pragerreview,pressbook,carlilereview}:

\begin{equation}
H_{\rm R}= -B\frac{\partial^2}{\partial\phi^{2}}+ V(\phi).
\label{eq:hr}
\end {equation}
In this equation $B=\hbar^{2}/2I$ is the rotational constant of the rigid rotor,
with $I$ the moment of inertia of the methyl group around its threefold symmetry axis.
For a fully protonated methyl group $B =$ 0.655 meV. For a fully deuterated one,
$B =$ 0.328 meV. 

The rotational potential $V(\phi)$ must be invariant under $2n\pi/3$ rotations,
with $n =\pm 1, \pm 2, \pm 3...$, because its symmetry cannot be lower than
that of the rigid rotor, and moreover, both symmetries cannot
be uncompatible. Hence, $V(\phi)$ can be expanded as a Fourier series of $3n$-fold terms
of amplitude $V_{3n}$:

\begin{equation}
V(\phi) = \sum_{n=1}^{\infty}\frac{V_{3n}}{2}[1-\cos(3n\phi + \delta_{3n})],
\end{equation}
where $\delta_{3n}$ are constant angular offsets.
Indeed, it can be formally demonstrated that the rotor-lattice interaction
can be expressed in this way provided that $V(\phi)$ is given by a sum
of two-body additive potentials \cite{emid80}. 
As mentioned in the Introduction, in a first approximation only the threefold term of the expansion is retained:

\begin{equation}
H_{\rm R}= -\frac{\hbar^2}{2I}\frac{\partial^2}{\partial\phi^{2}}
+ \frac{V_{3}}{2}(1-\cos3\phi) .
\label{eq:HR3f}
\end{equation}
If higher-order terms are neccesary to account
for the experimental results, they often contribute just as small corrections
to the main threefold term \cite{pragerreview}.

%Figure \ref{fig:levels} shows the level scheme of the Hamiltonian in Eq. (\ref{eq:HR3f}) for
%a barrier height $V_{3}=500$ K (in the following $V_3$ will be given as $V_3/k_B$, with $k_B$
%the Boltzmann constant). The energy levels of the individual wells
%are named {\it torsional} or {\it librational} levels.
%In the following, transitions between the ground and the different excited librational levels
%will be denoted as $E_{0i}$. The lowest transitions $E_{01}$ and $E_{02}$
%take place typically in the range $\sim 10-30$ meV and are strongly superposed
%with the phonon spectrum \cite{bat77,prager83,cavagnat84,heidemann89acet}.

As mentioned in the Introduction, the coupling of the wavefunctions of the individual wells
splits the librational levels. The resulting sub-levels are denoted as $A$ and $E$, 
the latter consisting of the degenerate doublet ($E_{\rm a}$, $E_{\rm b}$). These labels correspond to
the irreducible representations of the symmetry group $C_3$ of the Hamiltonian $H_{R}$.
The eigenstate $A$ remains unchanged under a rotation $2\pi/3$, while such an operation 
adds a phase $\pm 2\pi/3$ to the spatial part of the wave functions of $E_{a}$ and $E_{b}$
respectively. An eigenstate of $H_{R}$ is characterized by two discrete quantum numbers, namely
the librational, $n$, and symmetry, $S$, indexes: $|nS\rangle$.
As shown in Fig. \ref{fig:levels}, the levels $A$ and $E$ alternate between consecutive 
librational levels.

%The energy splittings of the different librational levels
%have been denoted as $\Delta_{i}$ in Fig. \ref{fig:levels}.
At very low temperature, $T\sim 1$ K, $k_{B}T \ll E_{0i}$ and the occupation probability
of the excited librational states $i=1,2,3...$ is negligible. Hence, the rotor oscillates coherently
between the three wells with an oscillation frequency equal
to the tunnelling frequency $\omega_t$. 
If the latter takes a value between the highest resolution available in
BS spectrometers,  $\hbar\omega_{t} \approx$ 0.3 $\mu$eV, and the free rotor limit, $\hbar\omega_{t}=B$,
it is detected as two resolution-width
inelastic peaks at $\omega=\pm\omega_{t}$ (see Fig. \ref{fig:sodacetcry}a).

The incoherent scattering functions for rotational tunnelling in different symmetries
can be calculated by using symmetry adapted scattering operators \cite{pressbook}.
Hence, the incoherent scattering operator

\begin{equation}
W = \sum_{j=1}^{N}({\bf S \cdot I}_{j})\exp(-i{\bf Q\cdot r}_{j})
\end{equation}
where ${\bf r}_j$, ${\bf I}_j$ are the hydrogen positions and spins, and 
${\bf S}$ is the neutron spin, can be rewritten, for threefold symmetry, as:

\begin{equation}
W = W_{A}+W_{E_a}+W_{E_b} .
\label{eq:wsym}
\end{equation}
The eigenstates $|A\rangle$, $|E_a\rangle$ and $|E_b\rangle$ are unchanged
by $W_{A}$, and are permutated cyclically by $W_{E_a}$, and anticyclically
by $W_{E_b}$. Hence, the decomposition (\ref{eq:wsym}) allows for a straightforward calculation
of the $W$-matrix elements between the different eigenstates \cite{pressbook}. 
After averaging over all the orientations of the scattering vector ${\bf Q}$,
the following transition amplitudes are obtained:
\\\\
i) $f_{AE}(Q)=[1-j_{0}(Qr)]/3$ for $A \leftrightarrow E$ ($E_a, E_b$) transitions.
The corresponding contribution to the scattering function 
is $2f_{AE}(Q)\delta(\omega \pm \omega_{t})$,
since there are two inelastic transitions $\omega = +\omega_{t}$ and other 
two $\omega = -\omega_{t}$.
\\
ii) $f_{E_aE_b}(Q)=[1-j_{0}(Qr)]/3$ for $E_{a} \leftrightarrow E_{b}$ transitions.
These two elastic transitions contribute as $2f_{E_aE_b}(Q)\delta(\omega)$.
\\
iii) $f_{NSC}(Q)=[1+2j_{0}(Qr)]/3$ for transitions without symmetry change.
These three elastic transitions contribute as $3f_{NSC}(Q)\delta(\omega)$.
\\\\
In these expressions, $j_{0}(Qr)=\sin(Qr)/Qr$, with $r$ the H-H distance in the methyl group,
($r$ = 1.78 $\rm\AA$). After normalization to one-hydrogen scattering, the corresponding
incoherent scattering function for methyl group
tunnelling is \cite{pragerreview,pressbook,carlilereview}:

\begin{eqnarray}
\nonumber S_{\rm inc}^{\rm MG}(Q,\omega) = \frac{5+4j_0(Qr)}{9}
\delta(\omega) + \\  
\frac{2[1-j_0(Qr)]}{9}[\delta(\omega+\omega_{\rm t})+
\delta(\omega-\omega_{\rm t})] .
\label{eq:sqtun}
\end{eqnarray}
\\
The EISF, $A(Q)=[5+4j_0(Qr)]/9$, and $[1-A(Q)]/2=2[1-j_0(Qr)]/9$, which
modulate respectively the elastic and inelastic contributions, are shown in Fig. \ref{fig:eisf}.

\subsection{Crossover from rotational tunnelling to classical hopping}

\begin{figure}
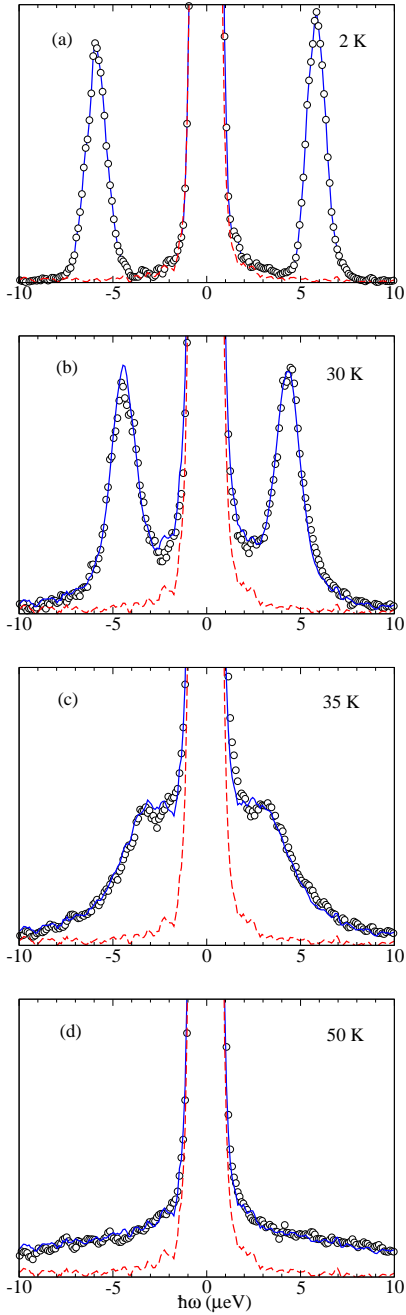

\begin{center}
\includegraphics[width=0.31\textwidth]{cry002.eps}
%\vspace{3 mm}
\newline
\newline
\includegraphics[width=0.31\textwidth]{cry030.eps}
%\vspace{3 mm}
\newline
\newline
\includegraphics[width=0.31\textwidth]{cry035.eps}
%\vspace{3 mm}
\newline
\newline
\includegraphics[width=0.31\textwidth]{cry050.eps}
\newline
\caption{Experimental spectra (circles) for methyl group dynamics in crystalline
sodium acetate trihydrate. Solid lines are fits to the model for crystalline systems exposed in
Section \ref{sec:theorcryst}.
Dashed lines correspond to the experimental 
resolution. $Q$= 1.8 ${\rm \AA}^{-1}$. Scales (referred to the maximum): (a) 8\%, (b)-(d) 5\%.}
\label{fig:sodacetcry}
\end{center}
\end{figure}

The usual temperature dependence of neutron scattering spectra for
methyl group dynamics in crystalline systems 
is illustrated in Fig. \ref{fig:sodacetcry} for crystalline sodium acetate trihydrate \cite{sodacetprb}.
At $T\sim 1$ K (Fig. \ref{fig:sodacetcry}a) the spectrum shows
a central elastic peak and two inelastic
peaks of resolution width, i.e., delta functions convoluted with the
instrumental resolution. According to the picture exposed above,
the inelastic peaks correspond to the two transitions $A \leftrightarrow E$
within the ground librational state. The elastic peak includes
the ground state transitions $E_a \leftrightarrow E_b$
and the transitions without symmetry change. It also includes the 
elastic incoherent contribution from other atoms different from methyl group hydrogens,
and the total coherent contribution (see below).

When increasing temperature 
\cite{sodacetprb,prager83,heidemann89acet,clough79,clough81acet,cockbain,cavagnat86jpc,%
prager86jcp,prager87jcp,prager88jcp,prager90jpcm}, 
the inelastic peaks progressively broaden and shift
to the elastic peak. At the same time, a quasielastic component appears
around the elastic peak, and broadens with increasing temperature
(Figs. \ref{fig:sodacetcry}b,c).
Finally, in a narrow temperature interval ($\Delta T \sim 7$ K), the 
quasielastic and inelastic components merge in a single 
quasielastic line (Fig. \ref{fig:sodacetcry}d).
In the following the width of this interval will be neglected and represented
as a unique temperature $T_c$, that will be denoted as the {\it crossover temperature}.
Above $T_c$, a classical picture of thermally activated hopping over the 
rotational barrier is commonly acepted \cite{pressbook,bee}.
The observed single quasielastic line is well described by a Lorentzian function.
The corresponding HWHM follows an Arrhenius-like temperature dependence:

\begin{equation}
\Gamma = \Gamma_{\infty}\exp(-E_{A}/k_{B}T) .
\label{eq:gcla} 
\end{equation}
where the classical activation energy $E_{A}$, as mentioned in the Introduction, 
is defined as the difference between the top of the barrier and the ground state
(see Fig. \ref{fig:levels}). In the following $E_A$ will be given in units
of $k_B$. The preexponential factor $\Gamma_{\infty}$ is temperature
independent and typically takes values of $\sim 5-10$ meV.

\begin{figure}
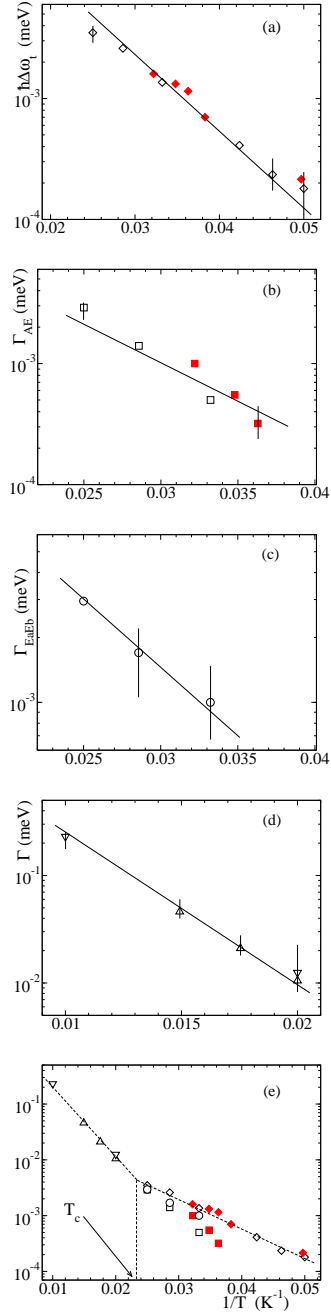

\begin{center}
%\vspace{-4 mm}
\includegraphics[width=0.24\textwidth]{invTshift.eps}
\newline
\newline
%\vspace{4 mm}
\includegraphics[width=0.25\textwidth]{invTgAE.eps}
\newline
\newline
%\vspace{3 mm}
\includegraphics[width=0.25\textwidth]{invTgEaEb.eps}
\newline
\newline
%\vspace{2 mm}
\includegraphics[width=0.24\textwidth]{invTcla.eps}
\newline
\newline
%\vspace{2 mm}
\includegraphics[width=0.24\textwidth]{invTall.eps}
\newline
%\vspace{3 mm}
\caption{Temperature dependence of the shift of the tunnelling frequency and of the crossover and
classical Lorentzian HWHM for crystalline sodium acetate trihydrate.
Solid lines in panels (a-d) are fits to respectively,
Eqs.(\ref{eq:shift}),(\ref{eq:gae}),(\ref{eq:gee}) and (\ref{eq:gcla}).
Panel (e) shows all the data in a common frame. 
The arrow indicates the crossover temperature $T_c$.
Dotted lines are a guide for the eyes.}
\label{fig:crosparam}
\end{center}
\end{figure}

While rotational tunnelling and classical hopping are well established pictures
respectively in the low and high temperature limits, no satisfactory theory for the
crossover regime between both dymanic limits has won general acceptance
\cite{huller80,clough82jpc,hewson1,hewson2,wurger89zp,clough93pra,%
allen,punkkinen,clough80jpc,clough85jpc,whitall,clough93pau}.
Every model for the crossover regime must give account for the following
phenomenology observed in a large amount of experiments 
\cite{sodacetprb,prager83,heidemann89acet,clough79,clough81acet,cockbain,cavagnat86jpc,%
prager86jcp,prager87jcp,prager88jcp,prager90jpcm}:
\\\\
i) Broadening of the quasielastic and inelastic crossover components
are well described by Lorentzian functions of the same intensity.
The relative intensities to the elastic peak are the same than in the purely
rotational tunnelling regime, i.e., $2f_{AE}(Q)/3f_{NSC}(Q)$ for
each of the inelastic lines
and $2f_{E_{a}E_{b}}(Q)/3f_{NSC}(Q)$ for the quasielastic line. The temperature 
dependence of the Lorentzian HWHM follows Arrhenius-like behavior. The observed 
activation energy corresponds approximately to the first librational energy:

\begin{equation}  
\Gamma_{i} = \gamma_{i}\exp(-E_{01}/k_{B}T) .
\label{eq:gi}
\end{equation}
\begin{equation}
\Gamma_{q} = \gamma_{q}\exp(-E_{01}/k_{B}T) .
\label{eq:gq}
\end{equation}
In these equations $\Gamma_{i}$ and $\Gamma_{q}$ are respectively the Lorentzian HWHM
for the inelastic and quasielastic components.
The preexponential factors $\gamma_{i}$ and $\gamma_{q}$ are temperature independent,
and typically take values in the range 0.05-1 meV, much lower than the observed values
for the classical $\Gamma_{\infty}$ (see above).
It is usually found that $\gamma_{q} < \gamma_{i}$, though both factors
take values of the same order of magnitude (typically $\gamma_{q}/\gamma_{i} \sim 0.5$). 
\\\\
ii) Shift of the tunnelling peaks also follows Arrhenius-like behavior:

\begin{equation}
\hbar\Delta\omega_{t} = \gamma_{s}\exp(-E_{S}/k_{B}T),
\label{eq:shift}
\end{equation} 
where $\Delta\omega_{t}$ is the shift (in absolute value) of the tunnelling peak and $\gamma_{s}$ 
a temperature independent preexponential factor, which takes values 
similar to $\gamma_{i}$ and $\gamma_{q}$.
The activation energy $E_{S}$ is generally smaller than $E_{01}$ though it takes a close value,
typically $E_{S}/E_{01} \sim 0.7$. Fig. \ref{fig:crosparam} shows all these features for 
crystalline sodium acetate trihydrate as a standard showcase \cite{sodacetprb}.
\\\\
All the theoretical models that aim to give account for these universal features
state that its physical origin is the coupling of the methyl group to the lattice phonons,
which progressively breaks the coherence of the wavefunction of the rotor,
leading to incoherent tunnelling 
\cite{huller80,clough82jpc,hewson1,hewson2,wurger89zp,clough93pra,%
emid80,allen,punkkinen,clough80jpc,clough85jpc,whitall,clough93pau,clough81cpl,emid81,%
clough81jpc,gunn,clough87jpc,huller88,wurger89kra,wurger89jpcm,%
clough89mp,huller90,wurger90zp,wurger91zp,braun93,vowe,clough93cpl,%
clough94zn,braun94,clough96phys,peternelj}.
The standard approach, and the only one that is able to give account quantitatively for
the observed behavior, at least below $T_{c}$, is to describe the system rotor-lattice by 
a Hamiltonian: 
\begin{equation}
H=H_{R}+H_{B}+H_{C} .
\label{eq:hcross}
\end{equation}
In this equation $H_{R}$ is the rigid rotor
Hamiltonian (\ref{eq:hr}), $H_{B}$ is a bath of harmonic oscillators
\cite{huller80,hewson1,hewson2,wurger89zp,whitall,huller88,wurger89kra,%
wurger89jpcm,huller90,wurger90zp,wurger91zp,braun93,vowe,braun94,peternelj}:

%\begin{eqnarray}
%\nonumber H = -B\frac{\partial^2}{\partial\phi^{2}}+\frac{V_{3}}{2}(1-\cos3\phi) 
%+\sum_{k=1}^{N}-\frac{\hbar^2}{2m_k}\frac{\partial^2}{\partial x_{k}^{2}} + \\
%\frac{1}{2}m_k\omega^{2}_{k}x^{2}_{k} +\sum_{k=1}^{N}g_{k}x_{k}\cos(3\phi +\delta_{k})
%\label{eq:coupling}
%\end{eqnarray}
\begin{equation}
H_{B}= \sum_{k=1}^{N}\left(-\frac{\hbar^2}{2m_k}\frac{\partial^2}{\partial x_{k}^{2}} +
\frac{1}{2}m_k\omega^{2}_{k}x^{2}_{k}\right) ,
\label{eq:hb} 
\end{equation}
and $H_{C}$ is a coupling term between the rotor and the bath:
\begin{equation}
H_{C}= \sum_{k=1}^{N}g_{k}x_{k}\cos(3\phi +\delta_{k}) .
\label{eq:hc}
\end{equation}
In these equations $x_{k}$, $m_{k}$, $\omega_{k}$,
$g_{k}$ and $\delta_{k}$ are respectively the displacements, 
masses and frequencies of the harmonic oscillators,
the coupling constants and the angular offsets relative to the rotor.

The total Hamiltonian (\ref{eq:hcross}) retains the threefold symmetry
of the interactions methyl group - environment.
Hence, an argumentation on the basis of transitions between different symmetry
states is still possible. The eigenvectors of the Hamiltonian (\ref{eq:hcross}) are
characterized by a large set of quantum numbers. However, since $H$ is, as $H_{R}$,
unvariant under the symmetry group $C_{3}$, the symmetry indexes $A$, $E_{a}$ and $E_{b}$
are still good quantum numbers for $H$, and the symmetry adapted formalism can still be used
to calculate the scattering function. In the limit $N\rightarrow \infty$, $H$ has
a continuous number of eigenstates. Hence, transitions with symmetry change will
take place not between two discrete eigenstates, but between two continuous sets of eigenstates.  
The delta functions for the low temperature limit [Eq. (\ref{eq:sqtun})], corresponding to transitions
between two eigenstates characterized by a symmetry index, are now substituted by Lorentzians,
corresponding to transitions between two large sets of eigenstates, 
each set characterized by a symmetry index, i.e.,

\begin{equation}
|\xi S\rangle \rightarrow |\xi 'S'\rangle ,
\end{equation}
where $S$ and $S'$ are the initial and final symmetry states, and $\xi$ and $\xi'$ a large set
of quantum numbers characterizing the initial and final state of the bath.
Within this picture, the inelastic Lorentzians involve transitions $A \leftrightarrow E$,
while the quasileastic Lorentzian involve transitions $E_{a} \leftrightarrow E_{b}$.
Hence, Eqs. (\ref{eq:gi},\ref{eq:gq}) can be rewritten as:

\begin{equation}  
\Gamma_{AE} = \gamma_{AE}\exp(-E_{01}/k_{B}T)
\label{eq:gae}
\end{equation}
\begin{equation}
\Gamma_{E_aE_b} = \gamma_{E_aE_b}\exp(-E_{01}/k_{B}T) .
\label{eq:gee}
\end{equation}
It can be demonstrated that elastic transitions without symmetry change do not lead to broadening
\cite{wurger89zp}, yielding an elastic line as in the low temperature limit.
Taking all this arguments into account, the incoherent scattering function
for the crossover regime is given by \cite{prager83,wurger91zp}:

\begin{eqnarray}
\nonumber  S_{\rm inc}^{\rm MG}(Q,\omega) = \frac{1+2j_0(Qr)}{3}\delta(\omega) + 
\frac{2[1-j_0(Qr)]}{9}\times \\  
\nonumber  [L(\omega;\Gamma_{\rm E_a E_b})
+ L(\omega+(\omega_{\rm t}-\Delta \omega_{\rm t});\Gamma_{\rm AE}) + \\
 L(\omega-(\omega_{\rm t}-\Delta \omega_{\rm t});\Gamma_{\rm AE})] , 
\label{eq:sqcross}
\end{eqnarray}
where $L(\omega-\nu;\eta)$ is a Lorentzian function of area unity, centered at $\omega=\nu$ and
HWHM = $\eta$, i.e.:

\begin{equation}
L(\omega-\nu;\eta)=\frac{1}{\pi}\frac{\eta}{(\omega-\nu)^2 +\eta^2 }.
\end{equation} 
It must be stressed that the functional form (\ref{eq:sqcross}) is a direct consequence of the
threefold symmetry of the Hamiltonian (\ref{eq:hcross}). It is the description of the temperature
dependence of shift and broadening of the Lorentzian lines which depends on the particular
model for the coupling of the rotor to the bath [Eqs. (\ref{eq:hb}, \ref{eq:hc})].
In the different models based on this approach,
Eqs. (\ref{eq:gae}, \ref{eq:gee}) are obtained as effective expressions from a sum over
complicated functions containing Bose factors resonant on the librational frequencies,
the density of states ${\cal Z}(\omega)$ and the coupling constants $\{g_{k}\}$:

\begin{equation}
\Gamma_{\xi\xi'} = \sum^{N}_{j=1}
{\cal G}_{\xi\xi'}[{\cal Z}(E_{0j}),\{g_{k}\},n(E_{0j})] ,
\label{eq:bosegamm} 
\end{equation}
with $\xi\xi'= AE$ or $E_aE_b$, and $n(\omega)$ the Bose occupation factor.
In practice, for the temperature range $T<T_c$, where the crossover regime is
experimentally observed, one has  $k_{B}T \ll E_{01}$. Hence,
the sum (\ref{eq:bosegamm}) is dominated by the term in $E_{01}$ and can be approximated
as $\Gamma_{\xi\xi'} = \gamma_{\xi\xi'}\exp(-E_{01}/k_{B}T)$, in agreement
with Eqs (\ref{eq:gae}, \ref{eq:gee}). 

For the case of the shift of the tunnelling frequency,
the sum is done over all the phonon spectrum:

\begin{equation}
\hbar\Delta\omega_{\rm t} = \int^\infty_0 d\omega
{\cal G}_{\rm s}[{\cal Z}(\omega),\{g_{k}\},n(\omega)] .
\label{eq:boseshift}
\end{equation}
Again for $k_{B}T \ll E_{01}$ the integral can be approximated as an effective Arrhenius
equation as (\ref{eq:shift}). Since the integration is done over all the
phonon spectrum, it is dominated by low-energy contributions, yielding
an effective activation energy $E_{S}$ smaller but close to $E_{01}$.

There is a strong controversy about the validity of the model
given by Eqs. (\ref{eq:hcross},\ref{eq:hb},\ref{eq:hc}) for arbitrarely high temperatures.
Some authors have claimed \cite{whitall,wurger91zp}
that it is a valid picture even above $T_{c}$, the temperature
above which spectra are well described by the classical hopping model.
According to this statement, the observed classical-like Arrhenius behavior (\ref{eq:gcla}) 
is also obtained as an effective expression of the sum (\ref{eq:bosegamm}), where
the highest order contributions would
lead, at high temperature, to the observed apparent activation energy $E_{A}$,
which is higher but close to the highest librational transition (see Fig \ref{fig:levels}).

\subsection{Classical hopping}
 
\begin{figure}
\begin{center}
\includegraphics[width=0.45\textwidth]{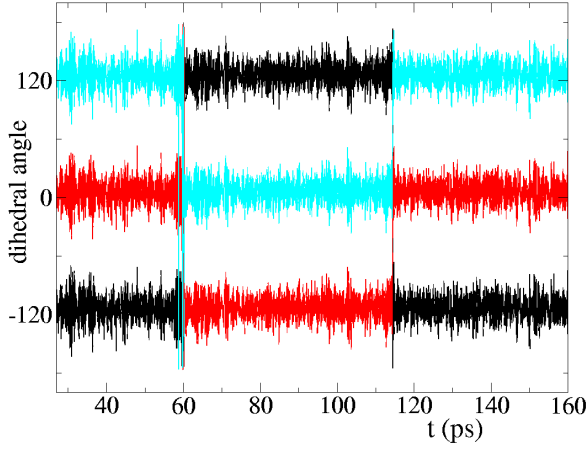}
%\vspace{3 mm}
\newline
\newline
%\vspace{2 mm}
\caption{Simulated time evolution of the orientation of a typical methyl group in 
polyisoprene at $T =$ 150 K. Trajectories of different hydrogens are represented
by different tones.}
\label{fig:anglesimul}
\end{center}
\end{figure}

The incoherent scattering function above $T_c$ can be derived in a 
classical hopping picture, disconnected from the low temperature quantum features
above described. A master equation \cite{pressbook,bee} is assumed for the classical probabilities 
(denoted as $p_{1}$, $p_{2}$, $p_{3}$)
of finding the rotor in each of the three equilibrium positions $0, \pm 2\pi/3$:

\begin{eqnarray}
\dot{p}_1 =\frac{1}{\tau}\left[ -p_1 + \frac{p_2}{2} + \frac{p_3}{2} \right] \hspace{1 mm}\nonumber \\
\dot{p}_2 =\frac{1}{\tau}\left[ -p_2 + \frac{p_3}{2} + \frac{p_1}{2} \right]  \hspace{1 mm}
\label{eq:master} \\
\dot{p}_3 =\frac{1}{\tau}\left[ -p_3 + \frac{p_1}{2} + \frac{p_2}{2} \right] \nonumber . 
\end{eqnarray}
\\
The hopping time between two equilibrium positions is assumed to be negligible
in comparison with the residence time, $\tau$, between consecutive jumps.
This assumption is confirmed (see e.g., \cite{nic97,sae00,simpip,ahumada})
by molecular dynamics simulations (MDS),
which allow one to observe the motion of each individual methyl group in the system.
Fig. \ref{fig:anglesimul} shows computed angular positions for a methyl group
hindered by a moderate barrier.
As can bee seen, sudden jumps occur between the three equilibrium positions,
with long intervals between consecutive jumps. 
It is worth emphasizing that $\tau$ is an average value over all the time intervals between
consecutive jumps {\it of a same methyl group}. The fact that the latters take different values
is a consequence of the stochastic nature of the hopping process. For a crystalline system
with a unique barrier, $\tau$ will be the same for all the individual methyl groups.
For a disordered system, $\tau$ will be different for each individual methyl group,
as will be discussed in Section \ref{sec:rrdm}.

After solving the master equation (\ref{eq:master}), averaging over all the orientations of ${\bf Q}$,
and Fourier transforming to the frequency domain, 
the following expression is obtained \cite{pressbook,bee}:

\begin{eqnarray}
\nonumber S_{\rm inc}^{\rm MG}(Q,\omega) = \frac{1+2j_0(Qr)}{3}\delta(\omega) +
\\ 
\frac{2[1-j_0(Qr)]}{3}L(\omega;\Gamma),
\label{eq:sqclass}
\end{eqnarray}
where $\Gamma = 3\tau/2$ is just assumed to follow the classical Arrhenius-like behavior
of Eq. (\ref{eq:gcla}). The EISF, $A(Q)=[1+2j_0(Qr)]/3$, and $1-A(Q)=2[1-j_0(Qr)]/3$, which
modulate respectively the elastic and quasielastic contributions, are shown in Fig. \ref{fig:eisf}.

\subsection{Additional remarks}

It must be noted that, as required by a consistent description of the temperature
dependence of the spectra, the scattering
function (\ref{eq:sqcross}) for the crossover regime
is reduced to Eqs. (\ref{eq:sqtun}) and (\ref{eq:sqclass}) respectively 
in the low and high temperature limits.
For very low temperature, shift and broadening in Eqs. (\ref{eq:shift},\ref{eq:gae},\ref{eq:gee})
tend to zero, recovering Eq. (\ref{eq:sqtun}). On the other hand, tunnelling is supressed
above the crossover temperature, i.e., $\Delta\omega_{t}=\omega_{t}$, and the three Lorentzians
merge into a single line if $\Gamma_{AE}=\Gamma_{E_aE_b}=\Gamma$. Hence, Eq.(\ref{eq:sqclass}) 
is recovered.

Finally, the total scattering function $S(Q,\omega)$ is obtained by adding the incoherent contribution
of the other atoms, and the total coherent contribution. If, except for methyl group
rotation and lattice vibrations, no other dynamics are present in the considered dynamic window,
the former contributions are elastic, and $S(Q,\omega)$ is given by:

\begin{eqnarray}
\nonumber S(Q,\omega) =  e^{-2W(Q)}[\sigma_{\rm coh}S(Q)\delta(\omega) +
\\
(\sigma_{\rm inc}-\sigma^{\rm MG}_{\rm inc})\delta(\omega) +
\sigma^{\rm MG}_{\rm inc}S_{\rm inc}^{\rm MG}(Q,\omega)], 
%\hspace{3 cm} 
\label{eq:sqtot}
\end{eqnarray} 
with $\sigma_{\rm coh}$ and $\sigma_{\rm inc}$ the total coherent
and incoherent cross-sections and $\sigma^{\rm MG}_{\rm inc}$ the 
incoherent cross-section of the three hydrogens of the methyl group.
$\sigma_{coh}S(Q)$ is the coherent static intensity. As mentioned above,
the Debye-Waller factor $e^{-2W(Q)}$
gives account for the intensity loss with increasing temperature due to vibrations.
Since the Debye-Waller factor only enters as a scaling factor in the total scattering function,
it can be removed in the analysis procedure of the experimental spectra.
Finally, it must be stressed that the function $\tilde{S}(Q,\omega)$ to be compared
with the experimental spectrum is given by the convolution
of the total scattering function (\ref{eq:sqtot}) with the instrumental resolution $R(Q,\omega)$:

\begin{equation}
\tilde{S}(Q,\omega) = \int S(Q,\omega)R(Q,\omega -\omega')d\omega'
\label{eq:convol}
\end{equation}
In the following, comparison between experimental spectra and any of the theoretical scattering 
functions given above, will be understood as done after convolution of the latter with the
instrumental resolution.

\section{Methyl group dynamics in polymer systems}
\label{sec:rrdm}
\subsection{Rotation Rate Distribution Model (RRDM)}

In this Section we summarize the grounds of the RRDM. Though the RRDM was initially developed
for polymer systems, it must be stressed that all its basic assumptions can, in principle, 
be also applied to non-polymeric disordered systems (see Section \ref{sec:origin}). 
In the framework of the RRDM 
\cite{chahid94,pvacprl,isotprb,crossprb,pmmamacro,tolujcp,sodacetprb,ahumada,mukhomacro,%
arrighi01,pillay,zorn,arrighi03macro,tolconf},
it is assumed that the only effect of the structural disorder
on methyl group dynamics is to introduce a distribution of rotational barriers $g(V_3)$,
originated from the different local environments felt by the individual methyl groups.
A non-trivial assumption is that the distribution of distances between neighbouring
methyl groups does not introduce significant coupling forces for the smallest distances,
which, in particular, would modify the single-particle Hamiltonian (\ref{eq:hr}) leading
to a more complex energy level scheme
\cite{prager86zphys,heidemann87zphys,heidemann89iso,fillaux90,fillaux94,fillaux98}.

\begin{figure}
\begin{center}
\includegraphics[width=0.38\textwidth]{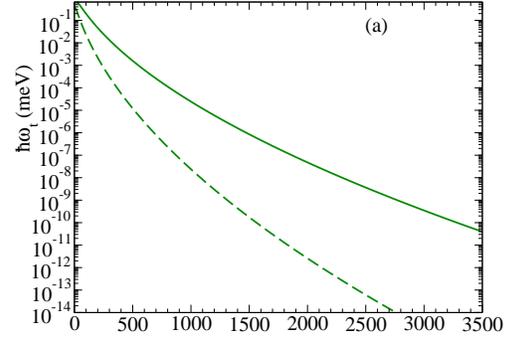}
%\vspace{3 mm}
\newline
\newline
\includegraphics[width=0.38\textwidth]{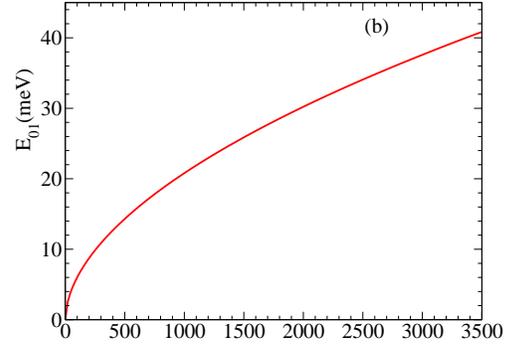}
%\vspace{2 mm}
\newline
\newline
\includegraphics[width=0.38\textwidth]{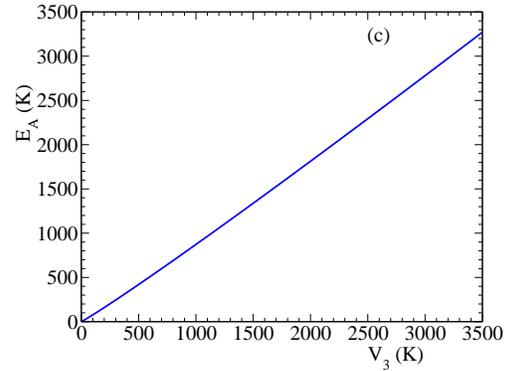}
\newline
%\vspace{2 mm}
\caption{Solid lines: dependence on the barrier height of the tunnelling 
frequency (a), first librational energy (b)
and classical activation energy (c) for a protonated methyl group in a purely threefold potential.
The dashed line in panel (a) corresponds to a deuterated methyl group.}
\label{fig:eneV3}
\end{center}
\end{figure}

According to this approach, the spectrum for the disordered system is simply
obtained as a superposition of crystal-like spectra weighted by the distribution $g(V_{3})$.
Due to the normalization condition $\int g(V_3)dV_3 = 1$,
the elastic terms in Eq. (\ref{eq:sqtot}) are not affected by the introduction of the distribution.
Hence, in the framework of the RRDM, $S(Q,\omega)$ for the glassy system is formally equal
to Eq. (\ref{eq:sqtot}). However, the introduction of $g(V_3)$
modifies the incoherent contribution for methyl group dynamics, which is given by: 

\begin{equation}
S_{\rm inc}^{\rm MG}(Q,\omega) = \int^\infty_0 g(V_{3})S_{\rm inc}^{\rm MG}(Q,\omega,V_3)dV_3 ,
\label{eq:sqrrdm}
\end{equation}
where each single crystal-like spectrum $S_{\rm inc}^{\rm MG}(Q,\omega,V_3)$ evolves with
temperature according to the description given in Section \ref{sec:theorcryst}, summarized
in Eqs. (\ref{eq:sqtun},\ref{eq:gcla},\ref{eq:shift},\ref{eq:gae},
\ref{eq:gee},\ref{eq:sqcross},\ref{eq:sqclass}).
The different parameters controlling this evolution are therefore,
{\it for each individual methyl group}:
\\\\
i) the tunnelling frequency $\omega_{t}$, first librational energy $E_{01}$,
and classical activation energy $E_{A}$. These three quantities
are direct functions of the potential barrier $V_{3}$, and are obtained by calculating the eigenvalues
of the Hamiltonian (\ref{eq:hr}).  Fig. \ref{fig:eneV3} shows the corresponding results
for the purely threefold case. Highly accurate numerical relations for the latters, in the range
$V_3 < 3500$ K, are given by:

\begin{equation}
E_A({\rm K}) = 0.598V_3^{1.05},
\label{eq:EAV3}
\end{equation}
\begin{equation}
E_{01}({\rm meV}) = 0.470V_3^{0.548},
\label{eq:E01V3}
\end{equation}
\begin{eqnarray}
\nonumber \hbar\omega_t ({\rm meV}) = \hspace{2.5 cm}\\
0.655\left(1+\frac{V_3}{2.67}\right)^{1.06}\exp[-\left(V_3/4\right)^{0.5}].
\label{eq:hwtV3}
\end{eqnarray}
In these equations $V_3$ is given in units of K.
\\
ii) the preexponential factors $\Gamma_{\infty}$, $\gamma_s$, $\gamma_{AE}$ and $\gamma_{E_aE_b}$
for the different Arrhenius-like equations (\ref{eq:gcla},\ref{eq:shift},\ref{eq:gae},\ref{eq:gee}).
\\
iii) the crossover temperature $T_{\rm c}(V_{3})$, that determines which
regime -crossover if $T < T_{\rm c}(V_{3})$ or classical hopping 
if $T > T_{\rm c}(V_{3})$- governs the dynamics
of a given methyl group of barrier height $V_3$ at moderate temperature $T$.
\\\\
Glassy systems show broad distributions $g(V_{3})$. For that reason, shift, broadening, 
and merging of the individual peaks with the quasielastic component,
cannot be solved from the spectrum for the glass. On the other hand, it is expected 
that the temperature dependence of individual spectra is highly sensitive to the barrier height.
In principle, methyl groups with high values of $V_{3}$
will reach the classical hopping regime at higher $T_{c}$ than those with low values of $V_{3}$.
Moreover, there is a large set of parameters, namely the preexponential factors for shift 
and broadening, which depend on the unknown constants $\{g_k\}$ describing the coupling between
the methyl group and the phonon bath [Eqs. (\ref{eq:bosegamm},\ref{eq:boseshift})].
Such preexponential factors cannot be easily deduced for a glassy system,
in contrast to the crystalline case
\cite{sodacetprb,prager83,heidemann89acet,clough79,clough81acet,cockbain,cavagnat86jpc,%
prager86jcp,prager87jcp,prager88jcp,prager90jpcm},
where shift and broadening of the single peaks can be directly observed and analyzed 
(see Figs. \ref{fig:sodacetcry}, \ref{fig:crosparam}). 
As mentioned above, the typical values of $\Gamma_{\infty}$ observed 
in crystalline systems are in the same order of magnitude (a few meV), but $\gamma_{s}$, $\gamma_{AE}$ 
and $\gamma_{E_aE_b}$ can take a wide range of values.

All these problems enormously complicate the analysis of the spectra in glassy systems.
Next it is shown that many of the unknown present paramaters, namely
$\gamma_{AE}$, $\gamma_{E_aE_b}$ and $\gamma_{s}$, controlling the temperature
dependence of the crystal-like spectrum for each individual methyl group,
can be removed without the need of carrying out complex calculations, as it would be required by the
introduction of a detailed model for the coupling of the methyl group to the glass vibrations.
Instead, a functional relation between the crossover temperature and the barrier height,
$T_c=T_c(V_3)$, is assumed \cite{crossprb,pmmamacro,tolujcp,sodacetprb}.
It is found that a consistent description of the spectra in all the temperature range
is possible in terms of only three parameters: the average and standard deviation 
of the barrier distribution $g(V_3)$, and the preexponential factor $\Gamma_{\infty}$ for classical hopping.

In the approach introduced by the {\it general version} of the RRDM
\cite{crossprb,pmmamacro,tolujcp,sodacetprb}, the crossover temperature $T_{\rm c}$
is operationally defined as the temperature where the rate for coherent quantum tunnelling
becomes equal to the rate
for incoherent classical hopping, i.e., 
$\Gamma(T_{\rm c}) = \hbar\omega_{\rm t}$, or from Eq. (\ref{eq:gcla}):

\begin{equation}
kT_{\rm c} = \frac{E_{\rm A}}{\ln(\Gamma_{\infty}/\hbar\omega_{\rm t})}.
\label{eq:Tc}
\end{equation}
As mentioned above, the classical activation energy $E_{\rm A}$ and the tunnelling frequency 
$\omega_{\rm t}$ are direct functions of $V_{3}$.
It has been shown (see Refs. \cite{chahid94,tolujcp} for a detailed discussion)
that $\Gamma_{\infty}$ can be taken as a barrier-independent quantity in a good approximation. 
Therefore, $T_{\rm c}$ defined in this way depends only on the barrier height $V_{3}$ and the
preexponential factor for classical hopping $\Gamma_{\infty}$. This latter material-dependent
parameter can be interpreted
as a measure of the strength of the coupling of the methyl groups to the glass vibrations.
For a given methyl group of barrier $V_3$, the larger $\Gamma_\infty$ 
---i.e., the stronger the coupling---,
the lower will be the $T_c$ for the onset of classical hopping.
Two technical approximations are introduced:
\\\\
i) For each individual methyl group, the preexponential factors $\gamma_{AE}$ and $\gamma_{E_aE_b}$
are taken as equal (in the following they will be denoted as $\gamma_{b}$).
Therefore,

\begin{equation}
\Gamma_{AE} = \Gamma_{E_aE_b} = \gamma_{b}\exp(-E_{01}/k_{B}T).
\label{eq:brorrdm}
\end{equation}
\\
ii) The activation energy $E_{S}$ for the shift of the tunnelling lines is assumed 
to be equal to $E_{01}$:

\begin{equation}
\hbar\Delta\omega_{t} = \gamma_{s}\exp(-E_{01}/k_{B}T).
\label{eq:shiftrrdm}
\end{equation}
\\
The general experimental evidence in crystalline systems for the values of
$\gamma_{AE}$, $\gamma_{E_aE_b}$ and $E_S$ (see above)
supports the reliability of these two approximations. In addition, two constraints are imposed
to Eqs. (\ref{eq:brorrdm},\ref{eq:shiftrrdm}) at the crossover temperature:
\\\\
i) The tunnelling frequency must be shifted to zero at the onset of the classical regime,
$\Delta\omega_{\rm t}(T_{\rm c}) = \omega_{\rm t}$, as required by the disappearence of quantum effects
above $T_c$. Hence, from Eq. (\ref{eq:shiftrrdm}):

\begin{equation}
\gamma_{\rm s} = \hbar\omega_{\rm t}\exp(E_{01}/kT_{\rm c})
\label{eq:shifttozero}
\end{equation}
ii) Continuity condition for the HWHM of the Lorentzians at the merging point $T_c$, i.e.,
$\Gamma_{AE}(T_{\rm c}) = \Gamma_{E_aE_b}(T_{\rm c}) = \Gamma(T_{\rm c})$. Hence,
from Eqs. (\ref{eq:gcla},\ref{eq:brorrdm}):

\begin{equation}
\gamma_{\rm b}=\Gamma_{\infty}\exp[(E_{01}-E_{\rm A})/kT_{\rm c}] .
\label{eq:continuity}
\end{equation}
A straightforward consequence of Eqs. (\ref{eq:Tc},\ref{eq:shifttozero},\ref{eq:continuity})
is that $\gamma_{s}=\gamma_{b}$, which is also in agreement with the generally 
close values found for both quantities
in crystalline systems.

All the quantities at the right sides of Eqs. (\ref{eq:Tc},\ref{eq:shifttozero},\ref{eq:continuity})
depend exclusively on $V_{3}$ and on the barrier-independent
factor $\Gamma_{\infty}$. In this way, the preexponential factors for the crossover,
$\gamma_{b}$ and $\gamma_{s}$, can be determined unambiguously for each value of $V_3$
in the distribution $g(V_3)$. Hence, the temperature dependence of the spectra for the glass
is modelled exclusively in terms of $\Gamma_{\infty}$ and $g(V_{3})$. 
Two-parameter (average barrier and standard deviation) simple distributions are
used, as Gaussian \cite{chahid94,pvacprl,isotprb,crossprb,pmmamacro,sodacetprb,ahumada,%
mukhomacro,arrighi01,pillay,zorn,arrighi03macro} functions.
In a few cases \cite{tolujcp,tolconf} two-parameter Gamma functions are more adequate.
In this way, the RRDM only introduces three independent parameters for the analysis
of experimental spectra in all the temperature range.

In the limit of low and high temperature, all the individual methyl groups will perform respectively
rotational tunnelling or classical hopping, and the corresponding crystal-like contributions
to the spectrum of the glass [Eq. (\ref{eq:sqrrdm})]
will be governed respectively by Eq. (\ref{eq:sqtun}) for tunnelling
and by Eqs. (\ref{eq:gcla},\ref{eq:sqclass}) for hopping.
These limits are referred to as the {\it tunnelling} and {\it hopping} limits of the RRDM.
At intermediate temperature the functional relation between
$T_{\rm c}$ and $V_{3}$ introduced in Eq. (\ref{eq:Tc}) allows one to select the corresponding
dynamic regime for each individual methyl group ---i.e., crossover, governed by Eqs. 
(\ref{eq:sqcross},\ref{eq:brorrdm},\ref{eq:shiftrrdm},\ref{eq:shifttozero},\ref{eq:continuity}), 
or classical hopping, governed by Eqs. (\ref{eq:gcla},\ref{eq:sqclass}). 

\begin{figure}
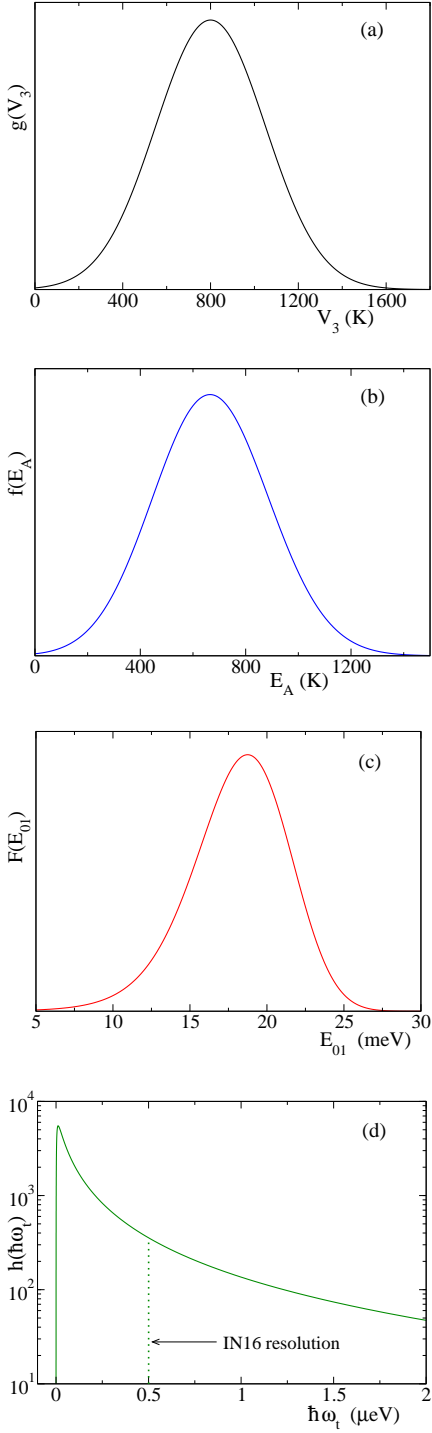

\begin{center}
\includegraphics[width=0.33\textwidth]{distV3.eps}
\newline
\newline
\includegraphics[width=0.33\textwidth]{distEA.eps}
\newline
\newline
\includegraphics[width=0.33\textwidth]{distE01.eps}
\newline
\newline
\includegraphics[width=0.33\textwidth]{distwt.eps}
\newline
\caption{Distributions $g(V_3)$, $f(E_A)$, $F(E_{01})$ and $h(\hbar\omega_{t})$
for an average barrier $\langle V_3 \rangle$ = 800 K  and a standard deviation $\sigma_V$ = 250 K.}
\label{fig:distrib}
\end{center}
\end{figure}

%Next we expose the data analysis procedure that provides
%the most reliable results when experimental spectra are analyzed
%in terms of the RRDM. 
In a first approximation, a Gaussian distribution of purely threefold rotational
barriers is assumed:

\begin{equation}
g(V_3) = \frac{1}{\sqrt{2 \pi}\sigma_{\rm V}}
\exp \left[ - \frac{(V_3-\langle V_3 \rangle)^{2}}{2\sigma_{\rm V}^2} \right] ,
\label{eq:gauss}
\end{equation}
with $\langle V_3 \rangle$ the average barrier and $\sigma_{\rm V}$ the standard
deviation of the distribution. See Fig. \ref{fig:distrib}a
for $\langle V_3 \rangle$ = 800 K, $\sigma_{\rm V}$ = 250 K, as an example
of parameters for which tunnelling and hopping features
are observable by neutron scattering (see below).

The functional relations ---e.g., Eqs. (\ref{eq:EAV3},\ref{eq:E01V3},\ref{eq:hwtV3})--- 
between $E_A$, $E_{0i}$, $\hbar\omega_{t}$, and $V_3$, 
allow for a straightforward transformation between $g(V_3)$ and the
corresponding distributions of activation energies for classical hopping, $f(E_{A})$, 
rotational tunnelling frequencies, $h(\hbar\omega_{t})$, and librational energies, $F(E_{0i})$:

\begin{eqnarray}
\nonumber g(V_3)dV_3 = f(E_A)dE_A = \\
F(E_{0i})dE_{0i} = -h(\hbar\omega_{t})d(\hbar\omega_{t}) .
\label{eq:transform}
\end{eqnarray}

\begin{figure}
\begin{center}
\includegraphics[width=0.45\textwidth]{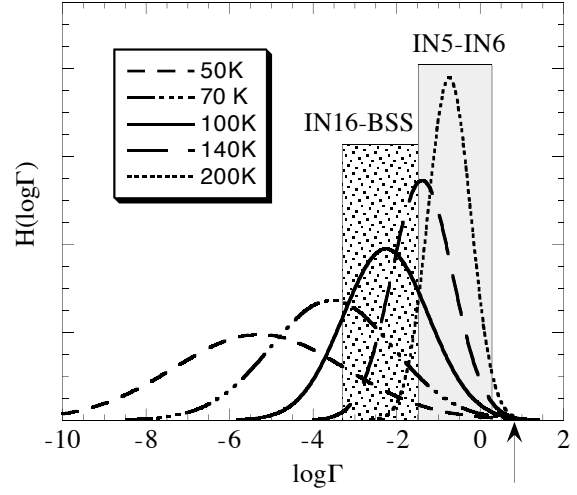}
\newline
\caption{Temperature dependence of the distribution $H(\log\Gamma)$,
for the set of parameters $\langle V_3\rangle$ = 800 K, $\sigma_V$ = 250 K and $\Gamma_{\infty}$ = 6 meV.
$\Gamma$ is given in meV. The value of $\Gamma_{\infty}$ is indicated by the arrow.
Shadowed areas correspond to the energy windows of some BS and TOF spectrometers.}
\label{fig:distHlogh}
\end{center}
\end{figure}

In a good approximation, the classical activation energy $E_{\rm A}$ depends linearly on $V_3$
(see Fig. \ref{fig:eneV3}c).
Hence, the distribution $f(E_{\rm A})$ is also Gaussian (see Fig. \ref{fig:distrib}b),
with average energy $\langle E_A\rangle$ and standard deviation $\sigma_{E}$.
%As mentioned above, the value of the classical activation energy is not very strongly
%dependent on the detailed shape of the rotational potential. However, in practice
%it is pure classical hopping which dominates methyl group dynamics in the widest
%temperature range of interest, and therefore it is convenient to first analyze
%high temperature data in order to have a first estimation of the parameters of the
%barrier distribution $g(V_3)$.
As mentioned above, the preexponential factor for classical hopping $\Gamma_\infty$ 
is assumed to be barrier-independent. This approximation allows one,
according to Eq (\ref{eq:gcla}), to transform the Gaussian distribution $f(E_{\rm A})$
into a log-Gaussian distribution of Lorentzian HWHM for classical hopping:

\begin{equation}
H(\log\Gamma) = \frac{1}{\sqrt{2 \pi}\sigma}
\exp \left[ - \frac{(\log\Gamma-\log\Gamma_0)^{2}}{2\sigma ^2} \right] ,
\label{eq:hlog}
\end{equation}
where
\begin{equation}
\Gamma_0 = \Gamma_\infty\exp(-\langle E_A\rangle/k_{B}T)
\label{eq:gamma0}
\end{equation}
is the corresponding Lorentzian HWHM for the average activation energy, and
\begin{equation}
\sigma = \sigma_{\rm E}\log(e)/k_{B}T
\label{eq:sigma}
\end{equation}
is the standard deviation for $H(\log \Gamma)$.
Fig. \ref{fig:distHlogh} shows the temperature dependence of this distribution
for the above parameters $\langle V_3 \rangle$ = 800 K, $\sigma_V$ = 250 K,
and for a typical value of $\Gamma_{\infty}$ = 6 meV. As can be seen, at high temperature 
the major part of the distribution
is accessible in the energy window of neutron scattering spectrometers.

The parameters $\Gamma_0$ and $\sigma$ at a given high temperature can be obtained from a 
fitting procedure of the experimental methyl group incoherent contribution 
$\nonumber S_{\rm inc}^{\rm MG}(Q,\omega)$ to the 
classical limit of Eq. (\ref{eq:sqrrdm}).
From the transformation $g(V_3)dV_3 = -H(\log\Gamma)d(\log\Gamma)$, and from the incoherent
scattering function (\ref{eq:sqclass}) for a single barrier, Eq. (\ref{eq:sqrrdm}) 
can be rewritten in the classical limit as:

\begin{eqnarray}
\nonumber S_{\rm inc}^{\rm MG}(Q,\omega) = \frac{1+2j_0(Qr)}{3}+\\
\frac{2[1-j_0(Qr)]}{3}\int_{-\infty}^{\infty} H(\log\Gamma)L(\omega ;\Gamma)d(\log\Gamma) .
\label{eq:sqrrdmcla}
\end{eqnarray}
The set of values $\Gamma_0 (T)$ and $\sigma (T)$ obtained for the different temperatures
is fitted to Eqs. (\ref{eq:gamma0},\ref{eq:sigma}) in order to obtain
$\Gamma_{\infty}$, $\langle E_A\rangle$ and $\sigma_{E}$. Finally, the knowledge of these two latter
quantities allows, by transformation, to determine $\langle V_3 \rangle$ and $\sigma_V$, i.e., the
parameters of the barrier distribution $g(V_3)$.
The latter, as obtained from the analysis of classical 
high temperature spectra, can next be transformed into the distribution
of tunnelling frequencies $h(\hbar\omega_{\rm t})$
and librational energies $F(E_{01})$  (see Figs. \ref{fig:distrib}c, \ref{fig:distrib}d).
The consistency of the RRDM requires that these two latter distributions
derived in this way, also reproduce, respectively, the corresponding experimental tunnelling
and librational spectra.

Due to the approximate exponential dependence of $\hbar\omega_{\rm t}$ 
on $V_3$ (see Eq. (\ref{eq:hwtV3}) or Fig. \ref{fig:eneV3}a), 
$h(\hbar\omega_{\rm t}$) takes an extremely asymmetric
shape (see Fig. \ref{fig:distrib}d) with the maximum shifted to low frequencies.
From the incoherent scattering function for a single barrier (\ref{eq:sqtun}),
the tunnelling limit of Eq. (\ref{eq:sqrrdm}) is given by:

\begin{eqnarray}
\nonumber S_{\rm inc}^{\rm MG}(Q,\omega) = \frac{5+4j_0(Qr)}{9}
\delta(\omega) + \\  
\frac{2[1-j_0(Qr)]}{9}[h(\omega_{\rm t})+h(-\omega_{\rm t})] .
\label{eq:sqrrdmtun}
\end{eqnarray}
For sufficiently high average barriers $\langle V_3 \rangle$ or broad distributions $g(V_3)$,
the maximum of $h(\pm \hbar\omega_{\rm t}$) is placed
beneath the instrumental resolution (see Fig. \ref{fig:distrib}d). Hence, the tunnelling
spectrum shows an apparently quasielastic contribution, which could be misinterpreted as
a signature of hopping events at very low temperature. However, its actually {\it inelastic} origin
finds a natural explanation, within the framework of the RRDM, in terms of 
a distribution of rotational tunnelling lines. 

Once the tunnelling and hopping limits of the RRDM are able to give account,
with the same set of parameters $\langle V_3\rangle$, $\sigma_{V}$ and $\Gamma_{\infty}$, respectively
for the experimental spectra at very low and high temperature, the consistency 
of the general version of the RRDM is tested, with such parameters, 
at intermediate temperature. It must be noted that this step
does not involve any further fitting procedure.
Instead, experimental spectra are directly compared with the theoretical ones, which are constructed
according to the procedure exposed above for the general version of the RRDM,
and by making use of the operational definition of the crossover temperature,
Eq. (\ref{eq:Tc}), in order to select the dynamic regime (crossover or classical) 
for each individual methyl group.

\begin{figure}
\begin{center}
\includegraphics[width=0.35\textwidth]{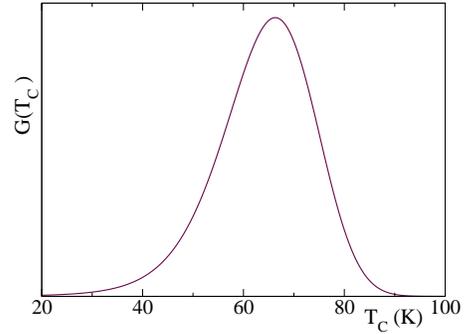}
\newline
\caption{Distribution of crossover temperatures $G(T_c)$,
for the same set of parameters $\langle V_3\rangle$ = 800 K, $\sigma_V$ = 250 K
and $\Gamma_{\infty}$ = 6 meV of Figs. \ref{fig:distrib} and \ref{fig:distHlogh}.}
\label{fig:distTc}
\end{center}
\end{figure}

Eq. (\ref{eq:Tc}) also introduces a distribution of crossover temperatures, $G(T_c)$.
Fig. \ref{fig:distTc} shows the latter, for the same parameters,
$\langle V_3\rangle$ = 800 K, $\sigma_{V}$ = 250 K and $\Gamma_{\infty}$ = 6 meV,
of Figs. \ref{fig:distrib} and \ref{fig:distHlogh}. 
As temperature $T$ increases, the fraction of methyl groups that reach
the classical hopping regime, i.e., the area of $G(T_c)$ for $T_c < T$, 
will also increase and will be negligible above some high temperature, where quantum 
effects will be unsolvable from the dominating hopping dynamics. In such conditions,
the system will be well described by the classical limit of the RRDM.

\subsection{Other approaches}

An alternative approach to the classical limit of the RRDM, also based on the idea of an
underlying distribution of rotational barriers, has been introduced by Arrighi {\it et al.}
\cite{arrighi95,arrighi96} for analyzing methyl group dynamics
in the high temperature hopping regime. By Fourier transforming into the time domain,
the intermediate incoherent scattering function for methyl group classical hopping is given by:

\begin{equation}
I_{\rm inc}^{\rm MG}(Q,t)= \frac{1+2j_0(Qr)}{3} + \frac{2[1-j_0(Qr)]}{3}{\cal F}(t) .
\label{eq:iqt}
\end{equation}
Obviously for a crystalline system $F(t) = \exp(-t/\tau)$, with $\tau$ the {\it single}
residence time for hopping. 

For a disordered system, the corresponding function ${\cal F}(t)$ in the framework of the RRDM is
a log-Gaussian distribution of time exponential functions:

\begin{equation}
{\cal F}(t)=\int_{-\infty}^{\infty} H(\log\tau)\exp(-t/\tau)d(\log\tau) ,
\label{eq:rrdmfttau}
\end{equation}
as trivially obtained by Fourier transformation of Eq. (\ref{eq:sqrrdmcla}).
In the approach introduced by Arrighi {\it et al.}, no functional form is assumed
for the distribution $H(\log\tau)$. On the contrary, ${\cal F}(t)$ is effectively represented
by a stretched exponential, or Kohlraush-Williams-Watts (KWW) function, $\exp[-(t/\tau_0)^{\beta}]$,
with $\tau_0$ a characteristic rotational relaxation time.
The stretching exponent $\beta$ takes values between 0 and 1.

\begin{figure}
\begin{center}
\includegraphics[width=0.44\textwidth]{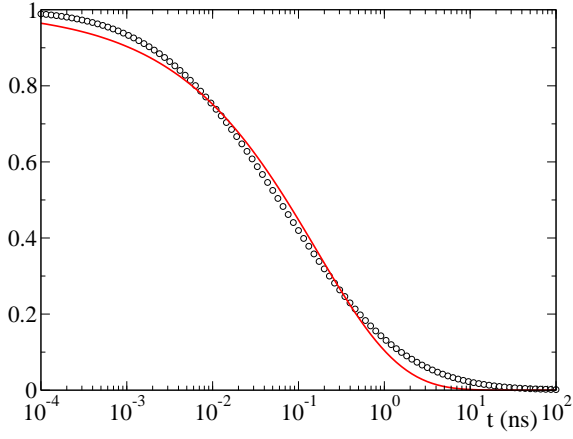}
\newline
\caption{Comparison between ${\cal F}(t)$ given by the RRDM expression
(\ref{eq:rrdmfttau}), for parameters $\tau =$ 0.1 ns and $\sigma =$ 1 (circles),
and a fit to a KWW function (solid line) in a wide dynamic range.} 
\label{fig:kwwfit}
\end{center}
\end{figure}

A KWW function can be effectively represented as a log-distribution of time exponential functions:

\begin{equation}
\hspace{-1.5 mm}\exp[-(t/\tau_0)^{\beta}] = \int_{-\infty}^{\infty} K(\log\tau)\exp(-t/\tau)d(\log\tau).
\end{equation}
The distribution function $K(\log\tau)$, which can be obtained by inverse Laplace transformation methods
\cite{alv95}, is asymmetric. Therefore it is clear that the KWW function cannot be equivalent to
the RRDM distribution ${\cal F}(t)$ if a log-Gaussian form is selected for $H(\log\tau)$ \cite{mukhomacro},
(note that for the same reason $\tau_0 \neq \tau)$. This fact is illustrated in Fig. \ref{fig:kwwfit}
for typical values $\tau$ = 0.1 ns and $\sigma =1$.

\begin{figure}
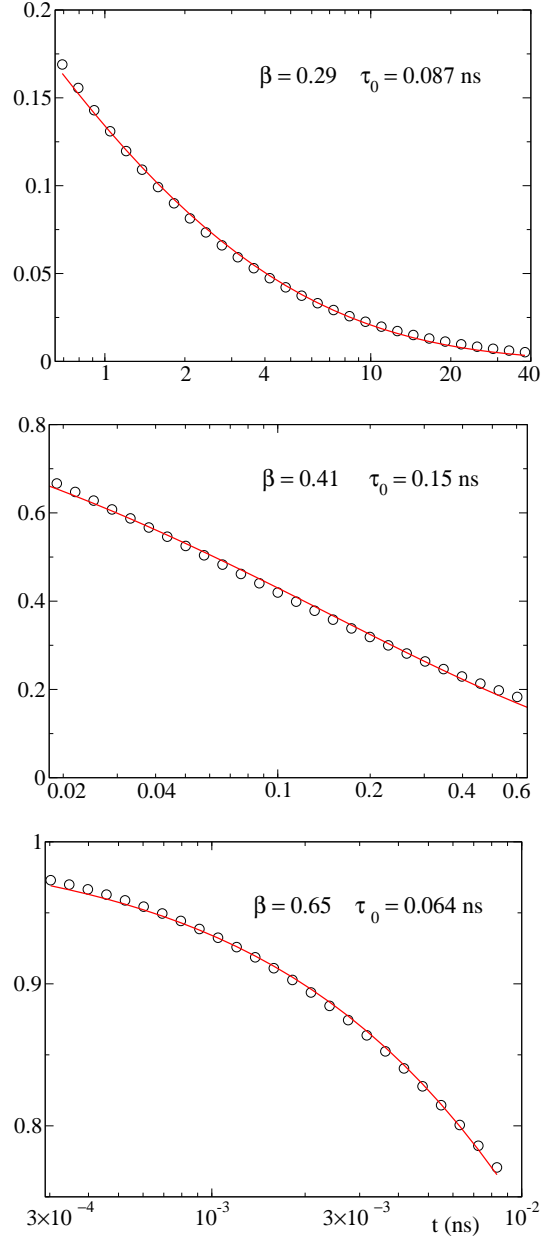

\begin{center}
\includegraphics[width=0.41\textwidth]{kwwnar1.eps}
\newline
\newline
\includegraphics[width=0.40\textwidth]{kwwnar2.eps}
\newline
\newline
\includegraphics[width=0.41\textwidth]{kwwnar3.eps}
\newline
\caption{As Fig. \ref{fig:kwwfit}, but KWW fits are done
in different narrow intervals covering, as in usual experimental
conditions, 1-2 decades in time. The obtained values
of $\beta$ and $\tau_{0}$
 for the KWW description are shown.} 
\label{fig:kwwbeta}
\end{center}
\end{figure}

Both pictures (KWW and RRDM) achieved descriptions of similar quality for experimental spectra
in a limited energy window, i.e., by using a single spectrometer.
The corresponding times $\tau$ and $\tau_0$ followed Arrhenius behavior with close values
of the respective activation energies \cite{arrighi95,arrighi96,mukhomacro}. 
However, an analysis in a wide dynamic range
\cite{mukhomacro}, by combining several spectrometers, showed that,
while an excellent agreement of the experimental ${\cal F}(t)$ with the RRDM function 
(\ref{eq:rrdmfttau}) was obtained in all the time interval,
a simultaneous fit to a KWW function was not satisfactory.
Moreover, independent fits in different experimental windows provided incompatible
values of $\tau_0$ and $\beta$. This result is illustrated in
Fig. \ref{fig:kwwbeta} for the same function ${\cal F}(t)$ of Fig. \ref{fig:kwwfit},
now analyzed in different narrow dynamic windows.

Another disadvantage of using a KWW functional form for ${\cal F}(t)$ is that,
in contrast to the unified picture introduced by the RRDM, 
it does not provide a direct physical connection between the classical
and quantum features observed respectively at high and low temperature.
It is worthy of remark that MDS support,
by following separately the behavior of each individual methyl group in the system, 
the general validity of the Gaussian approach for the distribution
of rotational barriers \cite{nic97,sae00,ahumada,karatasos02}.

\section{Application of the rrdm to a showcase: poly(vinyl acetate)}
\label{sec:pvac}

\begin{figure}
\begin{center}
\includegraphics[width=0.32\textwidth]{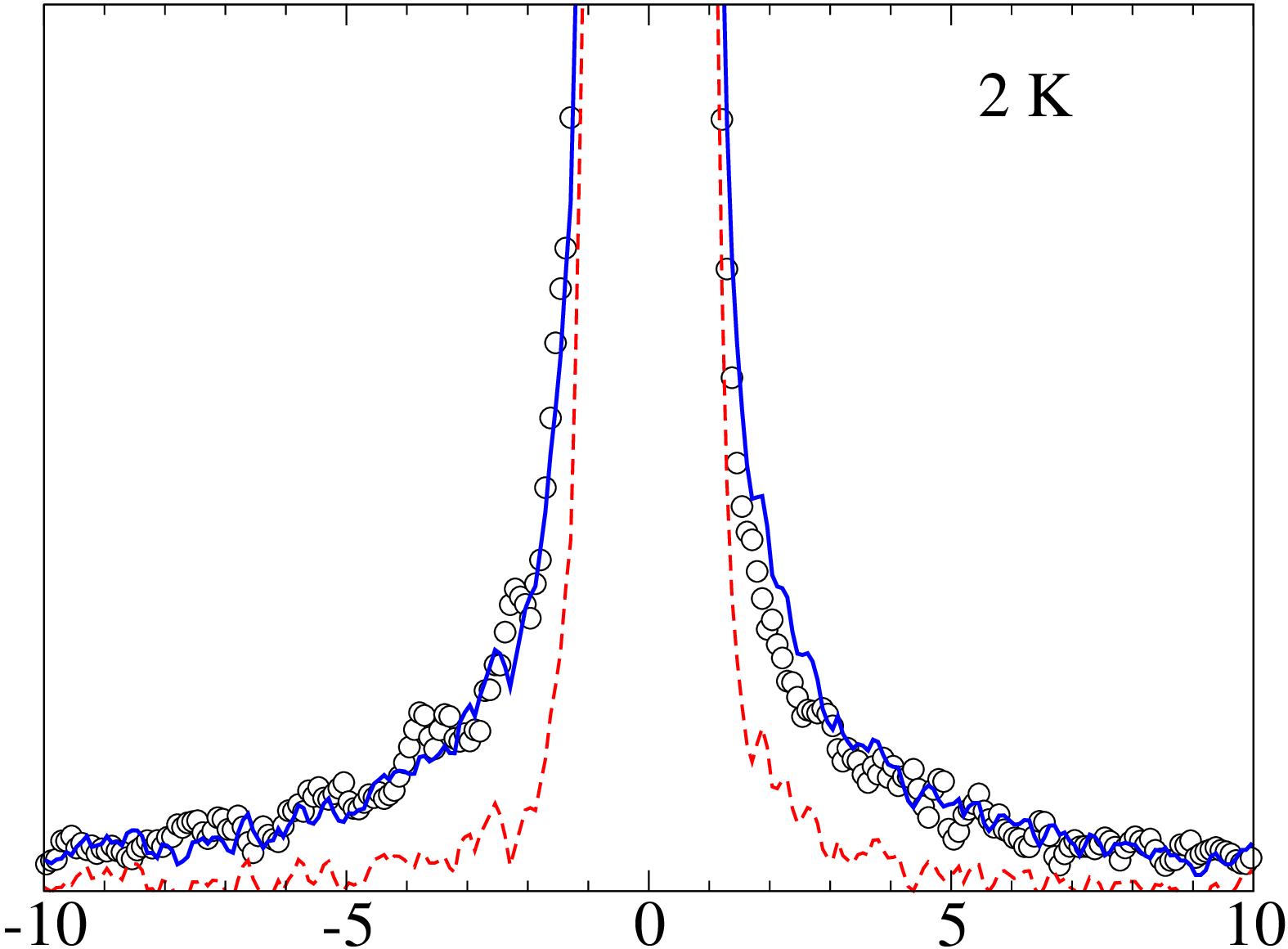}
\newline
\newline
\includegraphics[width=0.32\textwidth]{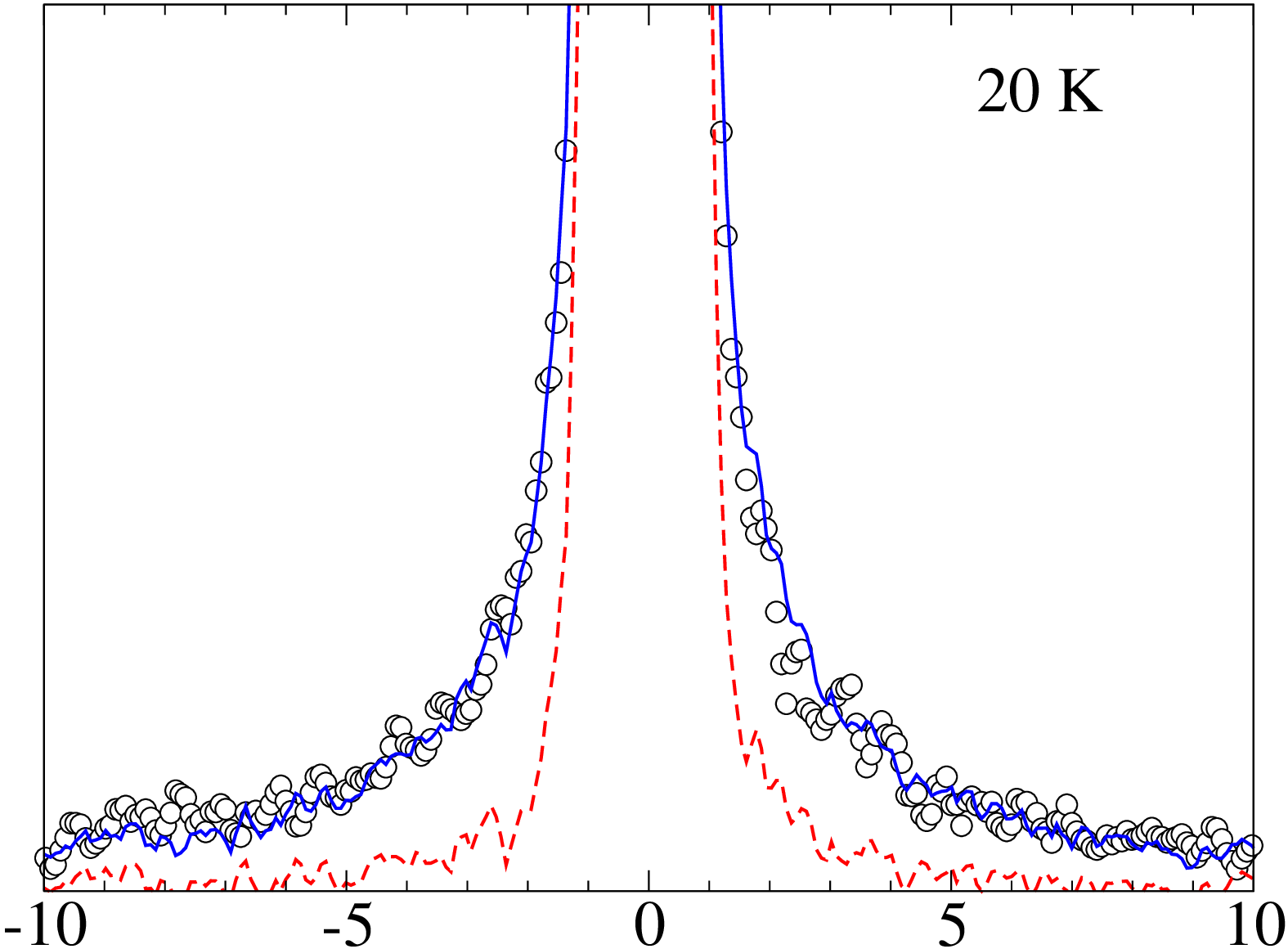}
\newline
\newline
\includegraphics[width=0.32\textwidth]{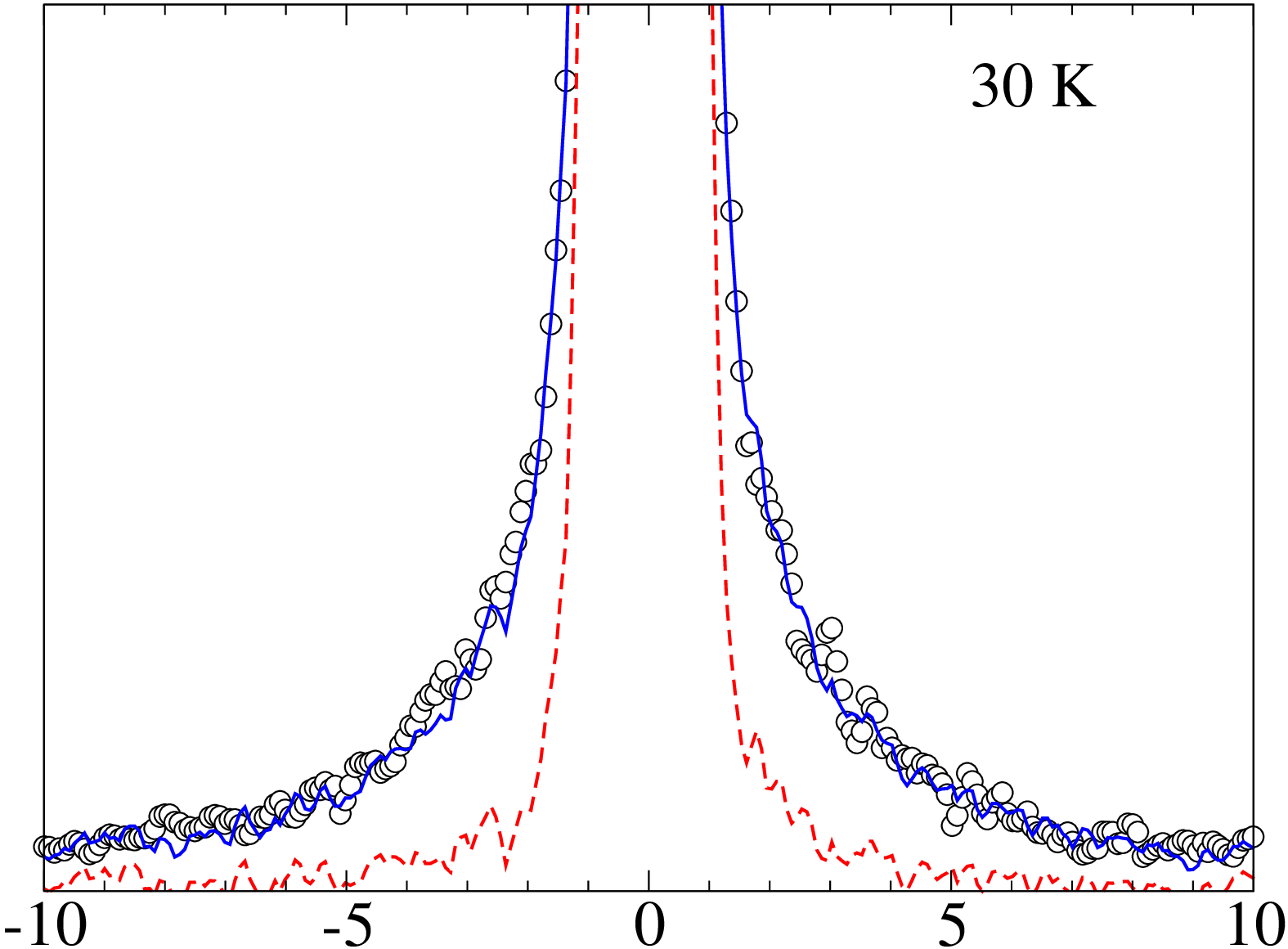}
\newline
\newline
\includegraphics[width=0.32\textwidth]{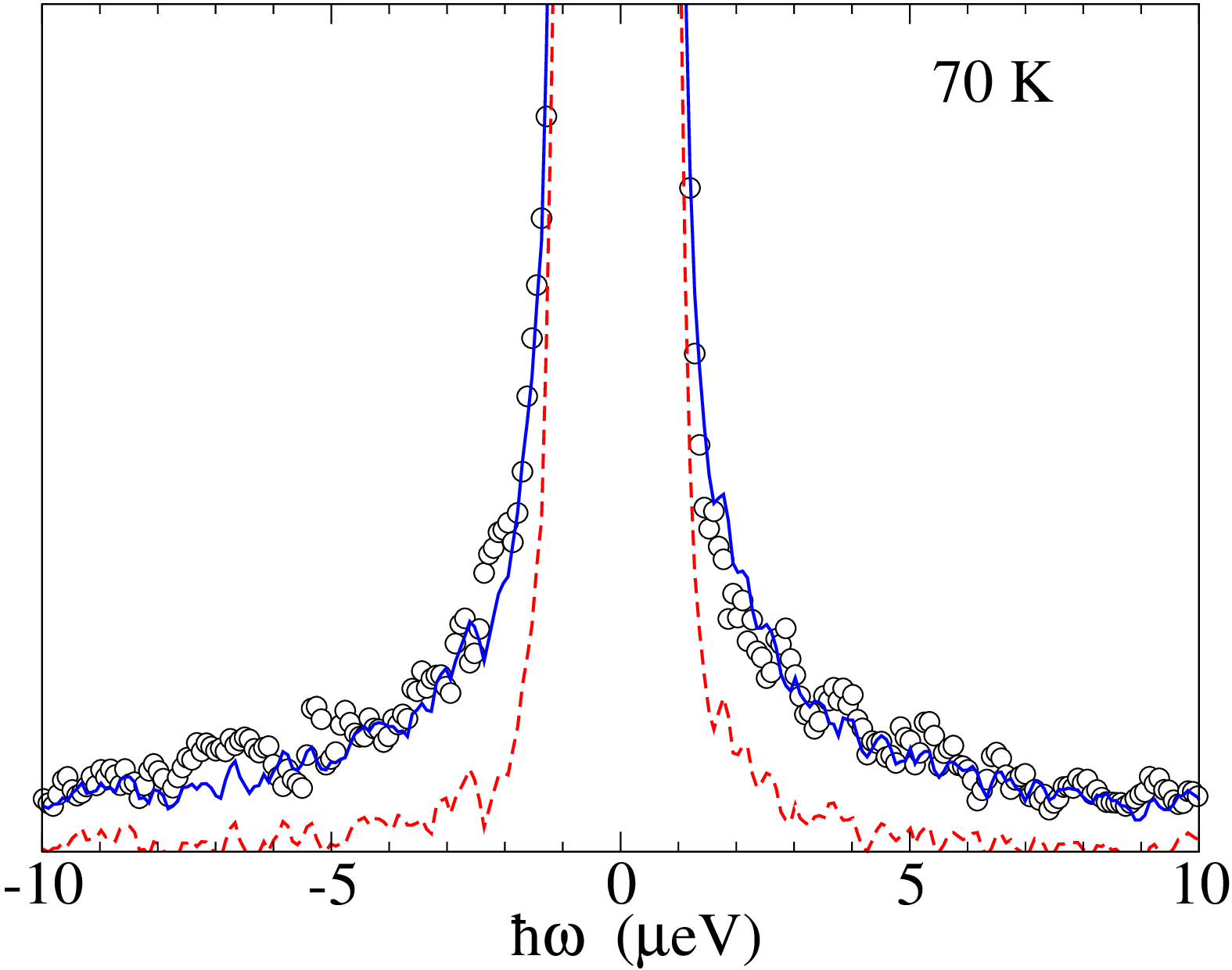}
\newline
\caption{IN16 spectra for PVAc at $Q=1.8 {\rm \AA}^{-1}$ and different temperatures.
Circles are experimental data. Solid lines
correspond to the RRDM description. Dashed lines are the resolution function.
The scale is a 5\% of the maximum.} 
\label{fig:pvacin16}
\end{center}
\end{figure}

PVAc was the first polymer ---and indeed the first highly disordered system---
where a complete investigation by neutron scattering was carried out
on all the features of methyl group rotation (tunnelling, crossover, hopping and librations).
It was also the first system where a fully consistent description of the experimental results
was achieved in terms of the RRDM \cite{pvacprl,crossprb,mukhomacro}.
Figs. \ref{fig:pvacin16}, \ref{fig:pvacin6} show respectively for a BS and a TOF
spectrometer, experimental spectra (circles) for PVAc, at several 
temperatures in the tunnelling ($T$ = 2 K), crossover ($T$ = 20 and 30 K) and classical hopping
($T$ = 70, 120, 160 and 200 K) regimes. All these temperatures are far below
the glass transition temperature ($T_g \approx 315$ K), where motions different from
methyl group rotation just contribute as vibrational dynamics via the Debye-Waller factor.

\begin{figure}
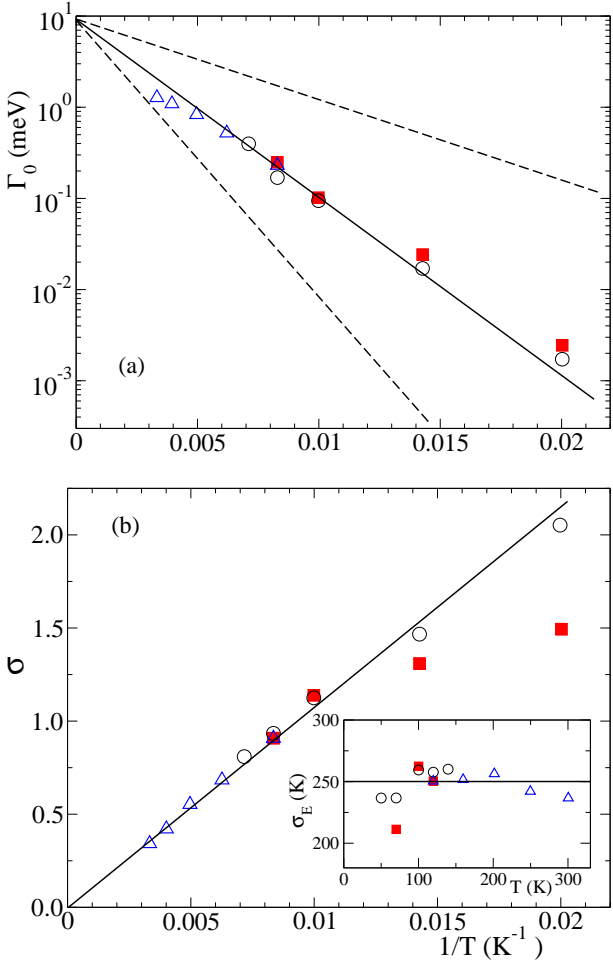

\begin{center}
\includegraphics[width=0.47\textwidth]{h0pvac.eps}
\newline
\newline
\includegraphics[width=0.47\textwidth]{sigpvac.eps}
\newline
\caption{Temperature dependence of $\Gamma_{0}$ (a) and $\sigma$ (b) for PVAc.
Different symbols indicate the different used spectrometers: 
IN16 (squares), IN5 (circles) and IN6 (triangles). Solid lines for $\Gamma_{0}$ and $\sigma$
are respectively fits to Eqs. (\ref{eq:gamma0}) and (\ref{eq:sigma}). 
The inset in panel (b) shows the values of $\sigma_{E}$ obtained 
at each temperature from the corresponding values of $\sigma (T)$ {\it by transformation}
through Eq. (\ref{eq:sigma}). The horizontal line corresponds to the value obtained
{\it by fitting} the latter values to Eq. (\ref{eq:sigma}).
Dashed lines in panel (a) correspond
to $\Gamma=\Gamma_{\infty}\exp[-(\langle E_A \rangle\pm\sigma_{E})/k_{B}T]$.} 
\label{fig:pvach0sig}
\end{center}
\end{figure}

An excellent agreement with the RRDM is observed in all the temperature range (solid lines). 
As in the usual analysis procedure, high temperature data of PVAc were
first analyzed in terms of the classical RRDM \cite{mukhomacro}.
Fig. \ref{fig:pvach0sig} shows the temperature dependence of the parameters $\log\Gamma_0$
and $\sigma$ of the distribution $H(\log \Gamma)$ (taken as Gaussian) obtained from the analysis in terms
of the hopping limit of the RRDM. The use of several spectrometers with different energy
resolutions and dynamic ranges reduced the uncertainties of the so 
determined parameters.  The observed deviations below $T \approx 70$ K
from the linear behavior predicted by Eqs. (\ref{eq:gamma0},\ref{eq:sigma}) show the 
relevance of quantum effects at low temperatures. The fact that the magnitude of such deviations
depend on the experimental dynamic window evidences the importance of using several
spectrometers in order to avoid, below some temperature, 
an analysis biased by the used instrumental resolution.
The results of the fits for $T > 70 $ K to Eqs. (\ref{eq:gamma0},\ref{eq:sigma}) provide 
a preexponential factor $\Gamma_{\infty} = 9.1$ meV, 
an average classical activation energy $\langle E_A \rangle = 450$ K, and
a standard deviation $\sigma_{E} = 250$ K. 
Transformation to $g(V_3)$ yields $\langle V_3 \rangle = 534$ K
and $\sigma_{V}=$ 274 K. The dashed lines in Fig. \ref{fig:pvach0sig},
corresponding to $\Gamma=\Gamma_{\infty}\exp[-(\langle E_A\rangle \pm\sigma_{E})/k_{B}T]$,
have been introduced in order to stress the presence of a wide distribution of hopping times,
which progressively spreads over several orders of magnitude as temperature decreases.
 
The corresponding distribution of tunnelling frequencies $h(\hbar\omega_{t})$ is obtained
by transformation of the barrier distribution $g(V_3)$ independently derived  
from the high temperature classical analysis. Once $h(\hbar\omega_{t})$ is known,
the theoretical $S_{\rm inc}^{\rm MG}(Q,\omega)$ is constructed according to Eq. (\ref{eq:sqrrdmtun}).
An excellent agreement between the tunnelling limit of the RRDM
and the experimental spectrum at 2 K is obtained (Fig. \ref{fig:pvacin16}).
The maximum of $h(\hbar\omega_{t})$ is at $\hbar\omega_t \approx$ 0.03 $\mu$eV \cite{pvacprl}, 
i.e., well beneath the instrumental resolution, leading to the apparently
quasielastic observed intensity . 

For the case of the librational distributions, a direct comparison with experimental results
is complicated, since librational energies are strongly superposed with the phonon
spectrum. This is particularly problematic in PVAc, which cannot be selectively deuterated
in order to attenuate the intensity scattered by nuclei different from methyl group protons.
Fig. \ref{fig:pvac-tosca} shows a comparison of the generalized VDOS of PVAc 
with the distribution of librational energies $F(E_{01})$ 
obtained by transformation of $g(V_3)$. A reasonable agreement between the maximum of the experimental peak
and the maximum of $F(E_{01})$ is achieved, supporting the validity of the previuosly obtained
distribution $g(V_3)$ and the consistency of the RRDM picture.
It must be stressed that, since the generalized VDOS is not
the real density of states (see Section \ref{sec:neutscat}), a rigorous comparison between the
widths of the experimental librational peaks and the theoretical librational
distributions cannot be made.

\begin{figure}
\begin{center}
%\vspace{-2 mm}
\includegraphics[width=0.42\textwidth]{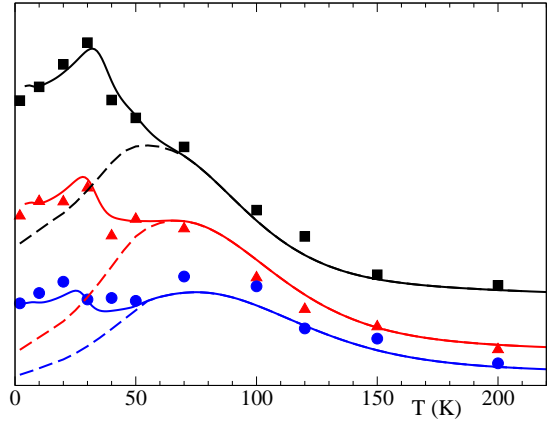}
\newline
\caption{Experimental integrated inelastic intensities for PVAc
at $Q=1.8$ ${\rm \AA}^{-1}$ in the intervals 1-2.5 $\mu$eV (squares),
2.5-6 $\mu$eV (triangles) and 6-10 $\mu$eV (circles). Solid lines
correspond to the theoretical description by the general version of the RRDM.
Dashed lines are an extrapolation of the classical limit of the RRDM to low temperatures.} 
\label{fig:pvacinteg}
\end{center}
\end{figure}

Once it has been checked that the distribution of potential barriers $g(V_3)$ can give 
account, by transformation to the other distributions in Eqs. (\ref{eq:transform}, \ref{eq:hlog}),
for the spectra  in the tunnelling and hopping dynamic limits, as well as for the librational
peak observed in the generalized VDOS, the consistency of the general RRDM is confirmed
by also reproducing the experimental spectra 
at intermediate temperatures (Fig. \ref{fig:pvacin16} for $T$ = 20 and 30 K). 
An excellent agreement is again achieved. A more detailed comparison 
between experiment and theory can be obtained by investigating the temperature
dependence of the integrated intensities in different inelastic windows, 
as shown in Fig. \ref{fig:pvacinteg}. Since Eq. (\ref{eq:sqrrdmtun}) is temperature-independent,
and the Debye-Waller factor is a decreasing function of $T$,
the initial increase of the integrated intensities when heating the system 
makes clear that the tunnelling limit of the RRDM is not appliable for $T \gtrsim 2$ K. 
It must be stressed that the observed double-peak structure is not an experimental artifact.
Indeed, it can also be observed in a crystalline system when the inelastic window
is properly selected, e.g., at lower energy but close to the single tunnelling line. 
As mentioned in Section \ref{sec:theorcryst},
when increasing temperature the tunnelling line broadens 
and shifts to lower energies until it merges, when reaching the hopping regime,
into a single quasielastic line. As a consequence, the intensity measured
in the selected inelastic window will pass through a first maximum. A second maximum will be
observed at higher temperature when broadening of the quasielastic line reaches the selected
inelastic window. In glassy PVAc, due to the broad distribution of tunnelling frequencies in comparison
with the energy window accesible by the spectrometer, the double-peak structure is 
observed at any inelastic window. 

As shown in Fig. \ref{fig:pvacinteg}, the double-peak structure
is nicely reproduced by the theoretical curves (solid lines) obtained by integration of
the RRDM scattering function in the corresponding inelastic windows. The theoretical curves
quantitatively reproduce the experimental intensities in all the temperature range. 
It must be noted that the formers have been modulated by a Debye-Waller factor
%$\exp(-2Q^2\langle u^2 \rangle/3)$, with $\langle u^2 \rangle = \theta T$,
$e^{-2Q^2\langle u^2 \rangle/3}$, with $\langle u^2 \rangle = \theta T$ 
and $\theta = 3\times 10^{-4}$ ${\rm \AA^2 K^{-1}}$ \cite{pvacprl}.
The theoretical curves corresponding to the classical hopping limit of the RRDM (dashed lines)
have been extrapolated to very low temperature in order to stress the relevance
of quantum effects below $T\approx 70$ K, as was pointed out above.

\section{Summary of experimental results by neutron scattering on polymers}
\label{sec:summexper}
\subsection{Classical hopping}

%Quasielastic neutron scattering (QENS) allows one to investigate rotational hopping,
%which manifests as a broad line around the elastic line \cite{bee}.
From early 80's, neutron scattering has been widely used to investigate
methyl group dynamics in the quasielastic energy range $|\hbar\omega| \lesssim 2$ meV,
or by using the FEW technique (see Section \ref{sec:neutscat}).
A huge collection of data has been obtained for several polymer systems, 
including standard materials as PVAc \cite{mukhomacro,tyagi},
PMMA \cite{gab84,flo92,arrighi95,arrighi96,pmmamacro,pillay,arr94},
poly(propylene oxide) (PPO) \cite{allen77}, poly(vinyl methyl ether)
(PVME) \cite{arrighi95,chahid94,chahid93,cha94c},
polyisoprene (PI) \cite{frickpip,zorn}, neat poly(dimethyl siloxane) (PDMS)
\cite{arrighi03macro,fri96,gra87,arr03prl},
and filled with silica nanoparticles \cite{arr98pol}, head-to-tail  \cite{ahumada,arrighi01},
and head-to-head (hh) \cite{perez} polypropylene (PP),
poly(ethylene propylene) (PEP) \cite{perez},
and poly(methyl phenyl siloxane) (PMPS)  confined in silica based nanoporous glasses  \cite{sch04},
polymer blends as solution chlorinated polyethylene (SCPE)/PMMA \cite{flo92,flo92phys}, 
polystyrene (PS)/ PVME \cite{arrblend,mukhoblend}, hh-PP/PEP\cite{perez}, 
or poly(ethylene oxide) (PEO)/PMMA \cite{genix,maranas}, thin film photoresist polymers \cite{soles03},
and other polymer based compounds as polymer electrolites
\cite{carl98,and99,karl03},
proteins \cite{fit95}, liquid crystalline
polyesters and polyethers \cite{arr97,arr99} or propylene glycol oligomers 
in the bulk state \cite{swe02} and confined in clay \cite{swe02clay,swe05clay}.
Very recently, Russina {\it et al.} have investigated methyl group classical hopping
in a neat, and doped with an inorganic salt, PEO-PPO random copolymer \cite{russina}.
Differently from the usual FEW technique, data have been acquired and analyzed
in an {\it inelastic} fixed energy window centered at $\hbar\omega \approx$ -15$\mu$eV with BS resolution. 

%Experimental results in some of the mentioned systems have been analyzed in terms of the RRDM.
%Table \ref{table:rrdm} shows the corresponding RRDM parameters for such systems.
A simple way to investigate the temperature dependence of methyl group rotation is by following
the temperature evolution of the FEW intensity, $I_{FEW}(Q,T)$ (see Section \ref{sec:neutscat}).
If the corresponding energy window for data acquisition
is $-\Lambda_{el} \le \hbar\omega \le \Lambda_{el}$, then integration
of Eq. (\ref{eq:sqrrdmcla}), together with the additional elastic contributions
(see Section \ref{sec:theorcryst}), provides the RRDM function
for the FEW intensity:
\\
\begin{eqnarray}
\nonumber I_{FEW}(Q,T)= e^{-2W(Q,T)}[\sigma_{\rm coh}S(Q) + \\
\sigma_{\rm inc}-\sigma^{\rm MG}_{\rm inc} + 
\sigma^{\rm MG}_{\rm inc}I_{FEW(inc)}^{\rm MG}(Q,T)],
\end{eqnarray}
where the incoherent contribution for methyl group classical
hopping is given by \cite{arrighi03macro}:

\begin{eqnarray}
\nonumber I_{FEW(inc)}^{\rm MG}(Q,T) = \frac{1+2j_0(Qr)}{3}+
\frac{2[1-j_0(Qr)]}{3}\times \\
\int_{-\infty}^{\infty} H(\log\Gamma)
\frac{2}{\pi}{\rm arctg}\left(\frac{\Lambda_{el}}{\Gamma}\right) d(\log\Gamma) .
\label{eq:sqrrdmfew}
\end{eqnarray}
If temperature is sufficiently low so that most of the methyl groups 
rotate with a hopping rate $\Gamma \ll \Lambda_{el}$, the limit 
$(2/\pi){\rm arctg}(\Lambda_{el}/\Gamma) \rightarrow 1$ 
is fulfilled and $I_{FEW(inc)}^{\rm MG}(Q,T) =1$. In such conditions the 
temperature dependence of the FEW intensity is just controlled by the
Debye-Waller factor, $I_{FEW}(Q,T)= e^{-2W(Q,T)}[\sigma_{\rm coh}S(Q) +\sigma_{\rm inc}]$.
It must be reminded that below some temperature the classical hopping picture is not valid,
and quantum effects dominate methyl group dynamics. However, 
from integration of Eqs. (\ref{eq:sqrrdmtun})
and (\ref{eq:sqcross},\ref{eq:sqrrdm}) in the FEW window, the latter result
can be extrapolated to very low temperature if, for most of the methyl groups, 
$\Gamma_{AE}$, $\Gamma_{E_a E_b}$ and $\omega_t$ are all smaller than $\Lambda_{el}$. 
  
\begin{figure}
\begin{center}
\includegraphics[width=0.41\textwidth]{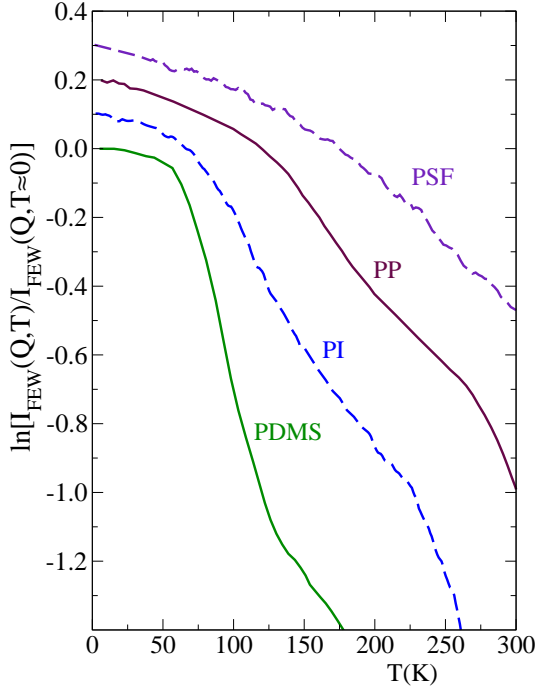}
\newline
\caption{Temperature dependence, for several polymers, 
of $\ln[I_{FEW}(Q,T)/I_{FEW}(Q,T\approx 0)]$, 
with $I_{FEW}(Q,T)$ the FEW intensity,
measured in BS spectrometers. For clarity, data are consecutively shifted by 0.1.
$Q=$ 1.8 ${\rm\AA^{-1}}$ for PDMS and PSF; $Q=$ 1.95 ${\rm\AA^{-1}}$ for PI and atactic PP.} 
\label{fig:few}
\end{center}
\end{figure}

FEW data for several polymers \cite{frickpip,arrighi01,arrighi03macro}, 
obtained by means of high resolution BS
spectrometers, are shown in Fig. \ref{fig:few}. Intensities are normalized
to their values at $T \rightarrow 0$.
The onset of methyl group dynamics with time scales shorter than that
corresponding to the instrumental resolution ($\sim 5$ ns for the
used spectrometers) manifests as a step-like decrease of the normalized intensity,
which, above some temperature, deviates from the exponential (linear in the 
logarithmic scale of Fig. \ref{fig:few}) decrease
corresponding to the Debye-Waller factor, i.e., to pure vibrational dynamics.
The second step at the very high temperature range corresponds
to the onset of much slower processes such as secondary relaxations or motions
involved in the glass transition process. 
%The temperature for the inflection point
%of the first step-like decrease allows one to obtain an estimate of the activation energy
%for classical hopping. 

The FEW technique requires much less acquisition time than QENS measurements, but obviuosly
quasielastic spectra are preferred for an accurate determination 
of the parameters of the distribution of hopping rates. Though for moderate average barriers,
analysis of FEW data have provided reliable results \cite{arrighi01},
this procedure is not recommended when the average
rotational barrier is very high ($V_{3} \gtrsim 2000$ K), since in
such cases the two step-like decays of the FEW intensity overlap.
Such a case is illustrated for polysulfone (PSF) in Figure \ref{fig:few}.

First quasielastic investigations on methyl group classical rotation in polymers
were analyzed in terms of the usual approach
for molecular crystals, i.e., by using a single Lorentzian function (single hopping rate)
to model the quasielastic component.
As mentioned in the Introduction, such a procedure fails when applied to polymers
(and in general to highly disordered sysems), providing inconsistent and unphysical results.
In polymeric materials, and in general in disordered systems \cite{tolujcp,sodacetprb}, the effect
of structural disorder has to be taken into account to describe properly methyl group dynamics.
%As it was also commented above, the more succesful approach was that referred to
%as RRDM \cite{cha94,mukhomacro,pmmamacro,mor02} in which a Gaussian
%distribution of potential barriers is considered in a first approach.
In the framework of the RRDM, it is assumed that the only effect
of the structural disorder on methyl group dynamics is to introduce a distribution of rotational
barriers $g(V_3)$, originating from the different local environments felt by the
individual methyl groups. The initial version of the RRDM ---the hopping limit---
was introduced for analyzing quasielastic data from PVME \cite{chahid94}. 
Nearly at the same time, a very close
approach based in a Gaussian distribution of activation energies was proposed
by Frick and Fetters \cite{frickpip} to describe the temperature dependence of FEW data from PI.
As mentioned in Section \ref{sec:rrdm}, an alternative approach
\cite{arrighi95,arrighi96}, first introduced to analyze time-domain data of PMMA and PVME,
assumed a KWW stretched exponential function for the intermediate scattering function $I(Q,t)$. 
Although this procedure was able to describe spectra obtained in a single
spectrometer, it failed to properly account for data
collected in a wide dynamic range by using several instruments \cite{mukhomacro}.

Classical hopping between three equivalent positions was assumed in all these approaches.
It must be noted that classical hopping is just driven by the
activation energy $E_{A}$, and that a purely threefold barrier and another one distorted by a small sixfold
contribution, $V_{6}(1-\cos (6\phi +\delta))/2$, can provide the same $E_{A}$ with an appropiate selection
of the parameters $V_3$, $V_6$ and $\delta$. Hence, an analysis of high temperature classical spectra
provides a first approach to the functional form of the rotational potential, but an unambiguous
determination of the latter is only possible by means of a complete analysis of hopping,
tunnelling and librational spectra.
For most of the polymers or molecular glasses for which at least two of these three features have been
investigated, it has been found that a distribution of purely threefold barriers
provides a good description of experimental spectra.
As shown below, small but significant corrections
to the threefold term have instead to be considered in order to 
consistently reproduce the different features of the ester-methyl group dynamics in PMMA.

Recent QENS investigations in PDMS \cite{arrighi03macro}, PVAc \cite{tyagi}
and in the blends hh-PP/PEP \cite{perez}, and PEO/PMMA \cite{genix} have tackled the problem
of methyl group classical hopping in the very high temperature range where secondary relaxations or motions
involved in the glass transition process enter the dynamic window of neutron scattering.
In that situation the time scales of such processes and methyl group hopping superpose.
A consistent description has been achieved by assuming that the mentioned motions and methyl group rotation
are statistically independent. This approach was first introduced in an investigation,
by means of molecular dynamics simulations, on methyl groups dynamics in PI above the
glass transition temperature \cite{alv00cp}.
Within this approach, the RRDM picture is maintained for methyl group rotation
and the total scattering function is just obtained as a convolution
of the RRDM scattering function (\ref{eq:sqrrdmcla}) and that 
corresponding to the other relaxational modes \cite{alv00cp}.
A unified description is achieved with the same values ---or small variations--- of the RRDM parameters
derived at lower temperatures, where the dynamic contribution of the polymer matrix 
just enters in the vibrational Debye-Waller factor. This result evidences that the barrier distribution
for methyl group rotation in polymers is not substantially changed by the structural
rearrangements produced when approaching the glass transition \cite{arrighi03macro,alv00cp}.

\subsection{Rotational tunnelling and crossover to hopping}

As mentioned in the Introduction,
the existence of quantum rotational tunnelling for methyl group dynamics in polymer systems 
was expected for a long time but it was not directly observed by neutron scattering
until late 90's \cite{pvacprl,isotprb}. In contrast to crystalline systems where,
due to the unique value of the rotational barrier, 
resolution-width tunnelling lines are observed at very low temperature, in amorphous systems,
a distribution of rotational barriers, and therefore of tunnelling frequencies, is expected.
Some examples are misoriented methyl groups in crystalline lattices
\cite{prager02cp} or absorbed in water clathrates \cite{prager04jpcm} or
crystalline zeolites \cite{nair04}, and other small molecular rotors in chemically
or geometrically disordered environments as 
non-stochiometric metal hexaammines \cite{lec90}, ammonia in metal alkali fullerides
\cite{margad}, diluted solutions of methane
in noble gases \cite{asmussen93jcp,pra89,pra94}, or of ammonium
in metal alkali halides \cite{bostoen89jcp,mukho95ssc}. 
In these systems, disorder effects provide narrow distributions of rotational
barriers and of the corresponding tunnelling frequencies, which just manifest as
weak broadening of the tunnelling peaks in neutron scattering spectra.
However, in the case of methyl groups in polymers \cite{pvacprl,isotprb,pmmamacro},
molecular glasses \cite{tolujcp,sodacetprb}, or close to pore walls
in mesoporous silicates \cite{tolconf,dimeo01prb,dimeomunich,dimeo02prb},
disorder leads to broad distributions of tunnelling frequencies.
Due to the typical values of the rotational barriers in the latter systems, specially in polymers,
most of the expected distribution of tunnelling lines corresponds
to energies below $\sim$ 1 $\mu$eV, i.e., beyond the dynamic limit explored
by BS spectrometers. This fact prevented for a long time the observation
of tunnelling features in low temperature spectra.

PVAc is the polymer having the lowest energy
barrier for methyl group rotation reported so far, and the most suitable one for the direct
observation of rotational tunnelling transitions by means of neutron scattering \cite{pvacprl,crossprb}.
%In this polymer, the low temperature behavior of PVAc has been
%investigated by means of INS, namely the TOF spectrometer IN5 \cite{pmmamacrod} and the BS
%spectrometer IN16.\cite{pvacprl}.
An analogous investigation has been carried out on rotational tunnelling
of the ester-methyl group of PMMA \cite{isotprb,pmmamacro}. Though having a higher barrier,
PMMA has the advantage of being selectively deuterable, in contrast to PVAc. In this way,
the observed scattered intensity excess for PMMA-d5 (i.e., with partial deuteration
of the $\alpha$-methyl group and the main-chain hydrogens)
over the elastic line at very low temperature can be unambiguously
assigned to the ester-methyl group dynamics. 

\begin{figure}
\begin{center}
\includegraphics[width=0.45\textwidth]{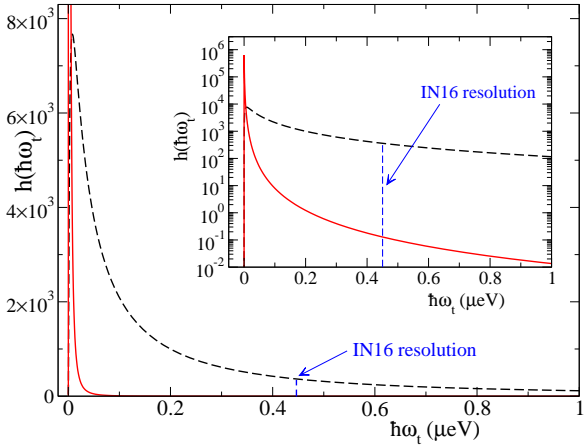}
\newline
\caption{Distribution of tunnelling frequencies for PMMA-d5 (dashed line), and PMMA-d8
(solid line). The inset shows both distributions in logarithmic scale. }
\label{fig:isotdist}
\end{center}
\end{figure}

\begin{figure}
\begin{center}
\includegraphics[width=0.42\textwidth]{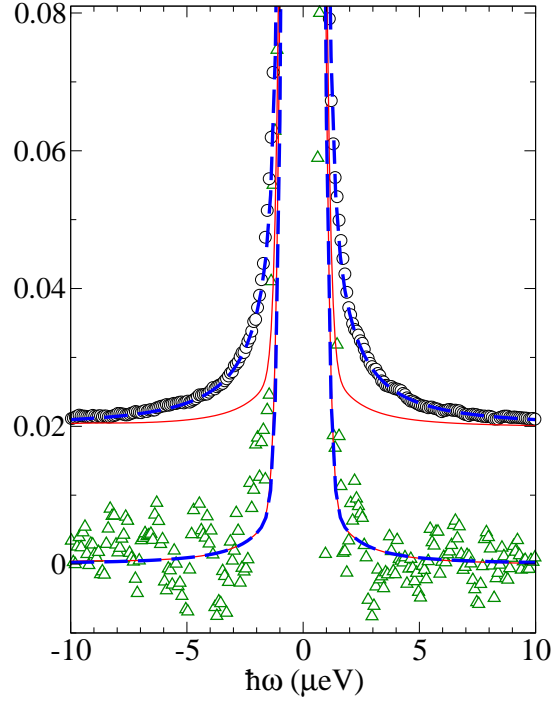}
\newline
\caption{Methyl group incoherent contribution, $S_{\rm inc}^{\rm MG}(Q,\omega)$,
normalized to unity, for PMMA-d5 (circles) and PMMA-d8 (triangles)
at $T =$ 2 K and $Q=$ 1.8 ${\rm \AA^{-1}}$. Dashed lines correspond to the
description by the tunnelling limit of the RRDM. Solid lines correspond
to the resolution function. Data for PMMA-d5 are shifted by 0.02 for clarity.} 
\label{fig:isotsqw}
\end{center}
\end{figure}

A further support of the interpretation of
the apparently quasielastic low-temperature intensity as the result of a distribution of rotational
tunnelling lines is provided by exploting the well-known isotope effect. Hence, deuteration
of the methyl group yields a rotational constant $B$ twice smaller than for the protonated
methyl group. The corresponding tunnelling frequency is reduced by several orders of magnitude
(dashed line in Fig. \ref{fig:eneV3}a), and consequently, the corresponding
distribution $h(\hbar\omega_{\rm t})$ is strongly shifted
to lower frequencies (see Fig.\ref{fig:isotdist} for the ester-methyl group of PMMA).
This effect predicts a strong supression of the scattered intensity excess in the experimental
spectrum, as it is indeed observed \cite{isotprb} in a fully deuterated sample (d8) 
of PMMA at the same temperature (see Fig. \ref{fig:isotsqw}). For a proper comparison
between the ester-methyl group dynamics in the d5- and d8-samples, the elastic contributions, corresponding
to the total coherent cross-section and to the incoherent cross-section of the other nuclei,
are substracted from the spectra. This procedure introduces large uncertainties in the so obtained spectrum
for the d8-sample, due to the large relative weight of such elastic contributions.
However, the supression in the d8-sample of the apparently quasielastic intensity is evident,
and consistent with the result predicted on the basis of the RRDM parameters
derived for the d5-sample \cite{isotprb}.

\begin{figure}
\begin{center}
\includegraphics[width=0.45\textwidth]{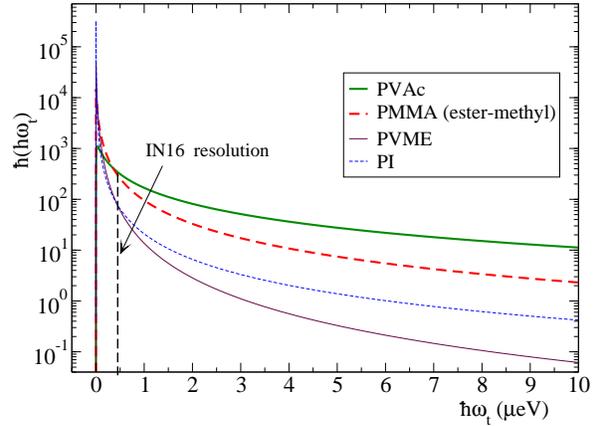}
\newline
\caption{Distribution of tunnelling frequencies for several polymers,
as derived from the RRDM analysis of hopping and/or librational data.} 
\label{fig:ghwt-comp}
\end{center}
\end{figure}

\begin{figure}
\begin{center}
\includegraphics[width=0.42\textwidth]{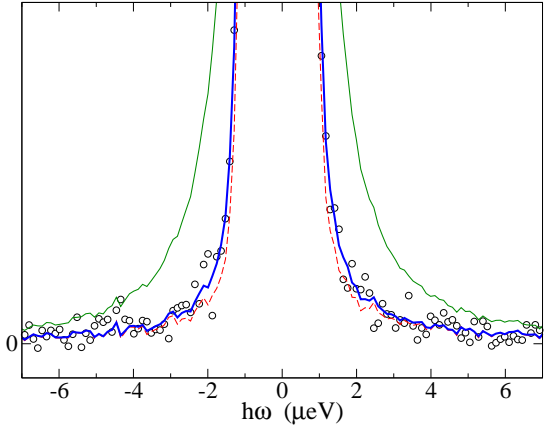}
\newline
\caption{Circles: Experimental spectrum at $T = 2$ K for PDMS. 
Thick and thin solid lines correspond to the theoretical spectra
calculated respectively from the RRDM parameters reported in
Refs. [\ref{mukhomacro}] and [\ref{arrighi03macro}].
The dashed line is the resolution function. The scale is a 3 \% of the maximum.} 
\label{fig:pdms-tun}
\end{center}
\end{figure}

Except for PVAc and for the ester-methyl group of PMMA, 
for the rest of the polymers investigated so far, 
the large values of the average barriers, $\langle V_3\rangle \gtrsim 1000$ K, 
determined from the analysis of high temperature hopping data
and/or of librational levels, prevent the observation of clear rotational tunnelling features
at very low temperature. This situation is illustrated in Fig. \ref{fig:ghwt-comp}
by a comparison  between the expected distributions of tunnelling frequencies for several polymers.
As can be seen, within the neutron scattering window, $h(\hbar\omega_t)$ for PVME and PI is
orders of magnitude smaller than for PVAc and for the ester-methyl group of PMMA. 
Convolution of $h(\hbar\omega_t)$ with the instrumental resolution provides
a scattering intensity excess unsolvable from the 
latter within the experimental noise, as indeed is experimentally confirmed.
An analysis of high temperature data for PDMS in terms of the hopping limit of the RRDM
\cite{arrighi03macro} provided a distribution of classical activation energies
$f(E_A)$, which, by transformation, would yield a distibution of tunnelling frequencies 
clearly observable by means of high-resolution BS spectrometers.
However, the measured intensity at $T=$ 2 K \cite{trivac} is hardly distinguishable
from the instrumental resolution, as can be seen in Fig. \ref{fig:pdms-tun}.
Moreover, the RRDM parameters given in Ref. \cite{arrighi03macro}
provide, by transformation, a theoretical librational maximum around $\approx 17$ meV,
far from the experimental value of 21.2 meV (see Fig. \ref{fig:lib-tosca}a).
On the contrary, the RRDM parameters in Ref. \cite{mukhomacro} reproduce the librational peak
and are compatible with the experimental spectrum at $T = 2$  K
(see Fig. \ref{fig:pdms-tun}). The reason for the mentioned uncompatibilities 
with parameters in Ref. \cite{arrighi03macro} remains to be understood.

Despite of the existence of very few polymer systems where methyl group rotational tunnelling
can be observed by neutron scattering, successful investigations have been carried
out in two molecular glasses ---toluene \cite{tolujcp} and sodium acetate
trihydrate \cite{sodacetprb}--- supporting the validity of the RRDM picture for rotational tunnelling.

As mentioned in Section \ref{sec:rrdm}, at the crossover intermediate temperature regime 
(typically 20 K $\lesssim T \lesssim$ 100 K) quantum effects are relevant for methyl group rotation,
and quasielastic spectra cannot be reproduced in terms of the classical limit of the RRDM.
However, the fact that spectra do not evolve with temperature 
according to the simple expectation for vibrational dynamics (see Section \ref{sec:pvac})
also evidences that the tunnelling limit of the RRDM cannot either be applied, 
and making use of the general version of the model is neccessary. 
The crossover regime has been investigated in detail
in the only two polymers (PVAc\cite{crossprb} and PMMA\cite{pmmamacro}) where
the distribution of rotational tunnelling lines can be (partially)
explored by neutron scattering. The success of the general version of the RRDM for
reproducing all the temperature evolution of the spectra for both polymers 
has been extended to the glassy states of toluene \cite{tolujcp}
and sodium acetate trihydrate \cite{sodacetprb}.

\subsection{Librational levels}

First neutron scattering experiments on methyl group dynamics
were focused on the determination of the librational levels.
Since early 70's INS investigations on methyl group librations have been performed
in a huge collection of polymeric materials as 
PPO \cite{hig72}, PP \cite{tak82,annis99}, PMMA \cite{hig72,allen74,gab84,pmmamacro},
poly($\alpha$-methylstyrene) (P$\alpha$MS)\cite{allen74}, PVAc\cite{ale05}, PDMS\cite{ale05},  
PI \cite{frickpip,ada05},  PIB \cite{annis99,ada05,frickpib},
bisphenol-A based engineering thermoplastics as PSF, polycarbonate (PC)
and phenoxy (PH) \cite{fer05}, and layered polytypes of hexatriacontane \cite{kubota05}. 

\begin{figure}
\begin{center}
\includegraphics[width=0.42\textwidth]{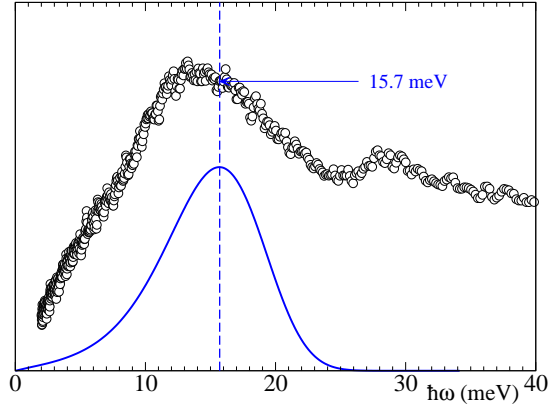}
\newline
\caption{Generalized VDOS (circles) for PVAc, measured at TOSCA at $T = 20$ K.
The curve corresponds to the RRDM theoretical distribution $F(E_{01})$.
The arrow indicates the maximum of the latter for comparison with experiment.}
\label{fig:pvac-tosca}
\end{center}
\end{figure}

\begin{figure}
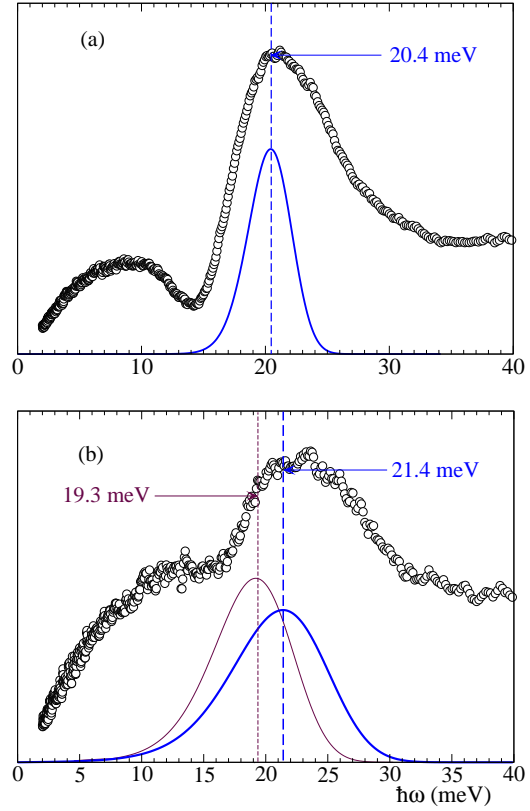

\begin{center}
\includegraphics[width=0.40\textwidth]{pdms-tosca.eps}
\newline
\newline
\includegraphics[width=0.40\textwidth]{pmma-lib3f6f.eps}
\newline
\caption{Generalized VDOS (circles) measured at TOSCA,
and RRDM theoretical distributions $F(E_{01})$ (solid lines). All data are obtained at $T= 20$ K.
(a): PDMS (b): PMMA-d5. Arrows indicate the maxima of the theoretical $F(E_{01})$
for comparison with experiment. In panel (b), the thin line corresponds to a purely
threefold approximation, while the thick line corresponds to a sixfold correction
$V_6/V_3 = 0.11$ (see text).}
\label{fig:lib-tosca}
\end{center}
\end{figure}

\begin{figure}
\begin{center}
\includegraphics[width=0.44\textwidth]{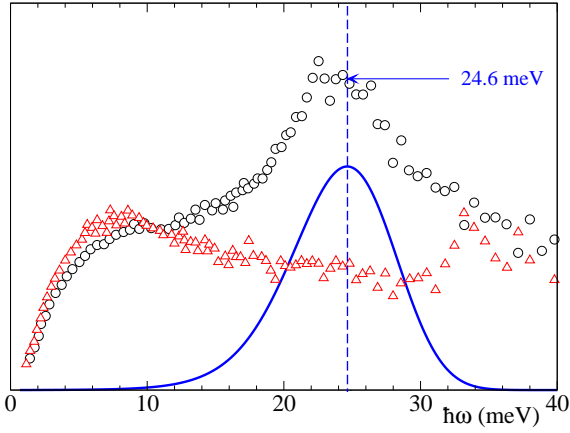}
\newline
\caption{Generalized VDOS for PI measured at IN6. $T= 140$ K.
Circles: PI-d5 (deuterated main chain - protonated methyl group).
Triangles: PI-d3 (protonated main chain - deuterated methyl group).
Curve: RRDM theoretical distribution $F(E_{01})$. 
The arrow indicates the maximum of the latter for comparison with experiment.}
\label{fig:lib-pi}
\end{center}
\end{figure}

\begin{figure}
\begin{center}
\includegraphics[width=0.47\textwidth]{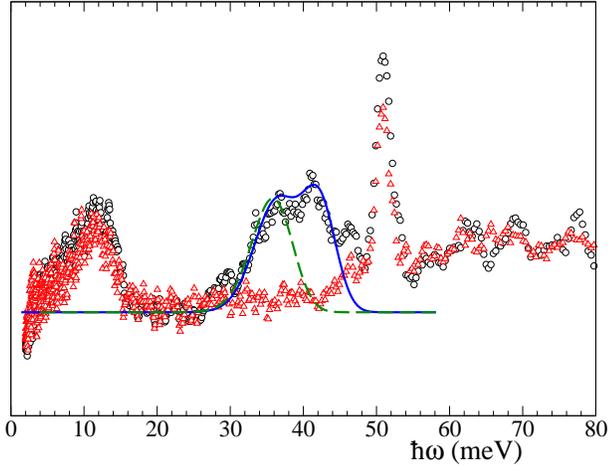}
\newline
\caption{Generalized VDOS for PSF measured at TOSCA. $T= 20$ K.
Circles: fully protonated sample. Triangles: deuterated methyl group.
Curves: RRDM theoretical distribution $F(E_{01})$ for a single-particle approach
(dashed) and for an additional methyl-methyl coupling interaction (solid, see text).}
\label{fig:lib-psf}
\end{center}
\end{figure}

Figs. \ref{fig:pvac-tosca}, \ref{fig:lib-tosca}, \ref{fig:lib-pi} and \ref{fig:lib-psf}
show some examples of the experimental generalized VDOS for PVAc \cite{ale05},
PDMS \cite{ale05}, PMMA (ester-methyl group) \cite{pmmamacro}, 
PI \cite{frickpip} and PSF \cite{fer05}, compared with the theoretical
librational distributions $F(E_{01})$ obtained within the RRDM approach.
For moderate and low energy barriers ($V_{3} \lesssim 500$ K), the lowest
librational transitions occur in the same energy range ($5-15$ meV) where
a broad shoulder in the generalized VDOS is observed for any disordered system \cite{fri88}. 
This strong superposition complicates the identification
and analysis of the corresponding librational peaks,
as in the case of PVAc \cite{ale05} (Fig. \ref{fig:pvac-tosca}). 
On the contrary, for moderate and high rotational barriers ($V_{3} \gtrsim 1000$ K) as in
PDMS, PI or PSF, librational peaks are well resolved from the generalized VDOS.

In order to solve methyl group librations, it is useful to carry out and compare
INS measurements in samples with both protonated and deuterated methyl groups
(see Figs. \ref{fig:lib-pi}  and \ref{fig:lib-psf} for PI and PSF respectively).
It must be stressed that even in the case of partial deuteration of all
the hydrogens not belonging to the methyl group, identification of librational peaks
is still a complicated task, since any other vibrational mode involving
the motion of the methyl group hydrogens will provide a peak in the INS spectrum.
A comparison with infrared and Raman measurements is helpful for a differentation
of the physical origin of the corresponding peaks. A nice example has been recently
reported by Adams {\it et al.} for PIB and PI \cite{ada05}.

A relevant point for the analysis of librational features in INS spectra
is the high sensitivity of the librational energies to the shape
of the rotational barrier. Hence, the introduction of moderate sixfold
corrections to the main threefold term of the rotational potential provides
a significant shift of the librational energy. Figure \ref{fig:lib-tosca}b shows
how the introduction of a small sixfold contribution
is sufficient to account for the first librational peak
of the ester-methyl group in PMMA\cite{pmmamacro}, 
by using the same RRDM parameters describing both the rotational
tunnelling and classical hopping regimes. It is noteworthy that such a consistent
description has been achieved by fixing a ratio $V_6/V_3 = 0.11$
for all the individual methyl groups. Hence, as in the
purely threefold case, the only {\it independent} parameter
affected by the distribution effects is the threefold term $V_3$.

Librational energies are also sensitive to coupling
between pairs of neighbouring methyl groups. In such cases, the single-particle
approach is modified by introducing a coupling Hamiltonian. In the simplest
approximation the coupling term is purely threefold and no angular offsets are included:

\begin{eqnarray}
\nonumber
H= -B\left(\frac{\partial^2}{\partial\phi^{2}} + \frac{\partial^2}{\partial\psi^{2}}\right) 
+ \frac{U_{3}}{2}(1-\cos 3\phi) \\ 
+ \frac{V_{3}}{2}(1-\cos 3\psi) + \frac{W^{c}_{3}}{2}(1-\cos 3(\phi-\psi)).
\label{eq:Hcoup}
\end{eqnarray}
In a recent INS investigation on methyl group librations in PSF \cite{fer05},
the double-peak structure in the range $\approx 34 - 44$ meV (see Fig. \ref{fig:lib-psf})
has been interpreted as the result of a distribution of splitted first librational energies,
such a splitting resulting from the coupling between the two methyl groups in each monomeric unit.
As in the single-particle approach, the terms $U_{3}$ and $V_{3}$ 
are taken as Gaussian distributed. Due to the highly disordered
character of the polymer matrix and the symmetric chemical arrangement
of both methyl groups in the monomeric unit, 
it is expected that, on average, they feel the same typical environment. 
Hence, the distributions $g(U_{3})$ and $g(V_{3})$ are assumed to be identical.
Since the coupling term $W_3^c$ is strictly due to the interaction
between the two methyl groups of the same monomeric unit, it is expected to be weakly affected
by local packing conditions. Hence for simplicity, distribution effects are neglected
for the coupling term. 

The classical hopping regime for PSF is not accesible by QENS
at low and moderate temperature, and at high temperature superposes
with secondary relaxations which complicate the analysis. Instead, it can be 
properly explored by NMR techniques. Hence, the temperature dependence
of the D-NMR line shape of PSF has been reproduced in
terms of a log-Gaussian distribution of classical hopping frequencies as Eq. (\ref{eq:hlog}),
with $\log \Gamma_0$ and $\sigma$ respectively following Eqs. (\ref{eq:gamma0}, \ref{eq:sigma}).
Analogously to the procedure exposed in Section \ref{sec:rrdm}, these results provide a
Gaussian distribution of classical activation energies $f(E_A)$,
%characterized by the parameters $\langle E_A\rangle$ and $\sigma_E$. 
%It must be noted that in the present case the equality
%$H(\log\Gamma)d(\log\Gamma) = -f(E_A)dE_A$ is just a formal transformation.
%However, if $f(E_A)$ is interpreted as a Gaussian distribution of physical activation energies for
%classical hopping, the latter transformation implicitely assumes that,
%despite of possible methyl-methyl coupling effects at low temperature,  
%in the classical hopping regime rotations of both methyl groups are sufficiently
%fast to be statistically independent. Taking this assumption for good,
which can be transformed into a distribution of rotational barriers $g(V_3)$
and librational energies $F(E_{01})$ via the functional relations $E_A = E_A(V_3)$,
$E_{01} = E_{01}(V_3)$. Such functional relations can be obtained by solving the eigenvalues
of the Hamiltonian (\ref{eq:Hcoup}) for the selected value of $W_3^c$ and a large
set of values $U_3 = V_3$ (see Ref. \cite{fer05} for numerical relations).

Fig. \ref{fig:lib-psf} shows the corresponding distributions
$F(E_{01})$ for the single-particle approach, and for coupling of methyl groups, 
---with the conditions above explained for the second case. While the single-particle
approach only gives account for the lowest experimental peak, the introduction
of a coupling term reproduces the double-peak structure. 
Analogous results are obtained for PC and PH \cite{fer05}. It is noteworthy that the introduction
of a coupling term does not substantially complicate the original RRDM, since the only
model parameter affected by the distribution effects is, as in the
single-particle case, the rotational barrier height $V_3$.

\section{Molecular origin of the barrier distribution for methyl group dynamics}
\label{sec:origin}

The results summarized in Table \ref{table:rrdm} show that the average 
barrier height $\langle V_3 \rangle$ for methyl group reorientation in amorphous polymers is highly
sensitive to the chemical structure of the monomeric unit.
This fact suggests that the rotational potential is
% determined to a certain extent by
only partially determined by intermolecular interactions.
On the other hand, there is not a clear correlation between
the width of the barrier distribution and the chemical structure of the monomeric unit.
Next we summarize the most relevant neutron scattering results 
relating the parameters of the barrier distribution with packing
conditions and intermolecular interactions.

\subsection{Influence of the chain conformation}

The relative orientation between identical side groups placed at
different monomeric units can be random or can adopt specific forms.
Hence, in the {\it isotactic} ($i$) and {\it sindiotactic} ($s$) conformations, 
side groups at adjacent monomeric units are respectively paralell and anti-paralell. 
If both relative orientations are randomly distributed, the conformation is {\it heterotactic} ($h$).
When there is no dominating tacticity in the ensemble of chains forming the system,
the latter is {\it atactic} ($a$). 
The ester-methyl group dynamics in PMMA have been analyzed in terms of the RRDM for
different degrees of tacticity of the ester group. The influence of local packing conditions on the
barrier distribution for samples chemically identical is evidenced by the small, but significant,
differences between the obtained RRDM parameters \cite{isotprb,pmmamacro,pillay,cereghetti} 
(see Table \ref{table:rrdm}).
Hence, the average activation energy for a purely syndiotactic sample
is $\langle E_A\rangle = 710$ K \cite{pmmamacro},
while for an atactic sample with relation $s$:$i$:$h$ = 50:10:40, 
$\langle E_A\rangle = 529$ K \cite{pillay}.

PMMA chains can be rearranged to form a stereocomplex form, where syndiotactic chains wrap around 
isotactic ones to form a double stranded helix structure. In Ref. \cite{pillay} 
stereocomplexed samples with complementary deuteration of chains with different tacticities
were investigated. In this way, the effect of stereocomplexation on methyl group dynamics
in different types of chains could be discriminated.
For the isotactic chains, stereocomplexation considerably reduced the width
of $f(E_A)$, yielding a standard deviation $\sigma_E =$ 205 K, as compared to the value
$\sigma_E =$ 313 K for the non-stereocomplexed sample \cite{pillay}. 

For the case of the $\alpha$-methyl group in PMMA, INS measurements showed a much higher
librational peak \cite{allen74}, and consequently a much higher average rotational barrier, for
the syndiotatic form as compared with the isotactic one.  
%However, a general conclusion on the effect of the chain tacticity on the 
%barrier distribution for methyl group dynamics in polymers cannot be made on the
%basis of the limited available information on this question. Hence,
On the contrary, INS spectra for P$\alpha$MS \cite{allen74} did not
show significant differences between a syndiotactic sample and an atactic one.
FEW scans reported in Ref. \cite{arrighi03jcp} for different tacticities of PP
did not show apparent differences, within the experimental noise, in the temperature
range dominated by methyl group dynamics. This result is supported by the
rather close values of the measured librational peaks for the isotactic \cite{tak82}
and atactic \cite{annis99} conformations.

There are very few systems where experimental data are available for both, the usually
investigated head-to-tail, and the head-to-head sequences of adjacent monomeric units.
For P$\alpha$MS \cite{allen74} and for PP \cite{perez} it has been found,
repectively by means of INS and QENS measurements, that the average barrier in the head-to-head
conformation is much lower than in the head-to-tail one.
In the case of hh-PP, the distribution of classical activation energies $f(E_A)$ has been
calculated from an analysis of high temperature data in terms of the hopping limit
of the RRDM \cite{perez}. Appart form the mentioned decrease of the
average barrier, the analysis provides a much narrower distribution than
that obtained for head-to-tail PP \cite{ahumada,arrighi01}. It is worthy of remark that 
the maximum of the librational distribution $F(E_{01})$ for hh-PP, 
obtained by transformation from $f(E_A)$ in a purely threefold approach
(see Section \ref{sec:rrdm}), is close to the experimental value reported in Ref. \cite{annis99},
though the agreement is not fully satisfactory \cite{perez}. A possible origin of such a difference
might be, similarly to the case of the ester-methyl group in PMMA (see Section \ref{sec:summexper}),
the presence of small higher-order corrections to the main threefold term of the rotational potential,
or small coupling effects between neighbouring methyl groups, which could
shift the librational energy to slightly higher values than the expected ones for
a purely threefold single-particle potential. This hypothesis is consistent 
with the proximity between methyl groups of adjacent units in the head-to-head conformation of PP.

\subsection{Influence of mixing with other materials}
 
A few works report comparisons between methyl group rotation in the neat polymer and blended
with other polymers. QENS investigations in the blends PVME/PS \cite{arrblend,mukhoblend}
and hh-PP/PEO \cite{perez} did not show, for similar concentrations of the two components, 
changes within the error bar on the barrier
distribution as compared respectively to neat PVME and hh-PP. High dilution of PVME in PVME/PS 
provided a significantly broader distribution, but again did not affect
to the average barrier \cite{mukhoblend}. On the contrary, progressive dilution of PMMA
in SCPE/PMMA increased the average barrier for the ester-methyl group even at moderate concentrations
of SCPE \cite{arrblend}.

The effect of doping with small molecules on methyl group dynamics has also been reported
in the literature, though a detailed quantitative analysis in terms of barrier distributions
is lacking. Hence, FEW scans in Ref. \cite{arr98pol} showed no significant differences
for neat PDMS and filled with silica nanoparticles. Some investigations on polymer electrolites
suggest that methyl group dynamics are not significantly affected by addition of the salt
\cite{carl98,and99,russina}, but the present information is rather limited
to make reliable conclusions at a quantitative level.

\subsection{Molecular glasses and comparison with the crystalline state}

\begin{figure}
\begin{center}
\includegraphics[width=0.44\textwidth]{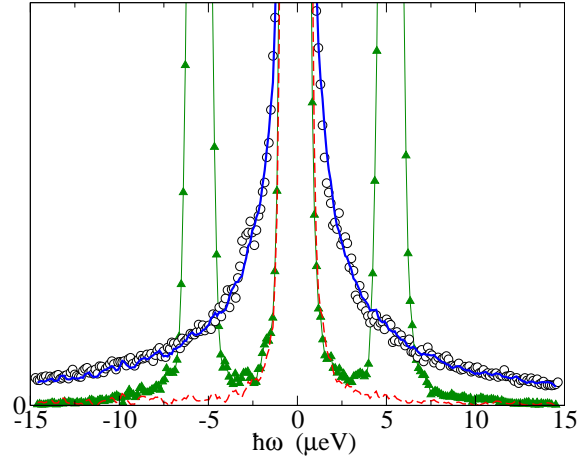}
\newline
\caption{Symbols: Experimental spectra for a same sample of sodium acetate trihydrate
in the crystalline (triangles) and glassy (circles) state at $T =$ 2 K 
and $Q=$ 1.8 ${\rm \AA^{-1}}$. The thick solid line is
the RRDM theoretical function. The thin solid line is a guide for the eyes. The dashed line
is the resolution function. Scale: 3\% of the maximum.}
\label{fig:sodacetcrygla}
\end{center}
\end{figure}

It is worthy of remark that, not only amorphous polymers, 
but also the low-molecular-weight glasses investigated until now
display a similar phenomenology for methyl group dynamics. In contrast to amorphous polymers, 
glass-forming molecular systems can be easily obtained in the crystalline state.
This possibility allows one to make a direct comparison between the methyl group dynamic features
in the crystalline and glassy states of a same system. 
This comparison was done {\it in situ} for a same sample of sodium acetate trihydrate \cite{sodacetprb}.
The analysis of the neutron scattering spectra, in the crystalline and glassy states,
unambiguously demonstrated that the apparently quasielastic intensity observed
for methyl group tunnelling in the glassy state is a consequence of the atomic disorder.
Fig. \ref{fig:sodacetcrygla} shows the corresponding tunnelling spectra for the crystalline
and glassy state of sodium acetate trihydrate.
For this system it was found that the average potential barrier
in the glassy state takes, within the experimental error, the same value as the unique barrier in the crystal.
A direct comparison between methyl group dynamics in the crystalline and glassy state
has also been reported for ethylbenzene \cite{fri95}. Disorder effects are evidenced by the broad structure
of the librational peak in the glassy state as compared to the resolution-width line in the crystalline case.
The small shift of the experimental maximum indicates that, as in sodium acetate trihydrate,
the distribution is centered around the unique barrier of the crystal.
On the contrary, an investigation on glassy toluene has reported an average rotational barrier
much larger than the unique one of the crystalline state \cite{tolujcp}.

\begin{figure}
\begin{center}
\includegraphics[width=0.44\textwidth]{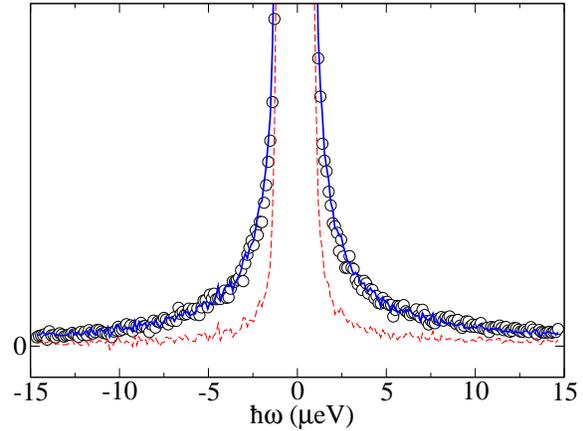}
\newline
\caption{Circles: Experimental tunnelling spectrum at $T =$ 2 K
and  $Q=$ 1.8 ${\rm \AA^{-1}}$ for tri(vinyl acetate).
Solid line: Theoretical spectrum calculated from the RRDM parameters for PVAc.
The dashed line is the resolution function. Scale: 3\% of the maximum.}
\label{fig:trivac}
\end{center}
\end{figure}

The widths of the barrier distributions obtained for the former molecular glasses
take values comparable to those obtained for polymer glasses.
All these results strongly suggest that the origin of the
distribution of potential barriers in polymers is directly related to the structural disorder and
packing conditions, and not with chain connectivity aspects which are specific of polymer glasses.
This hypothesis has been recently supported by measurements of
the tunnelling spectrum of tri(vinyl acetate) \cite{trivac}. As can be seen in Fig. \ref{fig:trivac},
the experimental spectrum at $T =$ 2 K is reproduced by using the same RRDM parameters as in PVAc,
suggesting that the barrier distribution is not affected by polimerization of the basic molecular unit.

\subsection{Molecular dynamics simulations}

\begin{figure}
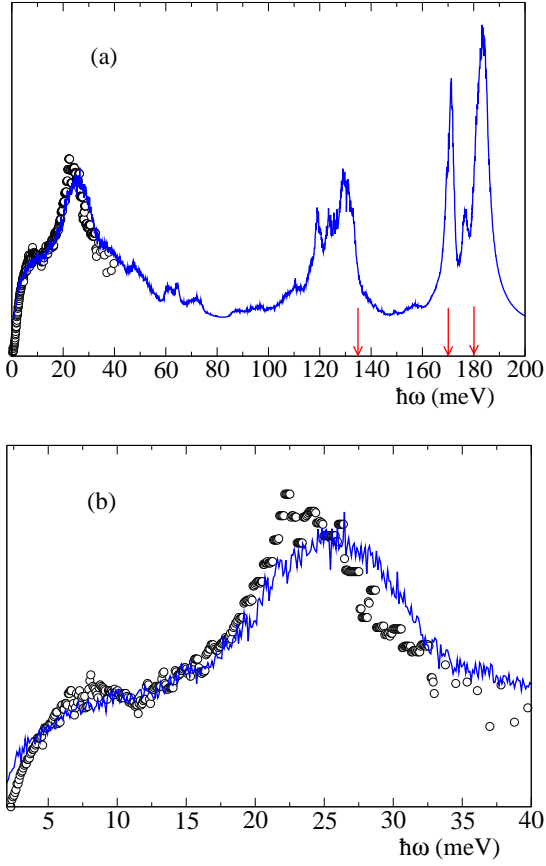

\begin{center}
\includegraphics[width=0.42\textwidth]{vdos-pi-tot.eps}
\newline
\newline
\includegraphics[width=0.42\textwidth]{vdos-pi-mg.eps}
\newline
\caption{Generalized VDOS obtained for PI-d5 by MDS
(solid line), and by INS (circles). Arrows in panel (a) indicate
the position of infrared bands. Panel (b) is an amplification
in the energy range around the librational peak.}
\label{fig:zw-mds}
\end{center}
\end{figure}

A deep understanding of the microscopic origin of the distribution of potential barriers is
not easily inferred from experiments. However, fully atomistic MDS are a 
suitable tool to gain insight into this problem. In particular, the calculation of the distribution
of librational energies is straightforward and does not require lengthy simulation runs.
As has been exposed in previous Sections, once this distribution is obtained, 
the corresponding distribution of potential barriers can be directly derived by transformation of the former.
In addition, MDS offer the possibility of varying the parameters of the employed force field,
and to investigate the influence of the modified force field on methyl group dynamics.
%With these ideas in mind, the microscopic origin of the distribution of potential barriers in
%PI was investigated in Ref. \cite{simpip} by means of MDS. 
%PI was choosen because it is a rather simple
%polymer and realistic models for MDS can be obtained by using commercial force-fields.
%There were also neutron scattering data available in the literature for methyl group
%torsional librations, allowing
%validation of MDS. The details of the simulations are described in detail in Ref.
A comparison between MDS and neutron scattering results on PI was reported in Ref. \cite{simpip}.
From the atomic trajectories computed in the simulations, 
the generalized VDOS, $Z(\omega)$, was calculated as the spectral density
of the velocity autocorrelation function:

\begin{equation}
Z(\omega) \propto \int_0^\infty e^{-i\omega t}\langle \varrho\ {\bf v}(t)\cdot {\bf v}(0) \rangle dt ,
\label{eq:zwmds}
\end{equation}
where brackets denote ensemble average over all the atoms.
%involved in the considered vibrational modes.
The weighting factor $\varrho$ contains, for each atomic species, the corresponding cross-section and mass,
in order to make a direct comparison with the experimental results.
Fig. \ref{fig:zw-mds} shows a comparison between the INS data reported
by Frick and Fetters \cite{frickpip} for PI-d5 (deuterated main-chain and protonated methyl group)
and the generalized VDOS, computed from the MDS as in Eq. (\ref{eq:zwmds}).
A good agreement is obtained at least in the energy range, $\hbar\omega \lesssim 40$ meV, experimentally
investigated in Ref. \cite{frickpip}, validating the employed force field.
 %The available neutron scattering results correspond to a PI sample (PI-d5), where the
%main chain hydrogens were deuterated, and another sample (PI-d3) where the methyl group
%hydrogens were deuterated. Therefore, in the PI-d5 case, the VDOS is dominated by the
%methyl group hydrogens, and in the case of PI-d3 by the main-chain hydrogens. Accordingly,
%$Z(\omega)$  was calculated from the simulations by considering either the methyl group hydorgens
%and the main-chain hydrogens. 
%Figure shows the good agreement obtained between simulations
%and neutron scattering results at least in the energy range ($\hbar\omega < 40$ meV) where
%experimental data are available. 
%The simulation results also show other maxima at high energies,
%which cannot be directly compared with inelastic neutron scattering results.
%However, 
Although quantum effects (not considered in the classical MDS) should certainly affect the 
high energy range of $Z(\omega)$, it is worthy of remark that the energies of the computed high energy maxima 
(see Fig. \ref{fig:zw-mds}) roughly correspond to the reported infrarred and Raman bands \cite{polhandbook}, 
as well as to recent INS results by Adams {\it et al} \cite{ada05}.
%The good agreement shown 
%in Figure between simulation and experimental results indicates that both the average methyl
%group rotation and the distribution of potential barriers, are well reproduced with the force
%field used, thereby validating the simulations. 

It must be reminded that, although the generalized VDOS for PI-d5
is dominated by motions of methyl group protons, it contains not only the contributions of the
librational modes, but also from any other mode involving the motion of the methyl group.
This is evidenced in Fig. \ref{fig:lib-pi}, where the librational peak
is superimposed to a broad low-energy hump similar to that present in
the generalized VDOS of PI-d3 (deuterated methyl group and protonated main-chain). However, 
the methyl group librational contribution can be separated from
the computed generalized VDOS by calculating the spectral density of the angular correlation
function \cite{simpip}:

\begin{equation}
F(E_{01}) \propto \int_0^\infty e^{-iE_{01} t}\langle \dot{\phi}(t) \dot{\phi}(0) \rangle dt ,
\end{equation}
where brackets denote ensemble average {\it over methyl group protons}. $F(E_{01})$ is formally
the librational density of states. Finally, the distribution of potential barriers $g(V_3)$ is
obtained by transformation from $F(E_{01})$ as exposed in Section \ref{sec:rrdm}. 

\begin{figure}
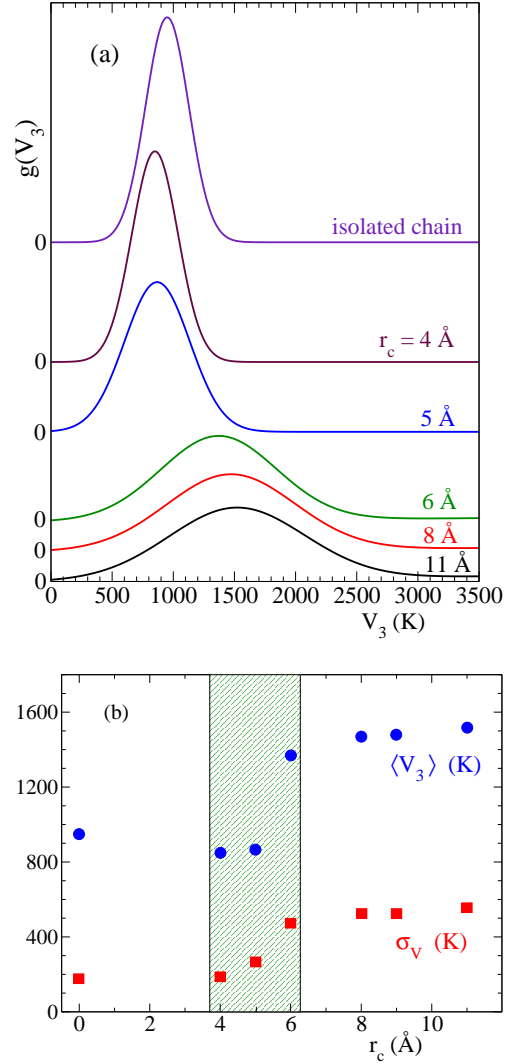

\begin{center}
\includegraphics[width=0.38\textwidth]{gVrc.eps}
\newline
\newline
\includegraphics[width=0.38\textwidth]{rcV3sig.eps}
\newline
\caption{Panel (a): Effect of the cutoff radius, $r_c$, for the nonbond interactions on the distribution
of potential barriers for methyl group reorientation in PI. Panel (b) shows the evolution 
with $r_c$ of the average potential barrier $\langle V_3 \rangle$
and the standard deviation $\sigma_V$. The shadowed area corresponds to the range of inter-chain distances.}
\label{fig:isolchain}
\end{center}
\end{figure}

In order to investigate the relation between the average and width of the barrier distribution
and the nonbond interactions, a series of MDS were carried out with different values
of the cutoff radius, $r_c$, for the nonbond interactions of the employed force field.
The cutoff $r_c$ was changed from 11 ${\rm \AA}$ (the value for realistic conditions)
to 4 ${\rm \AA}$. Simulations were also performed for the isolated chain.
The results reported in Ref. \cite{simpip} are summarized in Fig. \ref{fig:isolchain}.
Diminishing the cutoff radius yields a systematic decrease of the average barrier
and the width of the barrier distribution $g(V_3)$. An abrupt change of the latter parameters takes place 
in a narrow region around $r_c \approx 5$ {\rm \AA}, which roughly corresponds
to the inter-macromolecular distances in PI. This result suggests that only the nonbond interactions
at the inter-chain distances play a significant role in determining the width of the
distribution of rotational barriers for methyl group rotation. Moreover,
it is worthy of remark that these interactions also affect significantly to the average value
of the potential barrier, i.e., the latter is not only determined by the chemical structure of the
monomeric unit.

\section{Conclusions and outlook}
\label{sec:conclusions}

This paper gives a comprehensive review of neutron scattering investigations on methyl group dynamics in
glassy polymers of different chemical composition and microstructure. 
Considerable progress on this topic has been achieved over last 10 years mainly fueled by two facts:
(i) the proposal in 1994 of the Rotation Rate Distribution Model for methyl group dynamics in polymers;
(ii) the first direct experimental observation in 1998 by neutron scattering of quantum rotational tunnelling
in glassy polymers at very low temperature.

The experimental data presented in this paper
demonstrate the uniqueness, the power, and the efficiency of the combination of quasielastic and inelastic
neutron scattering for investigating this problem.
In particular, it is worthy of remark that neutron scattering is the only technique which provides
a direct experimental observation of quantum tunnelling processes of methyl groups.
There are, however, some limitations, mainly related to the current resolution
of the available high-resolution
backscattering spectrometers ($\approx $ 0.3 $\mu$eV), which
prevents the observation of clear rotational tunnelling features at very low temperature
in polymers when the average rotational barrier is higher than $\approx$ 1000 K.

On the other hand, in this article it is shown that distribution effects are essential to understand
methyl group dynamics ---from quantum tunnelling to classical hopping--- in polymeric materials and in other
highly disordered systems as well. The Rotation Rate Distribution Model (RRDM), which introduces a
distribution of potential barriers for methyl group rotation, provides a consistent description for the
experimental features observed at different temperatures, including the crossover from quantum to classical
behavior and the librational energies as well.

Taking into account the RRDM results on different polymer systems as well as on low molecular weight glasses,
it is now evident that the distribution of potential barriers is determined by the intermolecular disorder
and that the chain connectivity does not play any significant role. However, appart from these qualitative
results, it is clear that the actual origin of the distribution of potential barriers and its possible
relationship with the structure factor characterizing the packing conditions, is one of the main points that
still needs to be clarified. Concerning this question, fully atomistic molecular dynamics simulations can
play an important role in the near future. The results available by now suggest that the width of the
distribution of potential barriers is mainly determined by the nonbond interactions at the inter-chain
distances, at least in the case of polyisoprene. These results need to be confirmed by similar simulations in
other realistic model systems as for instance, PMMA.

Another way of addressing this question
experimentally would be to carry out neutron scattering measurements on methyl group dynamics in polymers
under confinement conditions, and to compare them with the corresponding results
for the bulk state. To the best of our knowledge, a quantitative 
determination of the barrier distribution for methyl group rotation has never been reported
for a confined polymer. However, this has been recently carried out for glassy toluene 
confined in mesoporous silicates \cite{tolconf}. From the knowledge
of the barrier distribution in the bulk glassy state \cite{tolujcp},
and by investigating the dependence of the
RRDM parameters on the monodisperse radius of the confining pores, 
a lower limit has been estimated for the distance beyond which methyl groups do not
feel the interaction with the pore wall and behave as bulk-like rotors. 
This procedure has confirmed that the interaction range relevant for
methyl group dymamics in glassy toluene extends beyond the typical
nearest-neighbour distance between centers-of-mass \cite{tolconf}.
Similar investigations on confined polymers would be helpful to improve
our understanding of the structural origin of the barrier distribution
for methyl group rotation in polymers.
\\\\
\begin{center}
{\bf ACKNOWLEDGEMENTS}
\end{center}

We are grateful to all colleagues with whom we have had the pleasure of collaborating 
on this topic over last years: C. Alba-Simionesco, J.M. Alberdi, F. Alvarez, A. Arbe, A. Chahid,
G. Dosseh, D. Fern\'{a}ndez, B. Frick, A.C. Genix,  H. Grimm, D. Morineau, R. Mukhopadhyay, 
R. P\'{e}rez, M. Prager, D. Richter, M. Tyagy, and L. Willner. We also thank M. Tyagy for
sharing his unpublished results on PDMS and tri(vinyl acetate).
We acknowledge experimental facilities from ILL (Grenoble, France), IFF (J\"{u}lich, Germany),
and ISIS (Chilton, United Kingdom).

\newpage
\newpage
%
%
%\newpage
%
%\vspace{-5 mm}
\begin{figure}
\begin{center}
\includegraphics[width=0.35\textwidth]{pvacsqw120.eps}
\newline
\newline
\includegraphics[width=0.35\textwidth]{pvacsqw160.eps}
\newline
\newline
\includegraphics[width=0.35\textwidth]{pvacsqw200.eps}
\newline
\caption{IN6 spectra for PVAc at $Q=1.5 {\rm \AA}^{-1}$ and different temperatures.
Circles are experimental data. Solid lines
correspond to the RRDM description. 
Dashed lines are the resolution function. The scale is a 3\% of the maximum.} 
\label{fig:pvacin6}
\end{center}
\end{figure}
\newpage
%
%
%
%
%\hsize\textwidth\columnwidth\hsize\csname@twocolumnfalse\endcsname
%\begin{figure}\begin{center}\includegraphics[width=0.45\textwidth]{few.eps}
%\caption{FEW experiments showing the step like decreasing of the elastic intensity
% associated to the methyl group hopping processes.} 
%\label{fig:few}\end{center}\end{figure}
%\begin{figure}\begin{center}\includegraphics[width=0.45\textwidth]{hwt.eps}
%\caption{Distribution of the methyl group tunnelling lines expected
%	 on the basis of the RRDM parameters determined from the description
%of the methyl
%	 group rotational hopping for several polymers.} 
%\label{fig:hwt}\end{center}\end{figure}%
%\begin{figure}\begin{center}\includegraphics[width=0.45\textwidth]{vdos.eps}
%\caption{Proton weighted VDOS for several polymers.} \label{fig: vdos}\end{center}\end{figure}
%\begin{figure}\begin{center}\includegraphics[width=0.45\textwidth]{fewdospsf.eps}
%\caption{Results from a FEW experiment (a) on fully protonated PSF and the proton weighted VDOS 
%(b)  of both  fully protonated and methyl group deuterated PSF.} \label{fig: fewdospsf}\end{center}\end{figure}
%\begin{figure}\begin{center}\includegraphics[width=0.45\textwidth]{vdospmma.eps}
%\caption{Results from a FEW experiment (a) on fully protonated PSF and the proton weighted VDOS 
%(b)  of both  fully protonated and methyl group deuterated PSF.} 
%\label{fig:vdospmma}\end{center}\end{figure}\newpage%%% =========
% = TABLES==
% =========
\vspace{-0.4 cm}
\begin{table}
\begin{tabular}{lcccccccc}
%\hline
\ &\ & H &\ & D &\ & C &\ & O\\
\hline
$\sigma_{inc}$ &\ & 80.3 &\ &2.0 &\ &0.0 &\ & 0.0\\
$\sigma_{coh}$ &\ & 1.8 &\ &5.6 &\ &5.5 &\ & 4.2\\
%\hline
\end{tabular}
\vspace{0.3 cm}
\caption{Scattering cross-sections of the most abundant nuclei in polymers.
Values are given in barn (1 barn = $10^{-24}$ cm$^{-2}$).}
\label{table:scs}
\end{table}

\onecolumn

\begin{table}[h]

\begin{tabular}{ cccccccccc }
%\hline
& $\langle E_A\rangle$ (K)&$\sigma_E$ (K)& $\Gamma_{\infty}$ (meV)& $\langle V_3 \rangle$ (K)&
$\sigma_V$ (K)&  $E_{01}^{max}$ (meV)\\\hline
\\
%Poly(vinyl methyl ether)% 
PVME \cite{chahid94,mukhoblend}&      1010&    276&   7.8&     &     &   \\\\
%Poly(vinyl acetate)% 
PVAc \cite{pvacprl,crossprb,mukhomacro,ale05}&     445&    253&   9.1&     534&     274&     14.4\\\\
%Polyisoprene% 
PI \cite{zorn,ada05} &     1170&    353&   62.1&     1364&     376&      23.4 \\\\
%Poly(dimethyl siloxane)% 
PDMS \\
\ \cite{mukhomacro,ale05} &     842 &   144&  18.5&   997  &  155  &       21.2\\
\ \cite{arrighi03macro} ($\ast$) &     602 &   132&  0.63&     &     &       \\\\
%Poly(ethylene propylene)% 
PEP \cite{perez} &    1288&    162&  1.6&     &   &       \\\\
%atactic poly(propylene)%
PP \\ 
atactic \cite{arrighi01,annis99}  &    1744&    421&  28.8&  1995  &  448  &       30.0\\
%isotactic poly(propylene)%
isotactic \cite{tak82} &     &    &   &     &   &   28.5\\
%head-to-head poly(propylene)% 
head-to-head  \cite{perez,annis99} &    1242 (\S)&    197 (\S)&  1.4&    &   &     28.0\\\\
%LCP1
DMB \cite{arr99} &  842 & 361 &   40 &   &  &    \\\\
%LCP2
DHMS \cite{arr99} &  1071 & 385 &   40 &   &  &   \\\\
%bisphenolA
PSF \cite{fer05} &  2553 & 396 & 4.1   &  2727 ; $W^c_3 = 522$& 406 &    36.7, 41.6 (\dag)\\\\
PH  \cite{fer05} &  2553 & 373 & 8.3   &  2727 ; $W^c_3 = 522$& 383 &    36.7, 41.6 (\dag)\\\\
PC  \cite{fer05} &  2553 & 379 & 10.4  &  2727 ; $W^c_3 = 522$& 390 &    36.7, 41.6 (\dag)\\\\
%Poly(methyl methacrylate) (ester)% 
PMMA (ester-methyl) \\
syndiotactic \cite{isotprb,pmmamacro} &  710&  241&   4.8&  820 ; $\langle V_6 \rangle = 90$&  265&  22.4\\
S1: (s:i:h) = (50:10:40) \cite{pillay} &     529&    192&   1.5&     &    &      \\
blend isotactic/S1  \\
non-sc (i-methyl) \cite{pillay} &     770&    313&   4.8&    &     &     \\
sc (i-methyl)  \cite{pillay} &     686&    205&   2.8&     &    &     \\
sc (S1-methyl) \cite{pillay} &     553&   180&   1.5&     &    &      \\\\
%Poly(methyl methacrylate) (alpha)%
PMMA ($\alpha$-methyl) \\
isotactic \cite{allen74} &  &   &  &     &    &         37.2\\
syndiotactic \cite{allen74} & &   &  &     &     &      44.6\\\\
%P-alpha-MS
P$\alpha$MS \\
syndiotactic \cite{allen74}  & &    &  &     &     &      47.1\\
atactic      \cite{allen74}  & &    &  &     &     &      47.1\\
head-to-head \cite{allen74}  & &    &  &     &     &      37.2\\\\
%poly(propylene oxide)% 
PPO \cite{hig72,russina} &    1684 (\ddag)&   &  &   &     &        28.3\\\\
%Polyisobutilene% 
PIB \cite{annis99,ada05,frickpib} &     &    &   &     &    &      40.0 \\\\\\
%blend pvme/ps
blend PVME/PS \\
(65:35) \cite{mukhoblend} &     1010 &   278&  7.8&     &     &      \\
(20:80) \cite{mukhoblend} &     1010 &   354&  7.8&     &     &      \\\\
%blend hhPP/PEP
blend hh-PP/PEP \\
(50:50; hh-PP-methyl) \cite{perez} & 1242&    197&   1.4&     &    &     \\\\
%polytypes hexatri
n-hexatriacontane \\
Mon (single-layered) \cite{kubota05} &      &    &   &     &     &      30.0 \\
Orth II (double-layered) \cite{kubota05} &  &    &   &     &     &      29.4 \\\\
%\hline
\end{tabular}
\vspace{4 mm}
\caption{Values of the RRDM parameters for the polymers investigated so far by neutron scattering techniques.
$E_A$, $\sigma_E$ and $\Gamma_\infty$ are obtained from the analysis of high temperature data.
Typical error bars are about a 10 \% for the two formers and a 30 \% for the latter.
$E_{01}^{max}$ denotes the maximum of the experimental librational peak.
The parameters $\langle V_3 \rangle$ and $\sigma_V$ of the barrier distribution $g(V_3)$ are given
in the cases where the quantities $E_A$, $\sigma_E$ and $E_{01}^{max}$ are known,
and a good correspondence between them
is possible by assuming a purely threefold potential. In the cases of PVAc and syndiotactic ($s$) PMMA
also the tunnelling and crossover regimes have been investigated. For PVAc a good description
of the latters is  possible with the given RRDM parameters, but for $s$-PMMA sixfold corrections
with ratio $V_6/V_3 = 0.11$ are necessary  (see Section \ref{sec:summexper}). $\langle V_6 \rangle$
is indicated. In the case of PSF, PH and PC, a non-distributed coupling term is necessary to reproduce
the librational double-peak (see Section \ref{sec:summexper}). The coupling term $W_c$ is indicated.
For the systems where hopping or librational data are absent, there is no sufficient information
to determine the shape of the potential, and the parameters of $g(V_3)$
are not given. \\
($\ast$): These parameteres, obtained from an analysis of hopping data, 
are uncompatible with both the experimental
librational peak and the tunnelling spectrum (see Section \ref{sec:summexper}).\\
($\S$): In a purely threefold approach, these parameters provide a theoretical librational maximum
of $\approx$ 26 meV, close but not fully satisfactory with the experimental value (see Section \ref{sec:origin}).\\
($\dag$): Double-peak structure. \\
($\ddag$): The standard deviation $\sigma_E$ is not reported. In a threefold approach,
the corresponding librational energy for the given $\langle E_A \rangle$
is 29.7 meV, to be compared with the experimental maximum. }
\label{table:rrdm}
\end{table}

%\begin{left}

%%
\twocolumn

\end{document}